\documentclass[12pt]{article}
\pdfoutput=1
\usepackage[a4paper]{geometry}
\usepackage{jheppub, amsmath,amssymb,amsfonts,amsxtra, mathrsfs, makeidx,graphics,graphicx,amsthm,epsfig, ytableau,bm,longtable,float, color,tikz,mathtools,xfrac,footnote,rotating, lscape, makecell, environ,mathtools, empheq}

\usetikzlibrary{positioning}
\usetikzlibrary{chains}
\usetikzlibrary{arrows,fit,decorations.pathreplacing}
\tikzstyle{every picture}+=[remember picture]
\tikzstyle{na} = [baseline]

\usetikzlibrary{arrows, decorations.markings, calc, fadings, decorations.pathreplacing, patterns, decorations.pathmorphing, positioning}

\def\node#1#2{\overset{#1}{\underset{#2}{{\color{gray} \bullet}}}}

\def\node#1#2{\overset{#1}{\underset{#2}{\circ}}}

\tikzstyle{every picture}+=[remember picture]
\tikzstyle{na} = [baseline=-.5ex]

\usepackage{bbm}
\usepackage{subfigure}

\newcommand{\eg}{\textit{e.g.}}

\newcommand{\ie}{\textit{i.e.}}

\numberwithin{equation}{section}

\newcommand{\be}{\begin{equation}} \newcommand{\ee}{\end{equation}}
\newcommand{\bea}{\begin{equation} \begin{aligned}} \newcommand{\eea}{\end{aligned} \end{equation}}

\def\tilde{\widetilde}

\def\bar{\overline}

\def\rt2{\sqrt{2}}

\def\tr{\mathop{\rm tr}}

\def\CC{{\cal C}}

\def\CH{{\cal H}}

\def\CN{{\cal N}}
\def\CO{{\cal O}}

\def\CT{{\cal T}}

\def\1{{\ds 1}}

\newcommand{\cC}{\mathcal{C}}

\newcommand{\cH}{\mathcal{H}}

\def\O{\mathrm{O}}

\def\repa{\raise4pt\hbox{$\square$}\mkern-14mu\raise-4pt\hbox{$\square$}}
\def\repab{\overline{\raise4pt\hbox{$\square$}\mkern-14mu\raise-4pt\hbox{$\square$}\mkern-1mu}}

\def\smileface{\ensuremath{\hbox{\large$\bigcirc$}\mkern-15mu\raise-1pt\hbox{\scriptsize$\smallsmile$}%
\mkern-10mu\raise4pt\hbox{..}\mkern4mu}}
\def\frownface{\ensuremath{\hbox{\large$\bigcirc$}\mkern-15mu\raise-1pt\hbox{\scriptsize$\smallfrown$}%
\mkern-10mu\raise4pt\hbox{..}\mkern4mu}}

\newcommand{\ba}{\begin{array}}
\newcommand{\ea}{\end{array}}
\newcommand{\bi}{\begin{itemize}}
\newcommand{\ei}{\end{itemize}}
\def\vec#1{\bm{#1}}
\def\bea#1\eea{\allowdisplaybreaks \begin{align}#1\end{align}}
 \newcommand{\ben}{\begin{enumerate}}
\newcommand{\een}{\end{enumerate}}
\newcommand{\bean}{\begin{eqnarray*}}
\newcommand{\eean}{\end{eqnarray*}}
\newcommand{\eref}[1]{(\ref{#1})}

\newcommand{\PE}{\mathop{\rm PE}}

\newcommand{\BC}{\mathbb{C}}

\newcommand{\BZ}{\mathbb{Z}}

\newcommand{\BH}{\mathbb{H}}

\newcommand{\comment}[1]{}

\definecolor{light-gray}{gray}{0.7}

\newcommand{\blue}{\color{blue}}

\newcommand{\red}{\color{red}}

\def\aup#1 {\overset{#1}{\uparrow} \, \overset{\tilde{#1}}{\downarrow}}

\newcommand{\Ot}{\mathrm{O3}}
\newcommand{\Ott}{\tilde{\mathrm{O3}}}
\newcommand{\Of}{\mathrm{O5}}
\newcommand{\ON}{\mathrm{ON}}
\tikzset{snake it/.style={decorate, decoration={snake, amplitude=.4mm, segment length=2mm,
                       post length=0mm,pre length=0mm}}}

\title{Variations on $S$-fold CFTs}
\author[a,b]{Ivan Garozzo,} 
\author[a,b]{Gabriele Lo Monaco,} 
\author[b,c]{and Noppadol Mekareeya}
\affiliation[a]{Dipartimento di Fisica, Universit\`a di Milano-Bicocca, \\ Piazza della Scienza 3, I-20126 Milano, Italy}
\affiliation[b]{INFN, sezione di Milano-Bicocca, \\Piazza della Scienza 3, I-20126 Milano, Italy}
\affiliation[c]{Department of Physics, Faculty of Science, \\
Chulalongkorn University, Phayathai Road, \\
Pathumwan, Bangkok 10330, Thailand}
\emailAdd{ivangarozzo@gmail.com}
\emailAdd{gabriele.lomonaco92@gmail.com}
\emailAdd{n.mekareeya@gmail.com}
\abstract{A local $SL(2, \mathbb{Z})$ transformation on the Type IIB brane configuration gives rise to an interesting class of superconformal field theories, known as the $S$-fold CFTs.  Previously it has been proposed that the corresponding quiver theory has a link involving the $T(U(N))$ theory.  In this paper, we generalise the preceding result by studying quivers that contain a $T(G)$ link, where $G$ is self-dual under $S$-duality.  In particular, the cases of $G = SO(2N)$, $USp'(2N)$ and $G_2$ are examined in detail. We propose the theories that arise from an appropriate insertion of an $S$-fold into a brane system, in the presence of an orientifold threeplane or an orientifold fiveplane.  By analysing the moduli spaces, we test such a proposal against its $S$-dual configuration using mirror symmetry. The case of $G_2$ corresponds to a novel class of quivers, whose brane construction is not available.  We present several mirror pairs, containing $G_2$ gauge groups, that have not been discussed before in the literature.}

\begin{document}
\maketitle
\section{Introduction}
Actions of the group $SL(2,\BZ)$ and dualities between three dimensional gauge theories have been a long-standing subject in quantum field theory.  A notable example of such dualities is mirror symmetry \cite{Intriligator:1996ex}, which corresponds to an operation of the generator $S$, such that $S^2=-1$, of $SL(2,\BZ)$ \cite{Kapustin:1999ha, Witten:2003ya}. From the string theoretic perspective, mirror symmetry has many interesting realisations \cite{deBoer:1996mp, Porrati:1996xi, Hanany:1996ie}.  One of which involves applying the $S$-transformation on a Type IIB brane system, known as the Hanany--Witten configuration, consisting of D3, NS5 and D5 branes preserving eight supercharges \cite{Hanany:1996ie}.  This realisation does not only allow for the construction of a number of interesting mirror pairs, it also provides for several variations of the models, such as the inclusion of an orientifold plane into the brane system \cite{Kapustin:1998fa, Hanany:1999sj, Feng:2000eq, Gaiotto:2008ak}.   Along with $S$, the group $SL(2,\BZ)$  has another generator, usually denoted by $T$, that obeys $(ST)^3=1$.  The operation by $T$ shifts the Chern--Simons level of the background gauge field \cite{Witten:2003ya, Gaiotto:2008ak}.  In terms of branes, the action $T^k$ transforms an NS5 brane into a $(1,k)$ fivebrane \cite{Bergman:1999na, Aharony:1997bh}.  

An $SL(2, \BZ)$ transformation can be applied {\it locally} on the Type IIB brane system in the following sense \cite{Gaiotto:2008ak, Gulotta:2011si, Assel:2014awa}.  For example, under the $S$-transformation, a $(p,q)$ fivebrane transforms into a $(-q,p)$ fivebrane.  It was pointed out in \cite{Gaiotto:2008ak, Gulotta:2011si} that we may trade a $(p,q)$ fivebrane in a given brane system for a $(-q, p)$ fivebrane with an $S$-duality wall on its right and $(S^{-1})$-duality wall on its left.  Indeed, the duality walls define the boundaries of the region of the local $SL(2,\BZ)$ action, and at the same time one may regard them as the new object in the brane configuration.  As suggested in \cite{Gaiotto:2008ak}, the intersection between an $S$-duality wall and a stack of $N$ D3 branes gives rise to a $T(U(N))$ theory\footnote{We shall not review about the $T(U(N))$ theory here.  The reader is referred to \cite{Gaiotto:2008ak, Assel:2014awa, Garozzo:2018kra} for further details.} coupling between two $U(N)$ groups, where $T(U(N))$ can be regarded as a 3d $\CN=4$ superconformal field theory on the Janus interface interface in 4d $\CN=4$ super--Yang--Mills \cite{Gaiotto:2008sd, Gaiotto:2008ak}.

This idea leads to a new class of conformal field theories (CFTs) in three dimensions, known as the $S$-fold CFTs \cite{Assel:2018vtq}.  From the brane perspective, we may insert such a duality wall into a D3 brane interval of the Hanany--Witten configuration.  For the duality wall associated with the $SL(2,\BZ)$ element $J=-ST^k$, the corresponding field theory can be described by a quiver diagram that contains the $T(U(N))$ theory connecting two $U(N)$ gauge nodes, with the Chern--Simons levels $k$ and $0$.

Such CFTs admit an interesting gravity dual.  The latter involves $\mathrm{AdS}_4 \times K_6$ Type IIB string solutions with monodromies\footnote{A similar solution in $\mathrm{AdS}_5$ was considered in \cite{Garcia-Etxebarria:2015wns, Aharony:2016kai}, and those in $\mathrm{AdS}_3$ were considered in \cite{Couzens:2017way, Couzens:2017nnr}.} in $K_6$ in the group $SL(2,\BZ)$. These solutions were obtained by applying the corresponding $SL(2,\BZ)$ quotient on the solutions associated with the holographic dual of Janus interfaces in 4d $\CN=4$ SYM \cite{DHoker:2007zhm, DHoker:2007hhe}.   In fact, such a construction for abelian gauge theories was studied in \cite{Ganor:2014pha}, and the supergravity solution corresponding to such a duality wall (dubbed the $S$-fold solution) was studied in \cite{Inverso:2016eet}.   Several related realisations of duality walls in 4d $\CN=4$ SYM with $SL(2,\BZ)$ monodromies can also be found in \cite{Martucci:2014ema, Gadde:2014wma, Assel:2016wcr, Lawrie:2018jut}.  Moreover, it is worth pointing out that quivers containing the $T(SU(N))$ theory as a component were discussed in \cite{Terashima:2011qi, Gang:2015wya}.  In this paper, we shall use the term $S$-fold and $S$-duality wall interchangeably.

The moduli space of three dimensional $S$-fold CFTs was studied extensively in \cite{Garozzo:2018kra}.  One of the main results is that the vector multiplets of the $U(N)$ gauge nodes, with zero Chern--Simons levels, that are connected by a $T(U(N))$ theory do not contribute the Coulomb branch.  We shall henceforth dub this result the ``freezing rule''.  In terms of branes, the freezing rule implies that the brane segment that intersects the $S$-fold cannot move along the Coulomb branch direction, but gets stuck at a given position.  This result has been tested using mirror symmetry, whereby the the mirror configuration was obtained by applying the $S$-operation on the original brane configuration.  We find that the Higgs branch ({\it resp.} Coulomb branch) of the original theory gets exchanged with the Coulomb branch ({\it resp.} Higgs branch) of the mirror theory, consistent with the freezing rule and mirror symmetry.

A natural question that arises from the study of \cite{Garozzo:2018kra} is whether we can replace the $T(U(N))$ link in the quiver by $T(G)$, where $G$ is a group that is not $U(N)$.  In order for $T(G)$ to be invariant under the $S$-action, $G$ has to be invariant under $S$-duality.  In this paper, we address this question by studying the cases in which $G$ is either $SO(2N)$, $USp'(2N)$ or $G_2$, and restrict the Chern--Simons levels of the gauge groups that are connected by $T(G)$ to zero.   

For $G$ being $SO(2N)$ and $USp'(2N)$, we propose that the corresponding theory can be realised from a brane construction that contains an intersection between an $S$-duality wall with the D3 brane segment on top of the orientifold threeplane of types $\Ot^-$ and $\tilde{\Ot}^+$ respectively.  In other words, the $S$-fold CFTs of this class can be obtained by inserting an $S$-duality wall into an appropriate D3 brane segment of the brane systems described in \cite{Feng:2000eq}. The mirror theory can be derived by first obtaining the $S$-dual configuration as discussed in \cite{Feng:2000eq}, and then insert an $S$-fold in the position corresponding to the original set-up.  We find that the moduli spaces of the original and mirror theories are consistent with the freezing rule and mirror symmetry.  This consistency also supports the existence of $S$-fold of the type $SO(2N)$ and $USp'(2N)$, and that the local $S$-operation can be consistently performed in the background of the $\Ot^-$ and $\tilde{\Ot}^+$ planes.

We also perform a similar analysis for the brane system that contains an orientifold fiveplane or its $S$-dual, which is also known as an $\ON$ plane.  In which case, the corresponding quiver may contain a hypermultiplet in the antisymmetric (or symmetric) representation, along with fundamental hypermultiplets, under the unitary gauge group, and the mirror quiver may contain a bifurcation \cite{Kapustin:1998fa, Hanany:1999sj}.  We find that the results are consistent with the freezing rule and mirror symmetry provided that the $S$-fold is not inserted ``too close'' to the orientifold plane and there must be a sufficient number of NS5 branes that separate the $S$-fold from the orientifold plane.  This suggests a consistency condition for the local $S$-action to be performed under the background of an orientifold fiveplane.

The class of theories that contain $G_2$ gauge groups is completely new and interesting.  To the best of our knowledge, the Type IIB brane construction for such theories is not available and mirror theories of this class of models have not been discussed in the literature.  In particular, we consider a family of quivers that contain alternating $G_2$ and $USp'(4)$ gauge groups, possibly with fundamental flavours under $USp'(4)$.  We propose that one can ``insert an S-fold'' into the $G_2$ and/or $USp'(4)$ gauge groups in the aforementioned quivers.  This results in the presence of the $T(G_2)$ link connecting two $G_2$ gauge groups, and/or the $T(USp'(4))$ link connecting two $USp'(4)$ gauge groups.  We also find that the mirror theory is also a quiver containing the $G_2$, $USp'(4)$ and possibly $SO(5)$ gauge groups if the original theory contains fundamental matter under $USp'(4)$.   We test, using the Hilbert series, that the moduli spaces of such theories are consistent with the freezing rule and mirror symmetry.  This, again, provides strong evidence for the existence of an ``$S$-fold of the type $G_2$''.

The paper is organised as follows. In section \ref{sec:Notations} we briefly review $T(G)$ theories and fix the notations that are adopted in the subsequent parts of the paper. In section \ref{sec:couplenilcone}, we study the hyperK\"ahler spaces that arise from coupling a nilpotent cone associated with a group $G$ to matter in the fundamental representation of $G$.  Such spaces have some interesting features and this notion turns out to be useful in the later sections because the nilpotent cone arises from the Higgs or Coulomb branch of the $T(G)$ theory. In section \ref{sec:O5}, we investigate quiver theories that arise from brane configuration with an $S$-fold in the background of the $\Of^-$ or the $\ON^-$ plane.  We provide the consistency conditions for the relative positions between the $S$-fold and the orientifold plane such that the moduli spaces of theories in question obey the freezing rule and mirror symmetry.   In section \ref{sec:O3} we study various models involving S-folds in the background of the $\Ot^-$ or the $\tilde{\Ot}^+$ planes.  The corresponding quivers contain a $T(SO(2N))$ link or a $T(USp'(2N))$ link between gauge nodes.  In section \ref{sec:G2}, we propose a new class of mirror pairs involving $G_2$ gauge nodes, as well as those with $T(G_2)$ link.  Finally, in Appendix \ref{app:O5plus}, we investigate the quivers that arise from the brane systems with $\Of^+$ or its $S$-dual $\ON^+$.  One of the features of the latter is that the quiver contains a ``double lace'', in the same way as that of the Dynkin diagram of the $C_N$ algebra.  Although this part of the quiver does not have a known Lagrangian description, one can still compute the Coulomb branch Hilbert series using the prescription given in \cite{Cremonesi:2014xha}.  We find that such a Coulomb branch agrees with the Higgs branch with the original theory, and for the theory with an $S$-fold the former also respects the freezing rule.

\section{Notations and conventions} \label{sec:Notations}
Let us state the notations and conventions that will be adopted in the subsequent parts of the paper.
\paragraph{Gauge and global symmetries.} In a quiver diagram, we denote the 3d $\CN=4$ vector multiplet in a given gauge group by a circular node, and a flavour symmetry by a rectangular node.  A black node with a label $n$ denotes the symmetry group $U(n)$, a blue node with an even label $m$ denotes the symmetry group $USp(m)$, and a red node with a label $k$ denotes the symetry group $O(k)$ or $SO(k)$.
\be
\begin{split}
U(n): \qquad &\begin{tikzpicture}[font=\footnotesize, baseline] \begin{scope}[auto,%
  every node/.style={draw, minimum size=0.8cm}, node distance=1cm]; 
  \node[circle] (0, 0) {$n$}; \node[rectangle] at (2,0) {$n$};  
  \end{scope}
  \end{tikzpicture} \\
USp(m): \qquad &\begin{tikzpicture}[font=\footnotesize, baseline] \begin{scope}[auto,%
  every node/.style={draw, minimum size=0.8cm}, node distance=1cm]; 
  \node[circle, blue] (0, 0) {$m$}; \node[rectangle, blue] at (2,0) {$m$};  
  \end{scope}
  \end{tikzpicture}  \qquad \text{with $m$ even}\\   
\text{$O(k)$ or $SO(k)$}: \qquad &\begin{tikzpicture}[font=\footnotesize, baseline] \begin{scope}[auto,%
  every node/.style={draw, minimum size=0.8cm}, node distance=1cm]; 
  \node[circle, red] (0, 0) {$k$}; \node[rectangle, red] at (2,0) {$k$};  
  \end{scope}
  \end{tikzpicture}
\end{split}
\ee
We shall be explicit whenever we would like to emphasise whether the group is $O(k)$ or $SO(k)$  In this paper we also deal with the group known as $USp'(2M)$, arising in the worldvolume $M$ physical D3 branes on the $\Ott^+$ plane \cite{Witten:1998xy}.  Note that under $S$-duality, $USp'(2M)$ transforms into itself.  This is in contrast to the group $USp(2M)$, arising in the worldvolume $M$ physical D3 branes on the $\Ot^+$ plane, where under the $S$-duality transforms into $SO(2M+1)$.
We denote the algebra corresponding to $USp'(M)$, with $M$ even, in the quiver diagram by a blue node with the label $M'$.  In the case that the brane configuration does not give a clear indication whether the group is $USp(M)$ or $USp'(M)$, we simply denote the label in the corresponding blue node by $M$.

\paragraph{The $T(G)$ theory.} In the following, we also study the 3d $\CN=4$ superconformal theory, known as $T(G)$, arising from a half BPS domain wall in the 4d $\CN=4$ super-Yang-Mills theory with gauge group $G$ that is self-dual under $S$-duality \cite{Gaiotto:2008ak}.  In this paper, we focus on $G= U(N), \, SO(2N),\, USp'(2N), \, G_2$.  The quiver descriptions for $T(U(N))$ and $T(SO(2N))$ are given in \cite{Gaiotto:2008ak}, whereas that for $T(USp'(2N))$ are given by \cite[sec. 2.5]{Cremonesi:2014uva}. 
The $T(G)$ theory has a global symmetry $G \times G$.  The Higgs and the Coulomb branches are both equal to the nilpotent cones $\CN_g$, where $g$ is the Lie algebra associated with the group $G$.  We denote the theory $T(G)$ by a wiggly red line connecting two nodes, both labelled by $G$.  As an example, the diagram below denotes the $T(USp'(2N))$ theory, with the global symmetry $USp'(2N) \times USp'(2N)$ being gauged:
\be \label{Tuspprimedex}
\begin{tikzpicture}[font=\footnotesize, baseline] \begin{scope}[auto,%
  every node/.style={draw, minimum size=0.8cm}, node distance=1cm]; 
  \node[circle, blue] (n1) at (0, 0) {$2N'$}; \node[circle, blue] (n2) at (3,0) {$2N'$};  
  \end{scope}
  \draw[draw=red,snake it,thick,-]  (n1) to  (n2) node at (1.5,0.3) {\scriptsize \red $T(USp'(2N))$} ; 
  \end{tikzpicture}
\ee
Furthermore, we can couple this theory to half-hypermultiplets in the fundamental representations of such $USp'(2N)$ gauge groups.  For example, if we have $m_1$ and $m_2$ flavours of fundamental hypermultiplets under the left and the right gauge groups of \eref{Tuspprimedex} respectively, the corresponding flavour symmetry algebras are $so(2m_1)$ and $so(2m_2)$, and the quiver diagram reads
\be  \label{m1m2flv}
\begin{tikzpicture}[font=\footnotesize, baseline] \begin{scope}[auto,%
  every node/.style={draw, minimum size=0.8cm}, node distance=1cm]; 
  \node[rectangle, red] (f1) at (-3, 0) {$2m_1$}; \node[rectangle, red] (f2) at (6,0) {$2m_2$};  
  \node[circle, blue] (n1) at (0, 0) {$2N'$}; \node[circle, blue] (n2) at (3,0) {$2N'$};  
  \end{scope}
  \draw[draw=red,snake it,thick,-]  (n1) to  (n2) node at (1.5,0.3) {\scriptsize \red $T(USp'(2N))$}  ; 
  \draw[draw,solid,thick,-]  (n1) to  (f1) ; 
  \draw[draw,solid,thick,-]  (n2) to  (f2) ; 
  \end{tikzpicture}
\ee
  
\paragraph{Brane configurations.}  In this paper, we use brane systems involving D3, D5, NS5 branes, possibly with orientifold planes, that preserve eight supercharges \cite{Hanany:1996ie, Kapustin:1998fa, Hanany:1999sj, Feng:2000eq, Gang:2015wya}.  Each type of branes spans the following directions:
\be
\begin{tabular}{c||cccccc|c|cccc}
\hline
~     & 0 & 1 &2 &3 & 4 & 5 & 6 & 7 &8 &9 &\\
\hline
D3, $\mathrm{O3}$   & X & X & X&  &  &   & X&  & & & \\ 
NS5, $\mathrm{O5}$ & X & X & X& X & X & X &    &   & & &\\
D5   & X & X & X&  &  &  &  &X &X &X & \\
\end{tabular}
\ee
The $x^6$ direction can be taken to be compact or non-compact.

\section{Coupling hypermultiplets to a nilpotent cone} \label{sec:couplenilcone}
In this section we study the hyperK\"ahler space that arises from coupling hypermultiplets or half-hypermultiplets to nilpotent cone $\CN_g$ of the Lie algebra $g$ associated with a gauge group $G$.  We start from the nilpotent cone of $g$, and denote this geometrical object by
\be 
\begin{tikzpicture}[font=\footnotesize, baseline]
\begin{scope}[auto,%
  every node/.style={draw, minimum size=0.8cm}, node distance=1cm];
\node[rectangle] (USp) at (0, 0) {$G$};
\node[draw=none, right=of USp] (cr) {} ;
\end{scope}
\draw[draw=red,solid, snake it,thick,-]  (USp) to  node[pos=1] {\normalsize \red{$\times$}}  (cr) ; 
\end{tikzpicture}
\ee
Note that a subgroup of $G$ may acts trivially on $\CN_g$.  For example, we may take $G$ to be $U(N)$; since the symmetry of the corresponding nilpotent cone is really $SU(n)$, the $U(1)$ subgroup of $G=U(N)$ acts trivially on the nilpotent cone.

The symmetry $G$ can be gauged and can then be coupled to hypermultiplets or half-hypermultiplets, which give rise to a flavour symmetry $H$.  We denote the resulting theory by the quiver diagram:
\be  \label{generalgh}
\begin{tikzpicture}[font=\footnotesize, baseline]
\begin{scope}[auto,%
  every node/.style={draw, minimum size=0.8cm}, node distance=1cm];
\node[circle] (USp) at (0, 0) {$G$};
\node[rectangle, left=of USp] (SO) {$H$};
\node[draw=none, right=of USp] (cr) {} ;
\end{scope}
\draw[draw,solid,thick,-]  (USp) to   (SO) ; 
\draw[draw=red,solid, snake it,thick,-]  (USp) to  node[pos=1] {\normalsize \red{$\times$}}  (cr) ; 
\end{tikzpicture}
\ee
The hyperK\"ahler quotient $\CH_{\eref{generalgh}}$ associated with this diagram is
\be \label{HKgen}
\CH_{\eref{generalgh}} = \frac{\CH \left(  [H]-[G] \right) \times \CN_{g}}{G}
\ee
where $\CH \left(  [H]-[G] \right)$ denotes the Higgs branch of quiver $[H]-[G]$.  We emphasise that we do not interpret \eref{generalgh} as a field theory by itself.  Instead, we regard it as a notation that can be conveniently used to denote the hyperK\"ahler quotient \eref{HKgen}.  This notation will turn out to be very useful in the subsequent part of the paper.


\subsection{$G=U(N)$ and $H=U(n)/U(1)$}
We take $G=U(N)$ and couple $n$ flavours of hypermutiplets to $G$:
\be  \label{suNsun}
\begin{tikzpicture}[font=\footnotesize, baseline]
\begin{scope}[auto,%
  every node/.style={draw, minimum size=1cm}, node distance=1cm];
\node[circle] (USp) at (0, 0) {$N$};
\node[rectangle, left=of USp] (SO) {$n$};
\node[draw=none, right=of USp] (cr) {} ;
\end{scope}
\draw[draw,solid,thick,-]  (USp) to   (SO) ; 
\draw[draw=red,solid, snake it,thick,-]  (USp) to  node[pos=1] {\normalsize \red{$\times$}}  (cr) ; 
\end{tikzpicture}
\ee
The hyperK\"ahler quotient associated with this diagram is
\be \label{quotientu}
\CH_{\eref{suNsun}}  = \frac{\CH \left(  [U(n)]-[U(N)] \right) \times \CN_{su(N)}}{U(N)}
\ee
where $\CH \left([U(n)]-[U(N)] \right)$ denotes the Higgs branch of the quiver $[U(n)]-[U(N)]$. The quaternionic dimension is
\be \label{dimquotientu}
\dim_{\BH} \CH_{\eref{suNsun}} = \frac{1}{2}N(N-1) + n N - N^2~.
\ee
The flavour symmetry in this case is $H=U(n)/U(1)$, whose algebra is $h=su(n)$.

For $N=1$, $\CN_{su(N)}$ is trivial.  The quotient \eref{quotientu} becomes the Higgs branch of the $U(1)$ gauge theory with $n$ flavours. $\CH_{\eref{suNsun}}$, therefore, turns out to be the closure of the minimal nilpotent orbit of $su(n)$, denoted by $\bar{\CO}_{(2,1^{n-2})}$ \cite{Hanany:2016gbz, Cabrera:2018ldc}.  This space is also isomorphic to the Higgs branch of the $T_{(n-1,1)}(SU(n))$ theory of \cite{Gaiotto:2008ak}, and is also isomorphic to the reduced moduli space of one $su(n)$ instanton on $\BC^2$.  It is precisely $n-1$ quaternionic dimensional.

For $N=2$, it turns out that $\CH_{\eref{suNsun}}$ is the closure $\bar{\CO}_{(3,1^{n-3})}$ of the orbit $(3,1^{n-3})$ of $su(n)$.  This is isomorphic to the Higgs branch of the $T_{(n-2,1^{2})}(SU(n))$ theory, namely that of the quiver $[U(n)]-(U(2))-(U(1))$.  The quaternionic dimension of this is precisely $2n-3$.  This is indeed in agreement with \eref{dimquotientu}.

For a general $N$, such that $n \geq N+1$, we see that $\CH_{\eref{suNsun}}$ is in fact
\be \label{orbsuNsun}
\CH_{\eref{suNsun}} = \bar{\CO}_{(N+1,1^{n-N-1})}~,
\ee
and in the special case of $n=N$, we have the nilpotent cone of $su(N)$:
\be\label{orbsuNsunspecial}
\CH_{\eref{suNsun}}|_{n=N} = \bar{\CO}_{(N)} = \CN_{su(N)}~.
\ee

One way to verify this proposition is to compute the Hilbert series of $\CH_{\eref{suNsun}}$.  This is given by\footnote{The plethystic exponential (PE) of a multivariate function $f(x_1,x_2, \ldots, x_n)$ such that $f(0,0,\ldots, 0)=0$ is defined as $\PE[f(x_1,x_2, \ldots, x_n)] = \exp \left( \sum_{k=1}^\infty \frac{1}{k} f(x_1^k, x_2^k,\ldots, x_n^k) \right).$}
\be \label{HSorb}
\begin{split}
H[\CH_{\eref{suNsun}}](t; \vec x) &= \int {\mathrm{d}} \mu_{SU(N)} (\vec z)  \oint_{|q|=1} \frac{dq}{2\pi i q} \PE\Big[ \chi^{su(N)}_{[1,0,\ldots,0]} (\vec x) \chi^{su(N)}_{[0,\ldots,0,1]} (\vec z) q^{-1}t  \\
& \qquad +  \chi^{su(N)}_{[0,\ldots,0,1]} (\vec x) \chi^{su(N)}_{[1,0,\ldots,0]} (\vec z) q -  \chi^{su(N)}_{[1,0,\ldots,0,1]} t^2\Big]  H[ \CN_{su(N)}](t, \vec z)\\
\end{split}
\ee
where $\vec x$ denotes the flavour fugacities of $su(N)$ and ${\mathrm{d}} \mu_{SU(N)} (\vec z)$ denotes the Haar measure of $SU(N)$.  We refer the reader to the detail of the characters and the Haar measures in \cite{Hanany:2008sb}.  The Hilbert series of the nilpotent cone of $su(N)$ was computed in \cite{Hanany:2011db} and is given by
\be
 H[ \CN_{su(N)}](t, \vec z) = \PE \left[ \chi^{su(N)}_{[1,0,\cdots,0,1]} (\vec z) t^2 - \sum_{p=2}^N t^{2p}  \right]~.
\ee
The Hilbert series \eref{HSorb} can then be used to checked against the results presented in \cite{Hanany:2016gbz}.  In this way, the required nilpotent orbits in \eref{orbsuNsun} and \eref{orbsuNsunspecial} can be identified.  This technique can also be applied to other gauge groups, as will be discussed in the subsequent subsections.  For the sake of brevity of the presentation, we shall not go through further details.

We remark that for $n\geq 2N+1$, the hyperK\"ahler space \eref{orbsuNsun} is isomorphic the Higgs branch of the $T_{(n-N, 1^{N})}(SU(n))$ theory\footnote{The partition $(n-N, 1^{N})$ is indeed the transpose of the partition $(N+1,1^{n-N-1})$ in \eref{orbsuNsun}.}, which corresponds to the quiver \cite{Gaiotto:2008ak}:
\be \label{TnN1}
T_{(n-N, 1^{N})}(SU(n)): \quad [U(n)]-(U(N))-(U(N-1))-\cdots-(U(1))~.
\ee
Note that quiver \eref{suNsun} can be obtained from \eref{TnN1} simply by replacing the wiggly line by the quiver tail as follows:
\be 
\begin{tikzpicture}[font=\footnotesize, baseline]
\begin{scope}[auto,%
  every node/.style={draw, minimum size=0.6cm}, node distance=1cm];
\node[circle] (USp) at (0, 0) {$N$};
\node[draw=none, right=of USp] (cr) {} ;
\end{scope}
\draw[draw=red,solid, snake it,thick,-]  (USp) to  node[pos=1] {\normalsize \red{$\times$}}  (cr) ; 
\end{tikzpicture}
\quad \longrightarrow \quad
(U(N))-(U(N-1))- \cdots - (U(1))~.
\ee

\subsection{$G=USp(2N)$ and $H=O(n)\,\,\text{or}\,\, SO(n)$}
We take $G=USp(2N)$ and couple $n$ half-hypermultiplets to $G$: 
\be  \label{usp2Nson}
\begin{tikzpicture}[font=\footnotesize, baseline]
\begin{scope}[auto,%
  every node/.style={draw, minimum size=1cm}, node distance=1cm];
\node[circle, blue] (USp) at (0, 0) {$2N$};
\node[rectangle, left=of USp, red] (SO) {$n$};
\node[draw=none, right=of USp] (cr) {} ;
\end{scope}
\draw[draw,solid,thick,-]  (USp) to   (SO) ; 
\draw[draw=red,solid, snake it,thick,-]  (USp) to  node[pos=1] {\normalsize \red{$\times$}}  (cr) ; 
\end{tikzpicture}
\ee
The corresponding hyperK\"ahler quotient is
\be \label{HKGUSp2NHO}
\CH_{\eref{usp2Nson}} = \frac{\CH \left( [SO(n)]-[USp(2N)] \right) \times \CN_{usp(2N)} }{USp(2N)}~.
\ee
The dimension of this space is
\be
\begin{split}
\dim_\BH \CH_{\eref{usp2Nson}} &= nN +\frac{1}{2} \left[ \frac{1}{2}(2N)(2N+1) -N \right] -  \frac{1}{2}(2N)(2N+1) \\
&= N(n-N-1)~.
\end{split}
\ee
For $n \geq 2N+1$, the hyperK\"ahler quotient \eref{HKGUSp2NHO} turns out to be isomorphic to the closure of the nilpotent orbit $(2N+1, 1^{n-(2N+1)})$ of $so(n)$:
\be \label{clOson}
\CH_{\eref{usp2Nson}} = \bar{\CO}_{(2N+1, 1^{n-(2N+1)})}~.
\ee 

For even $n$, say $n=2m$, this is isomorphic to the Higgs branch of $T_{\rho}(SO(n))$, with $\rho = (n-2N-1,1^{2N+1})$,\footnote{Note that the partition $\rho=(n-2N-1,1^{2N+1})$ can be obtained from the partition $\lambda= (2N+1, 1^{n-(2N+1)})$ of \eref{clOson} by first computing the transpose of $\lambda$, and then performing the $D$-collapse.  For example, for $N=2$ and $m=4$ (or $n=8$),
\be
\lambda=(5,1^4) \quad \overset{\text{transpose}}{\longrightarrow} \quad (4,1^4) \quad \overset{\text{$D$-coll.}}{\longrightarrow} \quad \rho=(3,1^5)~.
\ee} 
whose quiver description is
\be 
\scalebox{0.8}
{\begin{tikzpicture}[font=\scriptsize, baseline] \begin{scope}[auto,%
  every node/.style={draw, minimum size=1.2cm}, node distance=1cm]; 
  \node[rectangle, red] (f1) at (0, 0) {$n$}; 
  \node[circle, blue] (n1) at (2, 0) {$2N$};
  \node[circle, red] (n2) at (4, 0) {$2N$};
  \node[circle, blue] (n3) at (6, 0) {$2N-2$};
    \node[circle, red] (n4) at (8, 0) {$2N-2$};
  \node[draw=none] (n5) at (10, 0) {{\Large $\cdots$}}; 
  \node[circle, blue] (n6) at (12,0) {$2$};
  \node[circle, red] (n7) at (14,0) {$2$};
  \end{scope}
  \draw[draw, solid,thick,-]  (f1) to  (n1) ; 
  \draw[draw,solid,thick,-]  (n1) to  (n2) ; 
  \draw[draw,solid,thick,-]  (n2) to  (n3) ; 
  \draw[draw,solid,thick,-]  (n3) to  (n4) ; 
  \draw[draw,solid,thick,-]  (n4) to  (n5) ;
  \draw[draw,solid,thick,-]  (n5) to  (n6) ; 
  \draw[draw,solid,thick,-]  (n6) to  (n7) ; 
  \end{tikzpicture}}
\ee
For odd $n$, say $n=2m+1$, this is isomorphic to the Higgs branch of $T_{\rho}(SO(n))$, with $\rho= (n-2N-1, 2, 1^{2N-2})$ if $n>2N+1$ and $\rho=(1^{2N})$ if $n=2N+1$,\footnote{Note that the partition $\rho=(n-2N-1, 2, 1^{2N-2})$ can be obtained from the partition $\lambda= (2N+1, 1^{n-(2N+1)})$ of \eref{clOson} by first computing the transpose of $\lambda$, subtracting $1$ from the last entry, and then performing the $C$-collapse.  For example, for $N=3$ and $m=4$ (or $n=9$),
\be
\lambda= (7, 1^{2}) \quad \overset{\text{transpose}}{\longrightarrow} \quad  (3,1^6) \quad \longrightarrow \quad (3,1^5) \quad \overset{\text{$C$-coll.}}{\longrightarrow} \quad (2^2,1^4)~.
\ee} 
whose quiver description is
\be 
\scalebox{0.8}
{
\begin{tikzpicture}[font=\scriptsize, baseline] \begin{scope}[auto,%
  every node/.style={draw, minimum size=1.2cm}, node distance=1cm]; 
  \node[rectangle, red] (f1) at (0, 0) {$n$}; 
  \node[circle, blue] (n1) at (2, 0) {$2N$};
  \node[circle, red] (n2) at (4, 0) {$2N-1$};
  \node[circle, blue] (n3) at (6, 0) {$2N-2$};
  \node[draw=none] (n4) at (8, 0) {{\Large $\cdots$}}; 
  \node[circle, blue] (n5) at (10,0) {$2$};
  \node[circle, red] (n6) at (12,0) {$1$};
  \end{scope}
  \draw[draw, solid,thick,-]  (f1) to  (n1) ; 
  \draw[draw,solid,thick,-]  (n1) to  (n2) ; 
  \draw[draw,solid,thick,-]  (n2) to  (n3) ; 
  \draw[draw,solid,thick,-]  (n3) to  (n4) ; 
  \draw[draw,solid,thick,-]  (n4) to  (n5) ;
  \draw[draw,solid,thick,-]  (n5) to  (n6) ; 
  \end{tikzpicture}}
\ee


\subsection{$G=SO(N)\, \text{or} \, O(N)$ and $H=USp(2n)$}
Let us first take $G=SO(N)$ and take $H=USp(2n)$.
\be  \label{soNusp2n}
\begin{tikzpicture}[font=\scriptsize, baseline]
\begin{scope}[auto,%
  every node/.style={draw, minimum size=1cm}, node distance=1cm];
\node[circle, red] (USp) at (0, 0) {$SO(N)$};
\node[rectangle, left=of USp, blue] (SO) {$2n$};
\node[draw=none, right=of USp] (cr) {} ;
\end{scope}
\draw[draw,solid,thick,-]  (USp) to   (SO) ; 
\draw[draw=red,solid, snake it,thick,-]  (USp) to  node[pos=1] {\normalsize \red{$\times$}}  (cr) ; 
\end{tikzpicture}
\ee
This diagram defines the hyperK\"ahler quotient
\be \label{HKsoNusp2n}
\CH_{\eref{soNusp2n}} = \frac{\CH \left( [USp(2n)]-[SO(N)] \right) \times \CN_{so(N)} }{SO(N)}~.
\ee
The quaternionic dimension of this quotient is
\be \label{dimHKsoNusp2n}
\dim_{\BH} \, \CH_{\eref{soNusp2n}} = \begin{cases} m(2n-m)~,   & N=2m \\ 
m(2n-m-1)+n~,  & N= 2m+1
\end{cases}~.
\ee
It is interesting to examine \eref{HKsoNusp2n} for a few special cases.  For $N=2n$ or $N=2n+1$ or $N=2n-1$, we find that \eref{HKsoNusp2n} is in fact the nilpotent cone $\CN_{usp(2n)}$ of $usp(2n)$, whose quaternionic dimension is $n^2$:
\be \label{specialusp2n}
\CH_{\eref{soNusp2n}}|_{N=2n} = \CH_{\eref{soNusp2n}}|_{N=2n\pm 1} = \CN_{usp(2n)}~.
\ee
This statement can be checked using the Hilbert series:
\be \label{HSspecialcaseNusp2n}
\begin{split}
H[\CH_{\eref{soNusp2n}}](t; \vec x) &= \int \mathrm{d}\mu_{SO(N)} (\vec z)  \PE\Big[ \chi^{C_n}_{[1,0,\ldots,0]}(\vec x)  \chi^{so(N)}_{[1,0,\ldots,0]}(\vec z) t   \\
& \qquad\qquad\qquad -  \chi^{so(N)}_{[0,1,0,\ldots,0]}(\vec z) t^2 \Big] H[ \CN_{so(N)}](t, \vec z) \\
&= \PE \left[ \chi^{C_n}_{[2,0,\ldots,0]}(\vec x) t^2 - \sum_{j=1}^n t^{4j} \right]~, \,\, \text{if $N=2n$ or $2n\pm1$}~.
\end{split}
\ee
where the Haar measure and the relevant characters are given in \cite{Hanany:2008sb}. The last line is indeed the Hilbert series of the nilpotent cone $\CN_{usp(2n)}$ \cite{Hanany:2016gbz}.

It is important to note that the quotient \eref{HKsoNusp2n} is not the closure of a nilpotent orbit in general.  For example, let us take $n=4$ and $N=3$, \ie~ $G=SO(3)$ and $H=USp(8)$.  The Hilbert series takes the form
\be
H[\CH_{\eref{soNusp2n}}|_{n=4, N=3}](t; \vec x)  = \PE \left[ \chi^{C_4}_{[2,0,0,0]} t^2 + (\chi^{C_4}_{[0,0,1,0]} + \chi^{C_4}_{[1,0,0,0]}) t^3 - t^4 + \ldots \right]~.
\ee
Observe that there are generators with $SU(2)_R$-spin $3/2$ in the third rank antisymmetric representation $\wedge^3 [1,0,0,0] = [0,0,1,0]+[1,0,0,0]$ of $USp(8)$. These should be identified as ``baryons''.   Using Namikawa's theorem \cite{Namikawa2018}, which states that all generators of the closure of a nilpotent orbit must have $SU(2)_R$-spin $1$ (see also \cite{Cabrera:2016vvv}), we conclude that $\CH_{\eref{soNusp2n}}|_{n=4, N=3}$ is not the closure of a nilpotent orbit.  In general, these baryons can be removed by taking gauge group to be $O(N)$, instead of $SO(N)$.  The reason is because the $O(N)$ group does not have an epsilon tensor as an invariant tensor, whereas the $SO(N)$ group has one.

Let us now take $G=O(N)$ and take $H=USp(2n)$:   
\be  \label{oNusp2n}
\begin{tikzpicture}[font=\scriptsize, baseline]
\begin{scope}[auto,%
  every node/.style={draw, minimum size=1cm}, node distance=1cm];
\node[circle, red] (USp) at (0, 0) {$O(N)$};
\node[rectangle, left=of USp, blue] (SO) {$2n$};
\node[draw=none, right=of USp] (cr) {} ;
\end{scope}
\draw[draw,solid,thick,-]  (USp) to   (SO) ; 
\draw[draw=red,solid, snake it,thick,-]  (USp) to  node[pos=1] {\normalsize \red{$\times$}}  (cr) ; 
\end{tikzpicture}
\ee
This diagram defines the hyperK\"ahler quotient
\be
\CH_{\eref{oNusp2n}} = \frac{\CH \left( [USp(2n)]-[O(N)] \right) \times \CN_{so(N)} }{O(N)}~.
\ee
The dimension of this hyperK\"ahler space is the same as \eref{dimHKsoNusp2n}.
This quotient turns out to be isomorphic to the closure of the following nilpotent orbit of $usp(2n)$:
\be \label{orbitsUSp2n}
\CH_{\eref{oNusp2n}} = \begin{cases} \bar{\CO}_{(N,2,1^{2n-N-2})} & \text{$N$ even} \\ \bar{\CO}_{(N+1,1^{2n-N-1})}  & \text{$N$ odd} \end{cases}~.
\ee
In the special case where $N=2n$, $N=2n-1$ or $N=2n+1$, we have
\be \label{specialusp2no}
\CH_{\eref{oNusp2n}} |_{N=2n} = \CH_{\eref{oNusp2n}} |_{N=2n \pm 1} = \bar{\CO}_{(2n)} = \CN_{usp(2n)}~,
\ee
which is the same as \eref{specialusp2n}.    

For even $N=2m$, $\CH_{\eref{oNusp2n}}$ is isomorphic to the Higgs branch of $T_\rho(USp(2n))$ theory, with $\rho = (2n-N+1,1^{N})$, whose quiver description is
\be
\scalebox{0.75}{
\begin{tikzpicture}[font=\scriptsize, baseline]
\begin{scope}[auto,%
  every node/.style={draw, minimum size=1.2cm}, node distance=1cm];
\node[rectangle, blue] (f) at (0,0) {$2n$};
\node[circle, red] (n1) at (2, 0) {$2m$};
\node[circle, blue] (n2) at (4, 0) {$2m-2$};
\node[circle, red] (n3) at (6, 0) {$2m-2$};
\node[circle, blue] (n4) at (8, 0) {$2m-4$};
\node[circle, red] (n45) at (10, 0) {$2m-4$};
\node[draw=none] (n5) at (12, 0) {{\large $\cdots$}};
\node[circle, blue] (n6) at (14, 0) {2};
\node[circle, red] (n7) at (16, 0) {2};
\end{scope}
\draw[draw,solid,thick,-]  (f) to  (n1) to (n2) to (n3) to (n4) to (n45) to (n5) to (n6) to (n7) ; 
\end{tikzpicture}}
\ee
On the other hand, for odd $N=2m+1$, $\CH_{\eref{oNusp2n}}$ is isomorphic to the Higgs branch of $T_\rho(USp'(2n))$ theory, with $\rho = (2n-N+1,1^{N-1})$, whose quiver description is
\be
\scalebox{0.75}{
\begin{tikzpicture}[font=\scriptsize, baseline]
\begin{scope}[auto,%
  every node/.style={draw, minimum size=1.2cm}, node distance=1cm];
\node[rectangle, blue] (f) at (0,0) {$2n$};
\node[circle, red] (n1) at (2, 0) {$2m+1$};
\node[circle, blue] (n2) at (4, 0) {$2m$};
\node[circle, red] (n3) at (6, 0) {$2m-1$};
\node[circle, blue] (n4) at (8, 0) {$2m-2$};
\node[draw=none] (n5) at (10, 0) {{\large $\cdots$}};
\node[circle, blue] (n6) at (12, 0) {2};
\node[circle, red] (n7) at (14, 0) {1};
\end{scope}
\draw[draw,solid,thick,-]  (f) to  (n1) to (n2) to (n3) to (n4) to (n5) to (n6) to (n7) ; 
\end{tikzpicture}}
\ee

\subsection{$G=G_2$ and $H=USp(2n)$}
We take $G=G_2$ and $H=USp(2n)$:   
\be  \label{usp2ng2}
\begin{tikzpicture}[font=\footnotesize, baseline]
\begin{scope}[auto,%
  every node/.style={draw, minimum size=1cm}, node distance=1cm];
\node[circle] (g) at (0, 0) {$G_2$};
\node[rectangle, left=of g, blue] (h) {$2n$};
\node[draw=none, right=of g] (cr) {} ;
\end{scope}
\draw[draw,solid,thick,-]  (g) to   (h) ; 
\draw[draw=red,solid, snake it,thick,-]  (g) to  node[pos=1] {\normalsize \red{$\times$}}  (cr) ; 
\end{tikzpicture}
\ee
This diagram defines the hyperK\"ahler quotient
\be \label{HKG2}
\CH_{\eref{usp2ng2}} = \frac{\CH \left( [USp(2n)]-[G_2] \right) \times \CN_{g_2} }{G_2}~.
\ee
For $n\geq 2$, the quaternionic dimension of this space is
\be \label{dimG2}
\dim_{\BH} \,  \CH_{\eref{usp2ng2}} = 7n+\frac{1}{2}(14-2)-14 = 7n -8~,
\ee
and the Hilbert series of \eref{HKG2} is given by
\be
\begin{split}
H[\CH_{\eref{HKG2}}] (t, \vec x) &= \int \mathrm{d} \mu_{G_2} (\vec z) \PE \Big[ \chi^{G_2}_{[1,0]}(\vec z) \chi^{usp(2n)}_{[1,0,\ldots,0]} (\vec x) t \\
& \qquad - \chi^{G_2}_{[0,1]}(\vec z) t^2  \Big] H[\CN_{g_2}](t,\vec z)~,
\end{split}
\ee
where the relevant characters and the Haar measure is given in \cite{Hanany:2008sb}, and the Hilbert series of the nilpotent cone of $G_2$ can be obtained from \cite[Table 4]{Hanany:2017ooe}. 
The special case of $n=2$ is particularly simple. The corresponding space is a complete intersection whose Hilbert series is
\be \label{HSusp2ng2}
H[ \CH_{\eref{usp2ng2}}|_{n=2} ](t; x_1, x_2) =\PE\left[ \chi^{C_2}_{[2,0]}(x_1, x_2) t^2 +\chi^{C_2}_{[1,0]}(x_1, x_2) t^3 - t^8 -t^{12} \right]~.
\ee
Note that $\CH_{\eref{usp2ng2}}$ is not the closure of a nilpotent orbit, due to the existence of a generator at $SU(2)_R$-spin $3/2$ and Namikawa's theorem.

The case of $n=1$ needs to be treated separately, since \eref{dimG2} becomes negative.  We claim that
\be \label{C2Z2g2n1}
 \CH_{\eref{usp2ng2}}|_{n=1} = \BC^2/\BZ_2 = \CN_{su(2)}~.
\ee
The reason is as follows.  Let us denote by $Q^i_a$ the half-hypermultiplets in the fundamental representation of the $G_2$ gauge group\footnote{The three independent invariant tensors for $G_2$ can be taken as (1) the Kronecker delta $\delta^{ab}$, (2) the third-rank antisymmetric tensor $f^{abc}$ and (3) the fourth-rank antisymmetric tensor $\tilde{f}^{abcd}$. See \eg~\cite{Giddings:1995ns} for more details.}, where $i,j,k=1,2$ are the $USp(2)$ flavour indices and $a,b,c,d=1, \ldots, 7$ are the $G_2$ gauge indices.  Let us also denote by $X_{ab}$ the generators of the nilpotent cone of $G_2$.  Transforming in the adjoint representation of $G_2$, $X_{ab}$ is an antisymmetric matrix satisfying\footnote{Using the identity $f^{[abc} f^{cde]} = \tilde{f}^{abde}$ (see \cite[(A.13)]{Giddings:1995ns}), it follows immediately from this relation that $\tilde{f}^{abde} X_{ab} X_{de} =0$.}
\be
f^{abc}X_{ab}=0~;
\ee
this is because $\wedge^2 [1,0] = [0,1]+[1,0]$.
Moreover, being the generators of the nilpotent cone, $X_{ab}$ satisfy
\be \label{relationsg2n1}
\begin{split}
\tr(X^2) = \delta^{ad} \delta^{bc} X_{ab} X_{cd}  =0~, \qquad \tr(X^6) = 0~.
\end{split}
\ee
The moment map equations for $G_2$ read
\be \label{momentmap}
\epsilon_{ij} Q^i_a Q^j_b = X_{ab}~.
\ee
The generators of \eref{HKG2}, for $n=1$, are
\be
M^{ij} = \delta_{ab} Q^i_a Q^i_b
\ee
transforming in the adjoint representation of $USp(2)$.  Note that baryons vanish: 
\be
f^{abc} Q^i_a Q^j_b Q^k_c =0~, \qquad \tilde{f}^{abcd} Q^i_a Q^j_b Q^k_c Q^l_d=0~,
\ee
because $i,j,k,l=1,2$.  Other gauge invariant combinations also vanish; for example, $X_{ab} Q^i_a Q^j_b$
has one independent component and it vanishes thanks to \eref{relationsg2n1} and \eref{momentmap}.  Furthermore, the square of $M$ vanishes:
\be
\epsilon_{il} \epsilon_{jk} M^{ij}M^{kl} = (\epsilon_{il}  Q^i_a  Q^l_b)( \epsilon_{jk} Q^j_a Q^k_b) \overset{\eref{momentmap}}{=} \tr(X^2) \overset{\eref{relationsg2n1}}{=} 0~.
\ee 
Therefore, we reach the conclusion \eref{C2Z2g2n1}.

\section{Models with orientifold fiveplanes} \label{sec:O5}
In this section, we consider models that arise from brane systems involving an $S$-fold and orientifold 5-planes.  For the latter, we focus on the case of the $\O5^-$ plane and postpone to discussion about the $\Of^+$ plane to Appendix \ref{app:O5plus}.  In the absence of the $S$-fold, such models and the corresponding mirror theories were studied in detailed in \cite{Hanany:1999sj, Gaiotto:2008ak}.  We start this section by reviewing the latter and then discuss the insertion of an $S$-fold in the subsequent subsections.

\subsection{The cases without an $S$-fold}
 We consider three types of models, depending on the presence of NS5 branes and their positions relative to each $\Of^-$ plane \cite{Hanany:1999sj}.

\paragraph{\it The $USp(2N)$ gauge theory with $n$ flavours.}  The quiver diagram is
\be \label{USp2Nwnflv}
\begin{tikzpicture}[baseline,font=\scriptsize]
\begin{scope}[auto,%
  every node/.style={draw, minimum size=0.8cm}, node distance=0.6cm];
\node[draw, circle, blue] (node1) at (0,0) {$2N$};
\node[draw, rectangle,red] (sqnode) at (2,0) {$2n$};
\end{scope}
\draw (node1)--(sqnode);
\end{tikzpicture}
\ee
The brane system for this quiver is
\be 
\scalebox{1}{
\begin{tikzpicture} [baseline=0, scale=0.9, transform shape]
\draw [ultra thick, black!40!green] (0,0)--(0,2.5) node[black,midway, xshift =-0.3cm, yshift=-1.5cm] {\footnotesize $\Of^-$}; 
\node[midway, color=black, xshift=1.5cm, yshift=2cm] {\scriptsize $\bullet$};
\node[midway, color=black, xshift=2cm, yshift=2cm] {\scriptsize $\bullet$};
\node[midway, color=black, xshift=2.5cm, yshift=2cm] {$~\ldots~$};
\node[midway, color=black, xshift=3cm, yshift=2cm] {\scriptsize $\bullet$};
\draw [decorate, decoration={brace}](1.4,2.2)--(3.1,2.2) node[black,midway,yshift=0.5cm] {\scriptsize $n~\text{physical D5s}$};
\draw [ultra thick,black!40!green] (4.5,0)--(4.5,2.5) node[black,midway, xshift =0.3cm, yshift=-1.5cm] {\footnotesize $\Of^-$};
\draw (0,1)--(4.5,1) node[black,midway, yshift=0.2cm] {\scriptsize $2N$} node[black,near start, yshift=-0.2cm] {\scriptsize D3};
\end{tikzpicture}}
\ee
\paragraph{\it The $U(2N)$ gauge theory with one or two rank-two antisymmetric hypermultiplets and $n$ flavours in the fundamental representation.}  The quiver diagrams are
\be \label{U2Nwantisym}
\begin{tikzpicture}[baseline,font=\scriptsize]
\begin{scope}[auto,%
  every node/.style={draw, minimum size=0.8cm}, node distance=0.6cm];
\node[draw, circle, black] (node1) at (0,0) {$2N$};
\draw[black] (node1) edge [out=45,in=135,loop,looseness=5]  (node1) node[draw=none] at (0,1.2) {$A$} ;
\node[draw, rectangle,black] (sqnode) at (2,0) {$n$};
\end{scope}
\draw (node1)--(sqnode);
\end{tikzpicture}
\qquad\qquad\qquad
\begin{tikzpicture}[baseline,font=\scriptsize]
\begin{scope}[auto,%
  every node/.style={draw, minimum size=0.8cm}, node distance=0.6cm];
\node[draw, circle, black] (node1) at (0,0) {$2N$};
\draw[black] (node1) edge [out=45,in=135,loop,looseness=5]  (node1) node[draw=none] at (0,1.2) {$A$} ;
\draw[black] (node1) edge [out=-45,in=-135,loop,looseness=5]  (node1) node[draw=none] at (0,-1.2) {$A'$} ;
\node[draw, rectangle,black] (sqnode) at (2,0) {$n$};
\end{scope}
\draw (node1)--(sqnode);
\end{tikzpicture}
\ee
The brane systems for the cases with one adjoint and two adjoints are, respectively, as follows:
\be \label{braneonetwoanti}
\scalebox{0.97}{
\begin{tikzpicture} [baseline, scale=1, transform shape]
\draw [ultra thick,black!40!green] (0,0)--(0,2.5) node[black,midway, xshift =-0.3cm, yshift=-1.5cm] {\footnotesize $\overset{\Of^-}{{\tiny \text{with an NS5 on top}}}$}; 
\draw [thick,black] (0.02,0)--(0.02,2.5);
\node[midway, color=black, xshift=0.7cm, yshift=2cm] {\scriptsize $\bullet$};
\node[midway, color=black, xshift=1.2cm, yshift=2cm] {\scriptsize $\bullet$};
\node[midway, color=black, xshift=1.7cm, yshift=2cm] {$~\ldots~$};
\node[midway, color=black, xshift=2.2cm, yshift=2cm] {\scriptsize $\bullet$};
\draw [decorate, decoration={brace}](0.6,2.2)--(2.3,2.2) node[black,midway,yshift=0.5cm] {\scriptsize $n~\text{physical D5s}$};
\draw [thick,black] (3,0)--(3,2.5) node[black,midway, xshift =0.3cm, yshift=-1.5cm] {\footnotesize NS5};
\draw (0,1)--(3,1) node[black,midway, yshift=0.2cm] {\scriptsize $2N$} node[black,near start, yshift=-0.2cm] {\scriptsize D3};
\end{tikzpicture}}
\qquad \qquad 
\scalebox{0.97}{
\begin{tikzpicture} [baseline, scale=1, transform shape]
\draw [ultra thick,black!40!green] (0,0)--(0,2.5) node[black,midway, xshift =-0.3cm, yshift=-1.5cm] {\footnotesize $\overset{\Of^-}{{\tiny \text{with an NS5 on top}}}$}; 
\draw [thick,black] (0.02,0)--(0.02,2.5);
\node[midway, color=black, xshift=0.7cm, yshift=2cm] {\scriptsize $\bullet$};
\node[midway, color=black, xshift=1.2cm, yshift=2cm] {\scriptsize $\bullet$};
\node[midway, color=black, xshift=1.7cm, yshift=2cm] {$~\ldots~$};
\node[midway, color=black, xshift=2.2cm, yshift=2cm] {\scriptsize $\bullet$};
\draw [decorate, decoration={brace}](0.6,2.2)--(2.3,2.2) node[black,midway,yshift=0.5cm] {\scriptsize $n~\text{physical D5s}$};
\draw [ultra thick, black!60!green] (3,0)--(3,2.5) node[black,midway, xshift =-0.3cm, yshift=-1.5cm] {\footnotesize $\overset{\Of^-}{{\tiny \text{with an NS5 on top}}}$}; 
\draw [thick,black] (3-0.02,0)--(3-0.02,2.5);
\draw (0,1)--(3,1) node[black,midway, yshift=0.2cm] {\scriptsize $2N$} node[black,near start, yshift=-0.2cm] {\scriptsize D3};
\end{tikzpicture}}
\ee
\paragraph{\it The $USp(2N) \times U(2N)^{m} \times USp(2 N)$ gauge theory with $(n_1, f_1, \ldots, f_m, n_2)$ flavours in the fundamental representations under each gauge group.}    The quiver diagram is
\be \label{USp2NU2NmUSp2N}
\begin{tikzpicture}[baseline,font=\scriptsize]
\begin{scope}[auto,%
  every node/.style={draw, minimum size=1cm}, node distance=0.6cm];
\node[circle, blue] (USp1) at (0, 0) {$2N$};
\node[circle, right=of USp1] (U1)  {$2N$};
\node[draw=none, right=of U1] (dots) {$\cdots$};
\node[circle, right=of dots] (U2)  {$2N$};
\node[circle, blue, right=of U2] (USp2) {$2N$};
\node[rectangle, red, below=of USp1] (n1) {$2n_1$};
\node[rectangle, below=of U1]  (f1) {$f_1$};
\node[rectangle, below=of U2]  (fm) {$f_m$};
\node[rectangle, red,  below=of USp2] (n2){$2n_2$};
\end{scope}
  \draw  (USp1)--(U1)--(dots)--(U2)--(USp2);
  \draw  (USp1)--(n1);
  \draw  (U1)--(f1);
  \draw  (U2)--(fm);
    \draw  (USp2)--(n2);
\end{tikzpicture}
\ee
The brane system for this quiver is
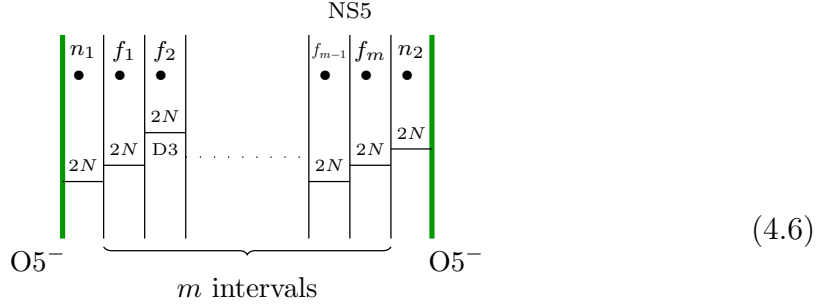
\be  \label{braneUSpUUUSp}
\scalebox{1.2}{
\begin{tikzpicture} [baseline=0, scale=0.9, transform shape]
\draw [ultra thick, black!40!green] (0,0)--(0,2.5) node[black,midway, xshift =-0.3cm, yshift=-1.5cm] {\footnotesize $\Of^-$}; 
\node[midway, color=black, xshift=0.2cm, yshift=2cm] {\scriptsize $\bullet$} node[black,midway, xshift=0.25cm, yshift=2.3cm] {\scriptsize $n_1$};
\node[midway, color=black, xshift=0.7cm, yshift=2cm] {\scriptsize $\bullet$} node[black,midway, xshift=0.75cm, yshift=2.3cm] {\scriptsize $f_1$};
\node[midway, color=black, xshift=1.2cm, yshift=2cm] {\scriptsize $\bullet$} node[black,midway, xshift=1.25cm, yshift=2.3cm] {\scriptsize $f_2$};
\node[midway, color=black, xshift=3.2cm, yshift=2cm] {\scriptsize $\bullet$} node[black,midway, xshift=3.25cm, yshift=2.3cm] {\scalebox{0.5}{$f_{m-1}$}} ;
\node[midway, color=black, xshift=3.7cm, yshift=2cm] {\scriptsize $\bullet$} node[black,midway, xshift=3.75cm, yshift=2.3cm] {\scriptsize $f_m$};
\node[midway, color=black, xshift=4.2cm, yshift=2cm] {\scriptsize $\bullet$} node[black,midway, xshift=4.25cm, yshift=2.3cm] {\scriptsize $n_2$};
\draw (0.5,0)--(0.5,2.5); \draw (1,0)--(1,2.5); \draw (1.5,0)--(1.5,2.5); \draw (3,0)--(3,2.5); \draw (3.5,0)--(3.5,2.5); \draw (4,0)--(4,2.5); 
\draw [ultra thick,black!40!green] (4.5,0)--(4.5,2.5) node[black,midway, xshift =0.3cm, yshift=-1.5cm] {\footnotesize $\Of^-$};
%
%
\draw (0,0.7)--(0.5,0.7) node[black,midway, yshift=0.2cm] {\tiny $2N$};
\draw (0.5,0.9)--(1,0.9) node[black,midway, yshift=0.2cm] {\tiny $2N$};
\draw (1,1.3)--(1.5,1.3) node[black,midway, yshift=0.2cm] {\tiny $2N$} node[black,midway, yshift=-0.2cm] {\tiny D3};
\draw (3,0.7)--(3.5,0.7) node[black,midway, yshift=0.2cm] {\tiny $2N$};
\draw node at (3.5,2.8) {\scriptsize NS5};
\draw (3.5,0.9)--(4,0.9) node[black,midway, yshift=0.2cm] {\tiny $2N$};
\draw (4,1.1)--(4.5,1.1) node[black,midway, yshift=0.2cm] {\tiny $2N$};
\draw [loosely dotted] (1.5,1)--(3,1);
\draw [decorate, decoration={brace, mirror}](0.5,-0.1)--(4,-0.1) node[black,midway,yshift=-0.5cm] {\footnotesize $m~\text{intervals}$};
\end{tikzpicture}}
\ee
where each black dot with a label $k$ denotes $k$ physical D5 branes, and each black vertical line denotes a physical NS5 brane.

Let us now discuss their mirror theories and the corresponding brane configurations.  Under the $S$-duality, each NS5 brane becomes a D5 brane and vice-versa, and an $\Of^-$ plane becomes an $\ON^-$ plane.  The following results can be obtained \cite{Hanany:1999sj}.

\paragraph{\it A mirror of \eref{USp2Nwnflv}.}  The brane system for this is
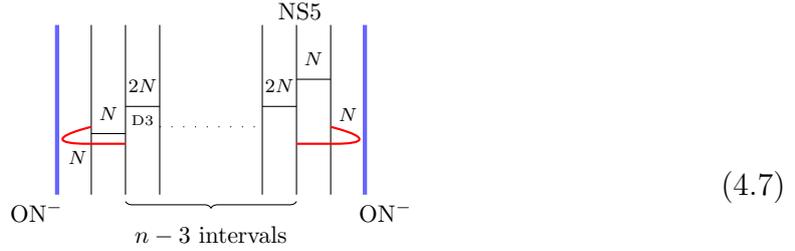
\be
\begin{tikzpicture} [baseline=0, scale=0.9, transform shape]
\draw [ultra thick,blue!60] (0,0)--(0,2.5) node[black,midway, xshift =-0.3cm, yshift=-1.5cm] {\footnotesize $\ON^-$};
\draw (0.5,0)--(0.5,2.5); \draw (1,0)--(1,2.5); \draw (1.5,0)--(1.5,2.5); \draw (3,0)--(3,2.5); \draw (3.5,0)--(3.5,2.5); \draw (4,0)--(4,2.5); 
\draw [ultra thick,blue!60] (4.5,0)--(4.5,2.5) node[black,midway, xshift =0.3cm, yshift=-1.5cm] {\footnotesize $\ON^-$};
%
\draw [thick, color=red, rounded corners=0.75cm](0.5,1)--(-0.4,0.75)--(1,0.75) node[black,midway, xshift=-0.0cm, yshift=-0.2cm] {\scriptsize $N$} ;
%
\draw (0.5,0.9)--(1,0.9) node[black,midway, yshift=0.3cm] {\scriptsize $N$};
\draw (1,1.3)--(1.5,1.3) node[black,midway, yshift=0.3cm] {\scriptsize $2N$} node[black,midway, yshift=-0.2cm] {\tiny D3}  ;
\draw [loosely dotted] (1.5,1)--(3,1);
\draw (3,1.3)--(3.5,1.3) node[black,midway, yshift=0.3cm] {\scriptsize $2N$} node[black,midway, xshift =0.3cm, yshift=1.4cm] {\footnotesize NS5};
\draw (3.5,1.7)--(4,1.7) node[black,midway, yshift=0.3cm] {\scriptsize $N$};
%
\draw [thick, color=red, rounded corners=0.75cm](3.5,0.75)--(4.9,0.75) --(4,1) node[black,midway, xshift=-0.2cm, yshift=0.3cm] {\scriptsize $N$};
%
\draw [decorate, decoration={brace, mirror}](1,-0.1)--(3.5,-0.1) node[black,midway,yshift=-0.5cm] {\footnotesize $n-3~\text{intervals}$};
\end{tikzpicture} 
\ee
Each of the left and the right boundaries contains an $\ON^-$ plane, which is an $S$-dual of the $\Of^-$ plane.  The combination of an $\ON^-$ plane and one NS5 brane is also known as $\ON^0$ and was studied in detail in \cite{Kapustin:1998fa, Sen:1998ii}.  The way that the D3-branes stretch between two NS5 branes at each boundary is depicted in red.  The corresponding theory can be represented by the following quiver diagram:
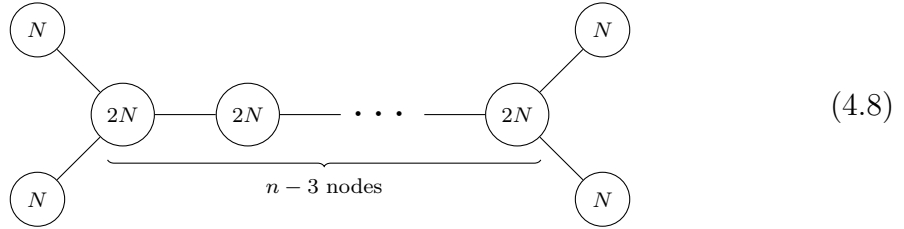
\be
\begin{tikzpicture}[font=\scriptsize,baseline]
\begin{scope}[auto,%
  every node/.style={draw, minimum size=0.5cm}, node distance=0.8cm];
\node[circle] (U2N1) at (1.5,0) {$2N$};
\node[circle,  right=of U2N1] (U2N2)  {$2N$};
\node[draw=none, right=of U2N2] (dots) {\Large $\cdots$};
\node[circle,  right=of dots] (U2N3)  {$2N$};
\node[circle, above right =of U2N3] (UN3)  {$N$};
\node[circle, below right =of U2N3] (UN4)  {$N$};
\node[circle, above left =of U2N1] (UN1)  {$N$};
\node[circle, below left =of U2N1] (UN2) {$N$};
\end{scope}
\draw (U2N1)--(U2N2)--(dots)--(U2N3);
\draw (UN1)--(U2N1)
(UN2)--(U2N1)
(U2N3)--(UN3)
(U2N3)--(UN4);
\draw [decorate, decoration={brace, mirror}](1.3,-0.6)--(7,-0.6) node[black,midway,below, yshift=-0.1cm] {$n-3~\text{nodes}$};
\end{tikzpicture}
\ee
This is indeed the affine Dynkin diagram of the $D_n$ algebra \cite{Kapustin:1998fa}.
\paragraph{\it Mirrors of \eref{U2Nwantisym}.} We consider two cases as follows:
\ben
\item The case of one antisymmetric hypermultiplet.  In this case the brane configuration of the mirror theory is
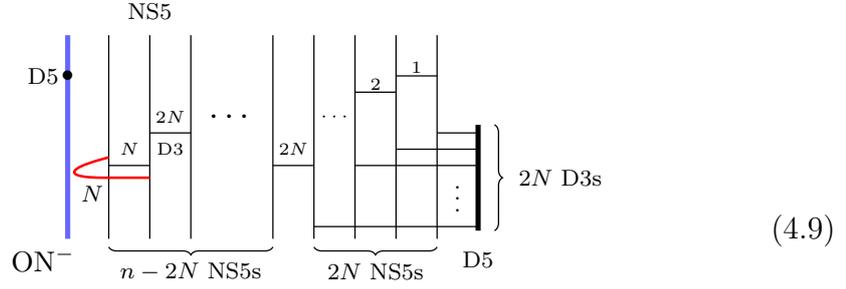
\be  \label{braneoneantisymm1}
\scalebox{1.2}{
\begin{tikzpicture} [baseline=0, scale=0.9, transform shape]
\draw [ultra thick, blue!60] (0,0)--(0,2.5) node[black,midway, xshift =-0.3cm, yshift=-1.5cm] {\footnotesize $\ON^-$}; 
\node[midway, color=black, xshift=0cm, yshift=2cm] {\scriptsize $\bullet$} node[midway, color=black, xshift=-0.3cm, yshift=2cm] {\scriptsize D5};
\draw (0.5,0)--(0.5,2.5); \draw (1,0)--(1,2.5); \draw (1.5,0)--(1.5,2.5); \draw (2.5,0)--(2.5,2.5); \draw (3,0)--(3,2.5); 
\draw (3.5,0)--(3.5,2.5); \draw (4,0)--(4,2.5); \draw (4.5,0)--(4.5,2.5); 
%
%
\draw [thick, color=red, rounded corners=0.75cm](0.5,1)--(-0.4,0.75)--(1,0.75) node[black,midway, xshift=-0.0cm, yshift=-0.2cm] {\scriptsize $N$} ;
\draw (0.5,0.9)--(1,0.9) node[black,midway, yshift=0.2cm] {\tiny $N$};
\draw (1,1.3)--(1.5,1.3) node[black,midway, yshift=0.2cm] {\tiny $2N$} node[black,midway, yshift=-0.2cm] {\tiny D3};
\draw (2.5,0.9)--(3,0.9) node[black,midway, yshift=0.2cm] {\tiny $2N$};
\draw (4,2)--(4.5,2) node[black,midway, yshift=0.1cm] {\tiny $1$};
\draw (3.5,1.8)--(4,1.8) node[black,midway, yshift=0.1cm] {\tiny $2$};
\draw node at (1,2.8) {\scriptsize NS5};
\draw (5,1.3)--(4.5,1.3) ;
\draw (5,1.1)--(4,1.1) ;
\draw (5,0.9)--(3.5,0.9);
\draw node at (4.75,0.6) {\scriptsize $\vdots$ };
\draw (5,0.15)--(3,0.15);
\draw node at (2,1.5) {$\cdots$ };
\draw node at (3.25,1.5) {\tiny $\mathbf{\cdots}$ };
\draw [ultra thick, black] (5,0.1)--(5,1.4) node[black,midway, yshift=-1cm] {\scriptsize D5}; 
\draw [decorate, decoration={brace, mirror}](3,-0.1)--(4.5,-0.1) node[black,midway,yshift=-0.3cm] {\scriptsize $2N~\text{NS5s}$};
\draw [decorate, decoration={brace, mirror}](0.5,-0.1)--(2.5,-0.1) node[black,midway,yshift=-0.3cm] {\scriptsize $n-2N~\text{NS5s}$};
\draw [decorate, decoration={brace, mirror}](5.2,0.1)--(5.2,1.4) node[black,midway,xshift=0.8cm] {\scriptsize $2N~\text{D3s}$};
\end{tikzpicture}}
\ee
One can then move the rightmost D5 brane into the interval and obtain
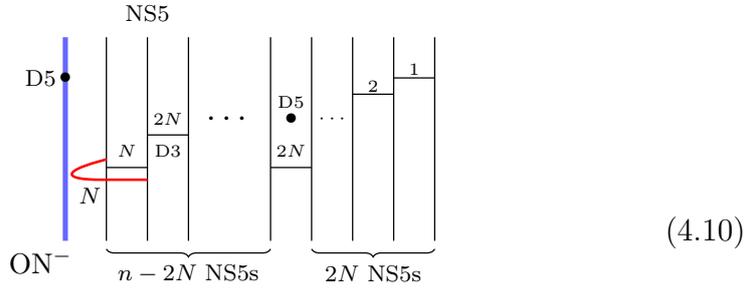
\be  \label{braneoneantisymm}
\scalebox{1.2}{
\begin{tikzpicture} [baseline=0, scale=0.9, transform shape]
\draw [ultra thick, blue!60] (0,0)--(0,2.5) node[black,midway, xshift =-0.3cm, yshift=-1.5cm] {\footnotesize $\ON^-$}; 
\node[midway, color=black, xshift=0cm, yshift=2cm] {\scriptsize $\bullet$} node[midway, color=black, xshift=-0.3cm, yshift=2cm] {\scriptsize D5};
\draw (0.5,0)--(0.5,2.5); \draw (1,0)--(1,2.5); \draw (1.5,0)--(1.5,2.5); \draw (2.5,0)--(2.5,2.5); \draw (3,0)--(3,2.5); 
\draw (3.5,0)--(3.5,2.5); \draw (4,0)--(4,2.5); \draw (4.5,0)--(4.5,2.5); 
%
%
\draw [thick, color=red, rounded corners=0.75cm](0.5,1)--(-0.4,0.75)--(1,0.75) node[black,midway, xshift=-0.0cm, yshift=-0.2cm] {\scriptsize $N$} ;
\draw (0.5,0.9)--(1,0.9) node[black,midway, yshift=0.2cm] {\tiny $N$};
\draw (1,1.3)--(1.5,1.3) node[black,midway, yshift=0.2cm] {\tiny $2N$} node[black,midway, yshift=-0.2cm] {\tiny D3};
\draw (2.5,0.9)--(3,0.9) node[black,midway, yshift=0.2cm] {\tiny $2N$};
\draw (4,2)--(4.5,2) node[black,midway, yshift=0.1cm] {\tiny $1$};
\draw (3.5,1.8)--(4,1.8) node[black,midway, yshift=0.1cm] {\tiny $2$};
\draw node at (1,2.8) {\scriptsize NS5};
\draw node at (2,1.5) {$\cdots$ };
\draw node at (3.25,1.5) {\tiny $\mathbf{\cdots}$ }; 
\node at (2.75,1.5) {\scriptsize $\bullet$} node at (2.75,1.7) {\tiny D5};
\draw [decorate, decoration={brace, mirror}](3,-0.1)--(4.5,-0.1) node[black,midway,yshift=-0.3cm] {\scriptsize $2N~\text{NS5s}$};
\draw [decorate, decoration={brace, mirror}](0.5,-0.1)--(2.5,-0.1) node[black,midway,yshift=-0.3cm] {\scriptsize $n-2N~\text{NS5s}$};
\end{tikzpicture}}
\ee
Hence the corresponding quiver is
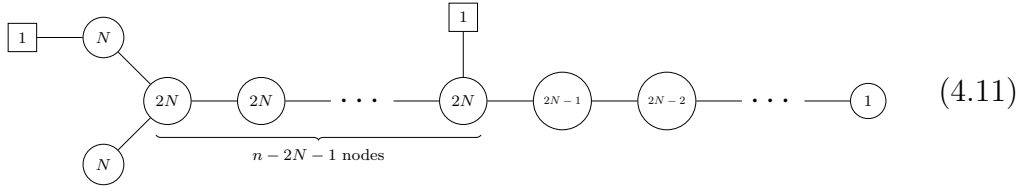
\be \label{mirroneantisymm}
\scalebox{0.75}{
\begin{tikzpicture}[font=\scriptsize,baseline]
\begin{scope}[auto,%
  every node/.style={draw, minimum size=0.5cm}, node distance=0.8cm];
\node[circle] (U2N1) at (0,0) {$2N$};
\node[circle,  right=of U2N1] (U2N2)  {$2N$};
\node[draw=none, right=of U2N2] (dots) {\Large $\cdots$};
\node[circle,  right=of dots] (U2N3)  {$2N$};
\node[circle, above left =of U2N1] (UN1)  {$N$};
\node[circle, below left =of U2N1] (UN2) {$N$};
\node[circle, right=of U2N3] (UN2Nm1) {\scalebox{0.7}{$2N-1$}};
\node[circle, right =of UN2Nm1] (UN2Nm2) {\scalebox{0.7}{$2N-2$}};
\node[draw=none, right =of UN2Nm2] (dots1) {\Large $\cdots$};
\node[circle, right =of dots1] (U1) {$1$};
\node[rectangle, left =of UN1] (f1)  {$1$};
\node[rectangle, above =of U2N3] (f2)  {$1$};
\end{scope}
\draw (U2N1)--(U2N2)--(dots)--(U2N3)--(UN2Nm1)--(UN2Nm2)--(dots1)--(U1);
\draw (UN1)--(U2N1)
(UN2)--(U2N1);
\draw (f1)--(UN1);
\draw (f2)--(U2N3);
\draw [decorate, decoration={brace, mirror}](1.3-1.5,-0.6)--(7-1.5,-0.6) node[black,midway,below, yshift=-0.1cm] {$n-2N-1~\text{nodes}$};
\end{tikzpicture}}
\ee
\item The case of two antisymmetric hypermultiplets.  In this case the brane configuration of the mirror theory is
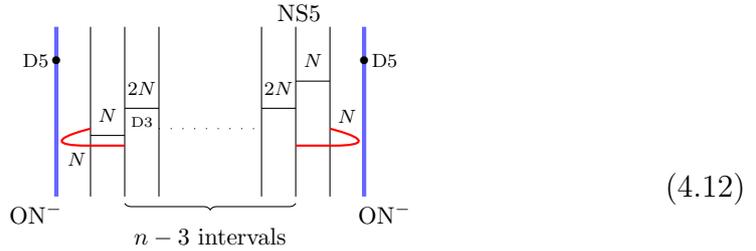
\be
\begin{tikzpicture} [baseline=0, scale=0.9, transform shape]
\draw [ultra thick,blue!60] (0,0)--(0,2.5) node[black,midway, xshift =-0.3cm, yshift=-1.5cm] {\footnotesize $\ON^-$};
\node[midway, color=black, xshift=0cm, yshift=2cm] {\scriptsize $\bullet$} node[midway, color=black, xshift=-0.3cm, yshift=2cm] {\scriptsize D5};
\draw (0.5,0)--(0.5,2.5); \draw (1,0)--(1,2.5); \draw (1.5,0)--(1.5,2.5); \draw (3,0)--(3,2.5); \draw (3.5,0)--(3.5,2.5); \draw (4,0)--(4,2.5); 
\draw [ultra thick,blue!60] (4.5,0)--(4.5,2.5) node[black,midway, xshift =0.3cm, yshift=-1.5cm] {\footnotesize $\ON^-$};
\node[midway, color=black, xshift=4.5cm, yshift=2cm] {\scriptsize $\bullet$} node[midway, color=black, xshift=4.8cm, yshift=2cm] {\scriptsize D5};
%
\draw [thick, color=red, rounded corners=0.75cm](0.5,1)--(-0.4,0.75)--(1,0.75) node[black,midway, xshift=-0.0cm, yshift=-0.2cm] {\scriptsize $N$} ;
%
\draw (0.5,0.9)--(1,0.9) node[black,midway, yshift=0.3cm] {\scriptsize $N$};
\draw (1,1.3)--(1.5,1.3) node[black,midway, yshift=0.3cm] {\scriptsize $2N$} node[black,midway, yshift=-0.2cm] {\tiny D3}  ;
\draw [loosely dotted] (1.5,1)--(3,1);
\draw (3,1.3)--(3.5,1.3) node[black,midway, yshift=0.3cm] {\scriptsize $2N$} node[black,midway, xshift =0.3cm, yshift=1.4cm] {\footnotesize NS5};
\draw (3.5,1.7)--(4,1.7) node[black,midway, yshift=0.3cm] {\scriptsize $N$};
%
\draw [thick, color=red, rounded corners=0.75cm](3.5,0.75)--(4.9,0.75) --(4,1) node[black,midway, xshift=-0.2cm, yshift=0.3cm] {\scriptsize $N$};
%
\draw [decorate, decoration={brace, mirror}](1,-0.1)--(3.5,-0.1) node[black,midway,yshift=-0.5cm] {\footnotesize $n-3~\text{intervals}$};
\end{tikzpicture} 
\ee
The corresponding quiver theory is
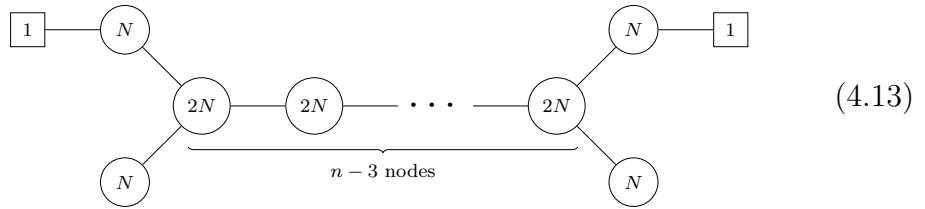
\be
\scalebox{0.9}
{\begin{tikzpicture}[font=\scriptsize,baseline]
\begin{scope}[auto,%
  every node/.style={draw, minimum size=0.5cm}, node distance=0.8cm];
\node[circle] (U2N1) at (1.5,0) {$2N$};
\node[circle,  right=of U2N1] (U2N2)  {$2N$};
\node[draw=none, right=of U2N2] (dots) {\Large $\cdots$};
\node[circle,  right=of dots] (U2N3)  {$2N$};
\node[circle, above right =of U2N3] (UN3)  {$N$};
\node[circle, below right =of U2N3] (UN4)  {$N$};
\node[circle, above left =of U2N1] (UN1)  {$N$};
\node[circle, below left =of U2N1] (UN2) {$N$};
\node[rectangle, left =of UN1] (f1)  {$1$};
\node[rectangle, right =of UN3] (f2)  {$1$};
\end{scope}
\draw (U2N1)--(U2N2)--(dots)--(U2N3);
\draw (UN1)--(U2N1)
(UN2)--(U2N1)
(U2N3)--(UN3)
(U2N3)--(UN4);
\draw (f1)--(UN1);
\draw (f2)--(UN3);
\draw [decorate, decoration={brace, mirror}](1.3,-0.6)--(7,-0.6) node[black,midway,below, yshift=-0.1cm] {$n-3~\text{nodes}$};
\end{tikzpicture}}
\ee
\een
\paragraph{\it A mirror of \eref{USp2NU2NmUSp2N}.}  The brane construction is
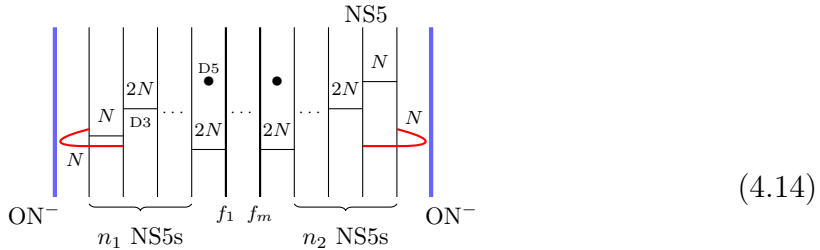
\be \label{mirrUSpUUUUSp}
\begin{tikzpicture} [baseline=0, scale=0.9, transform shape]
\draw [ultra thick,blue!60] (0,0)--(0,2.5) node[black,midway, xshift =-0.3cm, yshift=-1.5cm] {\footnotesize $\ON^-$};
\draw (0.5,0)--(0.5,2.5); \draw (1,0)--(1,2.5); \draw (1.5,0)--(1.5,2.5); \draw (2,0)--(2,2.5);
\draw[thick] (2.5,0)--(2.5,2.5) node[midway,yshift=-1.5cm] {\scriptsize $f_1$};  
\draw[thick] (3,0)--(3,2.5) node[midway,yshift=-1.5cm] {\scriptsize $f_m$};  
\draw (3.5,0)--(3.5,2.5); \draw (4,0)--(4,2.5); \draw (4.5,0)--(4.5,2.5); \draw (5,0)--(5,2.5); 
\draw [ultra thick,blue!60] (5+0.5,0)--(5+0.5,2.5) node[black,midway, xshift =0.3cm, yshift=-1.5cm] {\footnotesize $\ON^-$};
%
\draw [thick, color=red, rounded corners=0.75cm](0.5,1)--(-0.4,0.75)--(1,0.75) node[black,midway, xshift=-0.0cm, yshift=-0.2cm] {\scriptsize $N$} ;
\draw (0.5,0.9)--(1,0.9) node[black,midway, yshift=0.3cm] {\scriptsize $N$};
\draw (1,1.3)--(1.5,1.3) node[black,midway, yshift=0.3cm] {\scriptsize $2N$} node[black,midway, yshift=-0.2cm] {\tiny D3}  ;
\draw (2,0.7)--(2.5,0.7) node[black,midway, yshift=0.3cm] {\scriptsize $2N$};
\draw (3,0.7)--(3.5,0.7) node[black,midway, yshift=0.3cm] {\scriptsize $2N$};
\node at (1.75,1.25) {\tiny $\cdots$};
\node at (2.25,1.7) {\footnotesize $\bullet$} node[black,midway, xshift=2.25cm, yshift=1.9cm] {\tiny D5};
\node at (2.75,1.25) {\tiny $\cdots$};
\node at (3.25,1.7) {\footnotesize $\bullet$};
\node at (3.75,1.25) {\tiny $\cdots$};
\draw (3+1,1.3)--(3.5+1,1.3) node[black,midway, yshift=0.3cm] {\scriptsize $2N$} node[black,midway, xshift =0.3cm, yshift=1.4cm] {\footnotesize NS5};
\draw (3.5+1,1.7)--(4+1,1.7) node[black,midway, yshift=0.3cm] {\scriptsize $N$};
%
\draw [thick, color=red, rounded corners=0.75cm](3.5+1,0.75)--(4.9+1,0.75) --(4+1,1) node[black,midway, xshift=-0.2cm, yshift=0.3cm] {\scriptsize $N$};
%
\draw [decorate, decoration={brace, mirror}](0.5,-0.1)--(2,-0.1) node[black,midway,yshift=-0.5cm] {\footnotesize $n_1~\text{NS5s}$};
\draw [decorate, decoration={brace, mirror}](3.5,-0.1)--(5,-0.1) node[black,midway,yshift=-0.5cm] {\footnotesize $n_2~\text{NS5s}$};
\end{tikzpicture} 
\ee
where the boldface vertical line labelled by $f_j$ (with $j=1,\ldots, m$) denotes a set of $f_j$ NS5 branes, with $2N$ D3 branes stretching between two successive NS5 branes.  Note that there is also one D5 brane at the interval between each set.  For simplicity, let us present the quiver for the case of $m=1$:
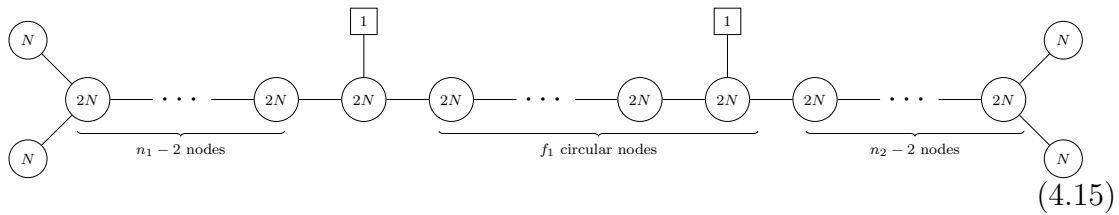
\be \label{mirrUSpUUUSp}
\scalebox{0.7}{
\begin{tikzpicture}[font=\scriptsize,baseline]
\begin{scope}[auto,%
  every node/.style={draw, minimum size=0.5cm}, node distance=0.8cm];
\node[circle] (U2N1) at (1.5,0) {$2N$};
\node[draw=none, right=of U2N1] (dots) {\Large $\cdots$};
\node[circle,  right=of dots] (U2N3)  {$2N$};
\node[circle,  right=of U2N3] (U2N4)  {$2N$};
\node[circle,  right=of U2N4] (U2N45)  {$2N$};
\node[draw=none,  right=of U2N45] (dots1)  {\Large $\cdots$};
\node[circle,  right=of dots1] (U2N5)  {$2N$};
\node[circle,  right=of U2N5] (U2N55)  {$2N$};
\node[circle,  right=of U2N55] (U2N555)  {$2N$};
\node[draw=none, right=of U2N555] (dots2) {\Large $\cdots$};
\node[circle,  right=of dots2] (U2N6)  {$2N$};
\node[circle, above right =of U2N6] (UN3)  {$N$};
\node[circle, below right =of U2N6] (UN4)  {$N$};
\node[circle, above left =of U2N1] (UN1)  {$N$};
\node[circle, below left =of U2N1] (UN2) {$N$};
\node[rectangle, above =of U2N55] (f2) {$1$};
\node[rectangle, above =of U2N4] (f1) {$1$};
\end{scope}
\draw (U2N1)--(dots)--(U2N3)--(U2N4)--(U2N45)--(dots1)--(U2N5)--(U2N55)--(U2N555)--(dots2)--(U2N6);
\draw (UN1)--(U2N1)
(UN2)--(U2N1)
(U2N6)--(UN3)
(U2N6)--(UN4)
(f1)--(U2N4)
(f2)--(U2N55);
\draw [decorate, decoration={brace, mirror}](1.3,-0.6)--(5.2,-0.6) node[black,midway,below, yshift=-0.1cm] {$n_1-2~\text{nodes}$};
\draw [decorate, decoration={brace, mirror}](8.1,-0.6)--(14.1,-0.6) node[black,midway,below, yshift=-0.1cm] {$f_1~\text{circular nodes}$};
\draw [decorate, decoration={brace, mirror}](15,-0.6)--(19.1,-0.6) node[black,midway,below, yshift=-0.1cm] {$n_2-2~\text{nodes}$};
\end{tikzpicture}}
\ee
This can be easily generalised to the case of $m>1$ by simply repeating the part under the second brace with $f_2, \, f_3, \, \ldots, \, f_m$ in a consecutive manner.

\subsection{The cases with an $S$-fold}
In this subsection, we insert an $S$-fold into a brane interval of the aforementioned configurations.  In general, the resulting quiver theory contains a $T(U(N))$ link connecting two gauge nodes corresponding to the interval where we put the $S$-fold. The mirror configuration can simply be obtained by inserting the $S$-fold in the same position in the $S$-dual brane configuration.  In the following, the moduli spaces of such a theory and its mirror are analysed in detail.  

We make the following important observation.  The Higgs ({\it resp.} Coulomb) branch of a given theory gets exchanged with the Coulomb ({\it resp.} Higgs) branch of the mirror theory in a ``regular way'', provided that 
\ben
\item the $S$-fold is not inserted ``too close'' to the orientifold plane; and
\item the $S$-fold is not inserted in the ``quiver tail'', arising from a set of D3 branes connecting a D5 brane with distinct NS5 branes.
\een
Subsequently, we shall give more precise statements for these two points using various examples.  In other words, we use mirror symmetry as a tool to indicate the consistency of the insertion of an $S$-fold to the brane system with an orientifold fiveplane.

\subsubsection{Models with one or two antisymmetric hypermultiplets}
In this subsection, we focus on the models with one antisymmetric hypermultiplet for definiteness.  The case for two antisymmetric hypermultiplets can be treated almost in the same way.
Let us insert an $S$-fold in the left diagram in \eref{braneonetwoanti} such that there are $n_1$ physical D5 branes on the left of the $S$-fold and there are $n_2$ physical D5 branes on the right.  The resulting theory is
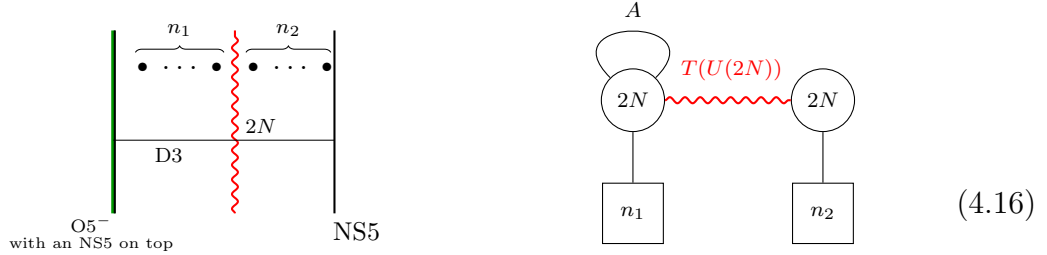
\be\label{SfoldU2Nantisym}
\scalebox{0.97}{
\begin{tikzpicture}[baseline, scale=1, transform shape]
\draw [ultra thick,black!40!green] (0,0)--(0,2.5) node[black,midway, xshift =-0.3cm, yshift=-1.5cm] {\footnotesize $\overset{\Of^-}{{\tiny \text{with an NS5 on top}}}$}; 
\draw [thick,black] (0.02,0)--(0.02,2.5);
\node[midway, color=black, xshift=0.4cm, yshift=2cm] {\scriptsize $\bullet$};
\node[midway, color=black, xshift=0.9cm, yshift=2cm] {$~\ldots~$};
\node[midway, color=black, xshift=1.4cm, yshift=2cm] {\scriptsize $\bullet$};
\draw [decorate, decoration={brace}](0.3,2.2)--(1.5,2.2) node[black,midway,yshift=0.3cm] {\scriptsize $n_1$};
\node[midway, color=black, xshift=1.9cm, yshift=2cm] {\scriptsize $\bullet$};
\node[midway, color=black, xshift=2.4cm, yshift=2cm] {$~\ldots~$};
\node[midway, color=black, xshift=2.9cm, yshift=2cm] {\scriptsize $\bullet$};
\draw [decorate, decoration={brace}](1.8,2.2)--(2.95,2.2) node[black,midway,yshift=0.3cm] {\scriptsize $n_2$};
\draw [thick,red,snake it] (1.65,0)--(1.65,2.5);
\draw [thick,black] (3,0)--(3,2.5) node[black,midway, xshift =0.3cm, yshift=-1.5cm] {\footnotesize NS5};
\draw (0,1)--(3,1) node[black,xshift=-1cm, yshift=0.2cm] {\scriptsize $2N$} node[black,near start, yshift=-0.2cm] {\scriptsize D3};
\end{tikzpicture}}
\qquad \qquad \quad
\begin{tikzpicture}[baseline,font=\scriptsize]
\begin{scope}[auto,%
  every node/.style={draw, minimum size=0.8cm}, node distance=0.6cm];
\node[draw, circle, black] (node1) at (0,1.5) {$2N$};
\node[draw, circle, black] (node2) at (2.5,1.5) {$2N$};
\draw[black] (node1) edge [out=45,in=135,loop,looseness=5]  (node1) node[draw=none] at (0,2.7) {$A$} ;
\node[draw, rectangle,black] (sqnode1) at (0,0) {$n_1$};
\node[draw, rectangle,black] (sqnode2) at (2.5,0) {$n_2$};
\end{scope}
\draw (node1)--(sqnode1);
\draw (node2)--(sqnode2);
\draw[draw=red,solid, snake it, thick,-] (node1)--(node2);
\node[draw=none] at (1.3,1.9) {{\red$T(U(2N))$}};
\end{tikzpicture}
\ee

\subsubsection*{The case in which $n_1 \geq 2$ and $n_2 \geq 2N$}
The corresponding mirror theory is 
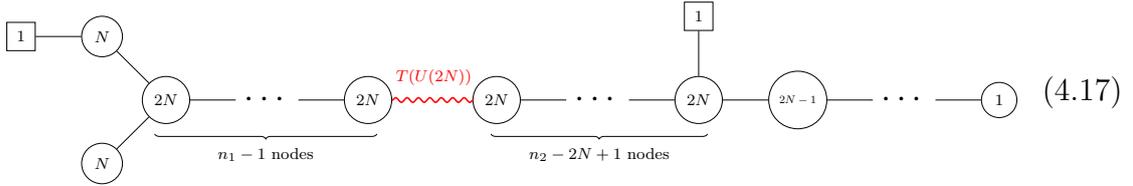
\be\label{SfoldU2NantisymMirr}
\scalebox{0.75}{
\begin{tikzpicture}[font=\scriptsize,baseline]
\begin{scope}[auto,%
  every node/.style={draw, minimum size=0.5cm}, node distance=0.8cm];
\node[circle] (U2N1) at (0,0) {$2N$};
\node[draw=none, right=of U2N1] (dots) {\Large $\cdots$};
\node[circle,  right=of dots] (U2N3)  {$2N$};
\node[circle] (U2N4) at (5.8, 0) {$2N$};
\draw[draw=red,solid, snake it, thick,-] (U2N3)--(5.4, 0);
\node[draw=none] at (4.7,0.4) {\red{$T(U(2N))$}};
\node[circle, above left =of U2N1] (UN1)  {$N$};
\node[circle, below left =of U2N1] (UN2) {$N$};
\node[draw=none, right=of U2N4] (dots1)  {\Large $\cdots$};
\node[circle,  right=of dots1] (U2N5)  {$2N$};
\node[circle, right=of U2N5] (UN2Nm1) {\scalebox{0.7}{$2N-1$}};
\node[draw=none, right=of UN2Nm1] (dots2)  {\Large $\cdots$};
\node[circle,  right=of dots2] (U1)  {$1$};
\node[rectangle, left =of UN1] (f1)  {$1$};
\node[rectangle, above =of U2N5] (f2)  {$1$};
\end{scope}
\draw (UN1)--(U2N1)
(UN2)--(U2N1)--(dots)--(U2N3);
\draw (f1)--(UN1);
\draw (f2)--(U2N5);
\draw (U2N4)--(dots1)--(U2N5)--(UN2Nm1)--(dots2)--(U1);
\draw [decorate, decoration={brace, mirror}](1.3-1.5,-0.6)--(5.2-1.5,-0.6) node[black,midway,below, yshift=-0.1cm] {$n_1-1~\text{nodes}$};
\draw [decorate, decoration={brace, mirror}](7.2-1.5,-0.6)--(11-1.5,-0.6) node[black,midway,below, yshift=-0.1cm] {$n_2-2N+1~\text{nodes}$};
\end{tikzpicture}}
\ee
The condition $n_1 \geq 2, \, n_2 \geq 2N$ ensures that the $T(U(2N))$ link in the mirror theory \eref{SfoldU2NantisymMirr} stay between the first $U(2N)$ gauge node and the $U(2N)$ gauge node with $1$ flavour.

The Higgs branch of theory (\ref{SfoldU2Nantisym}) has dimension
\be\label{HBSfoldU2antisym}
	\begin{split}
	\text{dim}_{\mathbb H}\,\cH_{(\ref{SfoldU2Nantisym})}&=2N n_1+\frac{1}{2}2N(2N-1)+2\cdot \frac{1}{2} (4N^2-2N)+2Nn_2 \\
	& \qquad -4N^2-4N^2 \\
	&=N(2n_1+2n_2-2N-3),
	\end{split}
\ee
while the Coulomb branch is empty because there are only two gauge nodes connected by a $T(U(2N))$-link
\be \label{zeroCoul}
\dim_{\mathbb H}\,\cC_{(\ref{SfoldU2Nantisym})}=0.
\ee

Since the moduli space of $T(U(2N))$ contains the Higgs and Coulomb branches, each of which is isomorphic to the nilpotent cone of $SU(2N)$, it follows that the Higgs branch of \eref{SfoldU2Nantisym} also splits into a product of two hyperK\"ahler spaces which can be written in the notation of section \ref{sec:couplenilcone} as
\be \label{prodHiggs}
\CH_{\eref{SfoldU2Nantisym}} =  \begin{tikzpicture}[baseline,font=\scriptsize]
\begin{scope}[auto,%
  every node/.style={draw, minimum size=0.8cm}, node distance=0.6cm];
\node[draw, circle, black] (node1) at (0,1.5-1) {$2N$};
\node[draw, circle, black] (node2) at (3,1.5-1) {$2N$};
\draw[black] (node1) edge [out=45,in=135,loop,looseness=5]  (node1) node[draw=none] at (0,2.7-1) {$A$} ;
\node[draw, rectangle,black] (sqnode1) at (0,0-1) {$n_1$};
\node[draw, rectangle,black] (sqnode2) at (3,0-1) {$n_2$};
\node[draw=none] at (1,1.5-1) {\large \red $\times$};
\node[draw=none] at (2,1.5-1) {\large \red $\times$};
\node[draw=none] at (1.5,0) {\Large $\times$};
\end{scope}
\draw (node1)--(sqnode1);
\draw (node2)--(sqnode2);
\draw[draw=red,solid, snake it, thick,-] (node1)--(1,1.5-1);
\draw[draw=red,solid, snake it, thick,-] (2,1.5-1)--(node2);
\end{tikzpicture}
\ee
The symmetry of $\CH_{\eref{SfoldU2Nantisym}}$ is $U(n_1) \times (U(n_2)/U(1))$, coming from the first and second factors respectively.
According to \eref{orbsuNsun} and below, the hyperK\"ahler space corresponding to the second factor is identified with $\bar{\CO}_{(2N+1,1^{n_2-2N-1})}$ for $n_2\geq 2N+1$ and $\bar{\CO}_{(2N)}$ for $n_2=2N$.

The mirror theory (\ref{SfoldU2NantisymMirr}) has the following Coulomb branch dimension
\be
\begin{split}
\dim_{\mathbb H}\,\cC_{(\ref{SfoldU2NantisymMirr})} &=N+N+(2N)(n_1+n_2-2N-2)+\sum_{i=1}^{2N-1}i \\
&= N(2n_1+2n_2-2N-3),
\end{split}
\ee
while the Higgs branch has dimension
\be
\begin{split}
	\text{dim}_{\mathbb H}\,\cH_{(\ref{SfoldU2NantisymMirr})}=&N+4N^2+4N^2(n_1+n_2-2N-1-1)+(4N^2-2N)\\
	&+2N+\sum_{i=1}^{2N-1}i(i+1)-2N^2-4N^2(n_1+n_2-2N)\\
	&-\sum_{i=1}^{2N-1}i^2=0
\end{split}
\ee
Indeed, we find an agreement for the dimensions of the Higgs and Coulomb branches under mirror symmetry, 
namely
\be
	\text{dim}_{\mathbb H}\,\cC_{(\ref{SfoldU2Nantisym})}=\text{dim}_{\mathbb H}\,\cH_{(\ref{SfoldU2NantisymMirr})}
	, \qquad 
	\text{dim}_{\mathbb H}\,\cC_{(\ref{SfoldU2NantisymMirr})}=\text{dim}_{\mathbb H}\,							\cH_{(\ref{SfoldU2Nantisym})}.
\ee

It should be pointed out the the Coulomb branch of \eref{SfoldU2NantisymMirr} is also a product of two hyperK\"ahler spaces.  The reason is that the nodes that are connected by the $T(U(2N))$ link do not contribute to the Coulomb branch and hence can be taken as flavours nodes.  Therefore, the Coulomb branch of \eref{SfoldU2NantisymMirr} is the product of the Coulomb branches of the following theories:
\be \label{Coulsplit}
\scalebox{0.75}{
\begin{tikzpicture}[font=\scriptsize,baseline]
\begin{scope}[auto,%
  every node/.style={draw, minimum size=0.5cm}, node distance=0.8cm];
\node[circle] (U2N1) at (0,0) {$2N$};
\node[draw=none, right=of U2N1] (dots) {\Large $\cdots$};
\node[rectangle,  right=of dots] (U2N3)  {$2N$};
\node[rectangle] (U2N4) at (5.8, 0) {$2N$};
\node[circle, above left =of U2N1] (UN1)  {$N$};
\node[circle, below left =of U2N1] (UN2) {$N$};
\node[draw=none, right=of U2N4] (dots1)  {\Large $\cdots$};
\node[circle,  right=of dots1] (U2N5)  {$2N$};
\node[circle, right=of U2N5] (UN2Nm1) {\scalebox{0.7}{$2N-1$}};
\node[draw=none, right=of UN2Nm1] (dots2)  {\Large $\cdots$};
\node[circle,  right=of dots2] (U1)  {$1$};
\node[rectangle, left =of UN1] (f1)  {$1$};
\node[rectangle, above =of U2N5] (f2)  {$1$};
\end{scope}
\draw (UN1)--(U2N1)
(UN2)--(U2N1)--(dots)--(U2N3);
\draw (f1)--(UN1);
\draw (f2)--(U2N5);
\draw (U2N4)--(dots1)--(U2N5)--(UN2Nm1)--(dots2)--(U1);
\draw [decorate, decoration={brace, mirror}](1.3-1.5,-0.6)--(5.2-1.5,-0.6) node[black,midway,below, yshift=-0.1cm] {$n_1-1~\text{nodes}$};
\draw [decorate, decoration={brace, mirror}](7.2-1.5,-0.6)--(11-1.5,-0.6) node[black,midway,below, yshift=-0.1cm] {$n_2-2N+1~\text{nodes}$};
\end{tikzpicture}}
\ee
Under mirror symmetry, each of the factor in the product \eref{prodHiggs} is mapped to the Coulomb brach of each of the above quiver.  Let us examine the symmetry of the Coulomb branch using the technique of \cite{Gaiotto:2008ak}.  In the left quiver, all balanced gauge nodes form a Dynkin diagram of $A_{n_1-1}$; together with the top left node which is overbalanced, these give rise to the global symmetry algebra $A_{n_1-1} \times u(1)$, corresponding to $U(n_1)$.  In the right quiver, all gauge nodes are balanced; these give rise to the symmetry algebra $A_{n_2-1}$, corresponding to $U(n_2)/U(1)$.  This is agreement of the symmetry of the Higgs branch $\CH_{\eref{SfoldU2Nantisym}}$.

It is worth commenting on the distribution of the flavours in theory (\ref{SfoldU2Nantisym}). It is clear from the computation of the dimension of the Higgs branch (\ref{HBSfoldU2antisym}) that one can change $n_1$ and $n_2$ 
keeping their sum $n=n_1+n_2$ constant, without changing the dimension of the Higgs branch. However, as can be clearly seen from \eref{prodHiggs}, the structure of the Higgs branch depends on $n_1$ and $n_2$. In addition, 
modifying the distribution of the flavour will change the position of the $T(U(2N))$ link in the mirror theory (\ref{SfoldU2NantisymMirr}). Let us focus the case of $N=1$ with $n_1=3, \, n_2=3$ and $n_1=4,\, n_2=2$. The theories and their mirrors are
\be\label{SfoldU2NantisymEx1}
\scalebox{0.9}{\begin{tikzpicture}[baseline,font=\scriptsize]
\begin{scope}[auto,%
  every node/.style={draw, minimum size=0.8cm}, node distance=0.6cm];
\node[draw, circle, black] (node1) at (0,0) {$2$};
\node[draw, circle, black] (node2) at (2.5,0) {$2$};
\draw[black] (node1) edge [out=45,in=135,loop,looseness=5]  (node1) node[draw=none] at (0,1.2) {$A$} ;
\node[draw, rectangle,black] (sqnode1) at (0,-1.5) {$3$};
\node[draw, rectangle,black] (sqnode2) at (2.5,-1.5) {$3$};
\end{scope}
\draw (node1)--(sqnode1);
\draw (node2)--(sqnode2);
\draw[draw=red,solid, snake it, thick,-] (node1)--(node2);
\node[draw=none] at (1.3,0.4) {{\red$T(U(2))$}};
\end{tikzpicture}} \qquad \quad
\scalebox{0.9}{
\begin{tikzpicture}[font=\scriptsize,baseline]
\begin{scope}[auto,%
  every node/.style={draw, minimum size=0.5cm}, node distance=0.8cm];
\node[circle] (U2N1) at (0,-0.8) {$2$};
\node[circle, above left =of U2N1] (UN1) {$1$};
\node[circle, below left =of U2N1] (UN2) {$1$};
\node[rectangle, left =of UN1] (f1)  {$1$};
\node[circle, right =of U2N1] (U2N2) {$2$};
\node[circle, right =of U2N2] (U2N3) {$2$};
\node[circle, right =of U2N3] (U2N4) {$2$};
\node[rectangle, above =of U2N4] (f2)  {$1$};
\node[circle, right =of U2N4] (UN3) {$1$};
\end{scope}
\draw (UN1)--(U2N1)
(UN2)--(U2N1);
\draw (f1)--(UN1);
\draw (U2N1)--(U2N2);
\draw (UN3)--(U2N4)--(f2);
\draw[draw=red,solid, snake it, thick,-] (U2N2)--(U2N3);
\draw (U2N3)--(U2N4);
\node[draw=none] at (2.2,-0.4) {{\red$T(U(2))$}};
\end{tikzpicture}}
\ee

\be\label{SfoldU2NantisymEx2}
\scalebox{0.9}{\begin{tikzpicture}[baseline,font=\scriptsize]
\begin{scope}[auto,%
  every node/.style={draw, minimum size=0.8cm}, node distance=0.6cm];
\node[draw, circle, black] (node1) at (0,0) {$2$};
\node[draw, circle, black] (node2) at (2.5,0) {$2$};
\draw[black] (node1) edge [out=45,in=135,loop,looseness=5]  (node1) node[draw=none] at (0,1.2) {$A$} ;
\node[draw, rectangle,black] (sqnode1) at (0,-1.5) {$4$};
\node[draw, rectangle,black] (sqnode2) at (2.5,-1.5) {$2$};
\end{scope}
\draw (node1)--(sqnode1);
\draw (node2)--(sqnode2);
\draw[draw=red,solid, snake it, thick,-] (node1)--(node2);
\node[draw=none] at (1.3,0.4) {{\red$T(U(2))$}};
\end{tikzpicture}} \qquad\quad
\scalebox{0.9}{
\begin{tikzpicture}[font=\scriptsize,baseline]
\begin{scope}[auto,%
  every node/.style={draw, minimum size=0.5cm}, node distance=0.8cm];
\node[circle] (U2N1) at (0,-0.8) {$2$};
\node[circle, above left =of U2N1] (UN1) {$1$};
\node[circle, below left =of U2N1] (UN2) {$1$};
\node[rectangle, left =of UN1] (f1)  {$1$};
\node[circle, right =of U2N1] (U2N2) {$2$};
\node[circle, right =of U2N2] (U2N3) {$2$};
\node[circle, right =of U2N3] (U2N4) {$2$};
\node[rectangle, above =of U2N4] (f2)  {$1$};
\node[circle, right =of U2N4] (UN3) {$1$};
\end{scope}
\draw (UN1)--(U2N1)
(UN2)--(U2N1);
\draw (f1)--(UN1);
\draw (U2N1)--(U2N2)--(U2N3);
\draw (UN3)--(U2N4)--(f2);
\draw[draw=red,solid, snake it, thick,-] (U2N3)--(U2N4);
\node[draw=none] at (3.6,-0.4) {{\red$T(U(2))$}};
\end{tikzpicture}}
\ee

As explained in \eref{prodHiggs}, the Higgs branch of the left diagram in each case splits into a product of two hyperK\"ahler spaces.  According to \eref{orbsuNsunspecial}, the second factor in each line is the Hilbert series for the closure of the nilpotent orbit $\bar{\CO}_{(3)}$ and $\bar{\CO}_{(2)}$, coincident with the Higgs branch of the theories $T(SU(3))$ and $T(SU(2))$ respectively.   The unrefined Hilbert series for the first factor is
\be
\begin{split}
&\oint_{|z|=1} \frac{d z}{2 \pi i z} (1-z^2) \oint_{|q|=1} \frac{d q}{2 \pi i q}  \PE\Big[ n_1 (z+z^{-1})(q+q^{-1}) \\
& \qquad + (q^2+q^{-2}) t + (z^2+1+z^{-2})t^2 - t^4 -(z^2+1+z^{-2}+1)t^2 \Big]\\
& \qquad \times \PE\left[(z^2 +1 +z^{-2})t^2 - t^4 \right]~.
\end{split}
\ee
We therefore arrive at the following results:
\be\label{DistrFlavEx}
\begin{split}
	&H[\cH_{(\ref{SfoldU2Nantisym})}^{n_1=3, n_2=3}\,]=\text{PE}\,[9 t^2 + 6 t^3 - t^4 - 6 t^5 - 10 t^6+\dots]\, \text{PE}\,[8 t^2 - t^4 - t^6], \\
	&H[\cH_{(\ref{SfoldU2Nantisym})}^{n_1=4, n_2=2}\,]=\text{PE}\,[16 t^2 + 12 t^3 - t^4 - 32 t^5 - 54 t^6+	\dots] \,\text{PE}\,[3 t^2 - t^4],
\end{split}
\ee
These indicate that the symmetry of the Higgs branch is $U(n_1) \times (U(n_2)/U(1))$.   

Of course, the above Hilbert series can also be obtained from the Coulomb branch of the corresponding mirror theory. As an example, as stated in \eref{Coulsplit}, for $n_1=4, n_2=2$, the Coulomb branch of the right quiver of \eref{SfoldU2NantisymEx2} is a product of the Coulomb branches of the following theories:
\be \label{splitexample}
\begin{tikzpicture}[font=\scriptsize,baseline]
\begin{scope}[auto,%
  every node/.style={draw, minimum size=0.5cm}, node distance=0.8cm];
\node[circle] (U2N1) at (0,0) {$2$};
\node[circle, above left =of U2N1] (UN1) {$1$};
\node[circle, below left =of U2N1] (UN2) {$1$};
\node[rectangle, left =of UN1] (f1)  {$1$};
\node[rectangle, right =of U2N1] (U2N2) {$2$};
\node[rectangle] at (3.5,0) (U2N3) {$3$};
\node[circle, right =of U2N3] (U2N4) {$2$};
\node[circle, right =of U2N4] (UN3) {$1$};
\end{scope}
\draw (UN1)--(U2N1)
(UN2)--(U2N1);
\draw (f1)--(UN1);
\draw (U2N1)--(U2N2);
\draw (UN3)--(U2N4);
\draw (U2N3)--(U2N4);
\end{tikzpicture}
\ee
The Coulomb branch Hilbert series of the left quiver can be computed as follows:
\be
\begin{split}
& \sum_{a_1 \geq a_2 > -\infty} \, \sum_{m \in \BZ} \,\, \sum_{n \in \BZ} t^{2\Delta(\vec a, m, n)}  P_{U(2)}(t, \vec a) P_{U(1)}(t,m) P_{U(1)}(t,n)\\
& = \text{PE}\,[16 t^2 + 20 t^3 - 12 t^5 - 32 t^6+\dots]~,
\end{split}
\ee
with $\vec a= (a_1, a_2)$,
\be
\begin{split}
\Delta(\vec a, m, n) &= \Delta_{U(2)-U(1)}(\vec a, m)+  \Delta_{U(2)-U(1)}(\vec a, n) +  \Delta_{U(2)-U(2)}(\vec a, 0) \\
&\qquad +\Delta_{U(1)-U(1)}(m, 0) - \Delta^{\text{vec}}_{U(2)}(\vec a)
\end{split}
\ee
and all of the other notations are defined in \eref{alldefU}.  This is indeed equal to the first factor in the first line of \eref{DistrFlavEx}.  The right quiver in \eref{splitexample} is the $T(SU(3))$ theory whose Coulomb and Higgs branch Hilbert series is equal to the second factor in the first line of \eref{DistrFlavEx}.

\subsubsection*{\it Issues regarding $S$-folding the quiver tail}
Let us consider the case in which $n_2< 2N$.  In this case, in the mirror theory \eref{mirroneantisymm}, the $T$-link appears on right of the $U(2N)$ node that is attached with one flavour.  Let us suppose that the $T$-link connects two $U(n_2)$ gauge nodes where $1 \leq n_2 \leq 2N-1$.  
\be \label{mirr1oneantisymm}
\scalebox{0.75}{
\begin{tikzpicture}[font=\scriptsize,baseline]
\begin{scope}[auto,%
  every node/.style={draw, minimum size=0.5cm}, node distance=0.8cm];
\node[circle] (U2N1) at (0,0) {$2N$};
\node[circle,  right=of U2N1] (U2N2)  {$2N$};
\node[draw=none, right=of U2N2] (dots) {\Large $\cdots$};
\node[circle,  right=of dots] (U2N3)  {$2N$};
\node[circle, above left =of U2N1] (UN1)  {$N$};
\node[circle, below left =of U2N1] (UN2) {$N$};
\node[draw=none, right =of U2N3] (dots1) {\Large $\cdots$};
\node[circle, right=of dots1] (UN2Nm1) {$n_2$};
\node[circle, right =of UN2Nm1] (UN2Nm2) {$n_2$};
\node[draw=none, right =of UN2Nm2] (dots3) {\Large $\cdots$};
\node[circle, right =of dots3] (U1) {$1$};
\node[rectangle, left =of UN1] (f1)  {$1$};
\node[rectangle, above =of U2N3] (f2)  {$1$};
\end{scope}
\draw (U2N1)--(U2N2)--(dots)--(U2N3)--(dots1)--(UN2Nm1);
\draw (UN2Nm2)--(dots3)--(U1);
\draw[draw=red,snake it,thick,-] (UN2Nm1) -- (UN2Nm2) ;
\draw (UN1)--(U2N1)
(UN2)--(U2N1);
\draw (f1)--(UN1);
\draw (f2)--(U2N3);
\draw [decorate, decoration={brace, mirror}](1.3-1.5,-0.6)--(7-1.5,-0.6) node[black,midway,below, yshift=-0.1cm] {$n_1+n_2-2N-1~\text{nodes}$};
\end{tikzpicture}}
\ee
The Higgs branch dimension of such theory is
\be
\dim_{\BH} \CH_{\eref{mirr1oneantisymm}} = \dim_{\BH} \CH_{ \eref{mirroneantisymm}} + (n_2^2-n_2) -n_2^2 = 2N -n_2~.
\ee
Observe that this is non-zero for $1 \leq n_2 \leq 2N-1$.  However, as in \eref{zeroCoul}, we have $\dim_{\mathbb H}\,\cC_{(\ref{SfoldU2Nantisym})}=0$ for any $n_2$, since the two gauge nodes are connected by a $T$-link.  Hence, this is inconsistent with mirror symmetry, based on our assumption that the gauge nodes connected by a $T$-link do not contribute to the Coulomb branch.  One possible explanation of this inconsistency is that, in the presence of the $S$-fold, when move the D5 brane into the interval between NS5 branes, as depicted in \eref{braneoneantisymm1}, such a D5 brane has to cross the $S$-fold. Since $S$-fold can be regarded as the duality wall, the aforementioned D5 brane turns into an NS5 brane, with fractional D3 branes ending on it. In this sense, the mirror theory is not \eref{mirr1oneantisymm}.  We postpone the study of such a brane configuration to the future.

Now let us consider the following possibility:
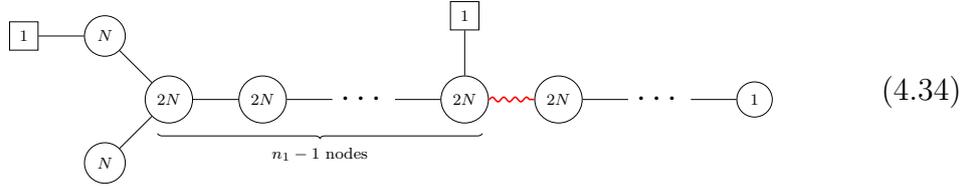
\be
\scalebox{0.75}{
\begin{tikzpicture}[font=\scriptsize,baseline]
\begin{scope}[auto,%
  every node/.style={draw, minimum size=0.5cm}, node distance=0.8cm];
\node[circle] (U2N1) at (0,0) {$2N$};
\node[circle,  right=of U2N1] (U2N2)  {$2N$};
\node[draw=none, right=of U2N2] (dots) {\Large $\cdots$};
\node[circle,  right=of dots] (U2N3)  {$2N$};
\node[circle, above left =of U2N1] (UN1)  {$N$};
\node[circle, below left =of U2N1] (UN2) {$N$};
\node[circle, right=of U2N3] (UN2Nm1) {$2N$};
\node[draw=none, right =of UN2Nm1] (dots3) {\Large $\cdots$};
\node[circle, right =of dots3] (U1) {$1$};
\node[rectangle, left =of UN1] (f1)  {$1$};
\node[rectangle, above =of U2N3] (f2)  {$1$};
\end{scope}
\draw (U2N1)--(U2N2)--(dots)--(U2N3);
\draw (UN2Nm1)--(dots3)--(U1);
\draw[draw=red,snake it,thick,-] (U2N3) -- (UN2Nm1) ;
\draw (UN1)--(U2N1)
(UN2)--(U2N1);
\draw (f1)--(UN1);
\draw (f2)--(U2N3);
\draw [decorate, decoration={brace, mirror}](1.3-1.5,-0.6)--(7-1.5,-0.6) node[black,midway,below, yshift=-0.1cm] {$n_1-1~\text{nodes}$};
\end{tikzpicture}}
\ee
In the brane picture \eref{braneoneantisymm}, this corresponds to putting the $S$-fold just next to the right of the D5 brane located in the the $(2N)$-th interval from the right.  This also corresponds to taking $n_2=2N$.  As before, the Higgs branch of this theory is expected to be a product of two hyperK\"ahler spaces, with one factor being
\be
\scalebox{0.75}{
\begin{tikzpicture}[font=\scriptsize,baseline]
\begin{scope}[auto,%
  every node/.style={draw, minimum size=0.5cm}, node distance=0.8cm];
\node[draw=none] (U2N3)  at (-2,0) {\red \large $\times$};
\node[circle] at (0,0) (UN2Nm1) {$2N$};
\node[draw=none, right =of UN2Nm1] (dots3) {\Large $\cdots$};
\node[circle, right =of dots3] (U1) {$1$};
\end{scope}
\draw (UN2Nm1)--(dots3)--(U1);
\draw[draw=red,snake it,thick,-] (-2,0) -- (UN2Nm1) ;
\end{tikzpicture}}
\ee
The Higgs branch dimension turns out to be negative if one assume that all gauge groups are completely broken:
\be
\frac{1}{2}(4N^2-2N)+\frac{1}{2}(2N-1)(2N)-(2N)^2 = -2N~.
\ee
Since the case of $n_2=2N$ has been discussed earlier, we shall not explore this possibility further.

\subsubsection*{\it Issues regarding putting the $S$-fold ``too close'' to the orientifold plane}
Consider the model with one rank-two antisymmetric hypermultiplet where we put an $S$-fold next to the $\text{O5}^-$ 
plane in the left diagram of \eref{braneonetwoanti}.  In this case we have $n_1=0$ and $n_2=n$ (with $n \geq 2N$).  The corresponding quiver diagram is
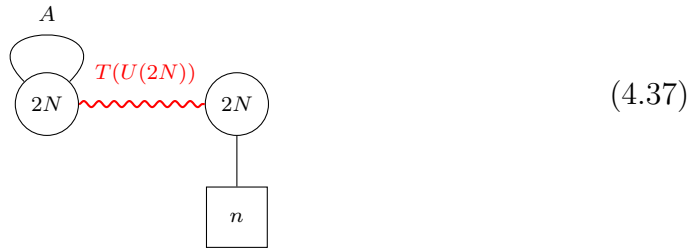
\be\label{NotWorkEx1A}
\begin{tikzpicture}[baseline,font=\scriptsize]
\begin{scope}[auto,%
  every node/.style={draw, minimum size=0.8cm}, node distance=0.6cm];
\node[draw, circle, black] (node1) at (0,0) {$2N$};
\node[draw, circle, black] (node2) at (2.5,0) {$2N$};
\draw[black] (node1) edge [out=45,in=135,loop,looseness=5]  (node1) node[draw=none] at (0,1.2) {$A$} ;
\node[draw, rectangle,black] (sqnode2) at (2.5,-1.5) {$n$};
\end{scope}
\draw (node2)--(sqnode2);
\draw[draw=red,solid, snake it, thick,-] (node1)--(node2);
\node[draw=none] at (1.3,0.4) {{\red$T(U(2N))$}};
\end{tikzpicture}
\ee
The dimension of the Higgs branch is
\be	
\begin{split}
	\text{dim}_\mathbb {H} \,\cH_{(\ref{NotWorkEx1A})}&=\frac{1}{2}(2N)(2N-1)+(4N^2-2N)+2Nn -4N^2 	-4N^2\\
	&=2Nn-2N^2-3N~,
\end{split}
\ee
assuming that the gauge symmetry is completely broken.
For a given $N$, this is positive for a sufficiently large $n$.  However, it is also worth pointing out that if we split the above Higgs branch into a product as in \eref{prodHiggs}, we see that the first factor 
\be
\scalebox{0.8}{
\begin{tikzpicture}[baseline,font=\scriptsize]
\begin{scope}[auto,%
  every node/.style={draw, minimum size=0.8cm}, node distance=0.6cm];
\node[draw, circle, black] (node1) at (0,1.5-1.5) {$2N$};
\draw[black] (node1) edge [out=45,in=135,loop,looseness=5]  (node1) node[draw=none] at (0,2.7-1.5) {$A$} ;
\end{scope}
\draw[draw=red,solid, snake it, thick,-] (node1)--(2,0);
\node[draw=none] at (2,0) {\large \red $\times$};
\end{tikzpicture}}
\ee
has a negative dimension, provided that the gauge symmetry $U(2N)$ is completely broken:
\be
\frac{1}{2}(4N^2-2N) + \frac{1}{2}(2N)(2N-1) - (2N)^2 = -2N~.
\ee
Since both gauge nodes are connected by the $T$-link, we expect that
\be	
	\text{dim}_\mathbb{H} \,\cC_{(\ref{NotWorkEx1A})}=0
\ee

The putative mirror theory can be obtained by inserting an $S$-fold next to the $\ON^-$ plane in \eref{braneoneantisymm}.  The corresponding quiver is
\be\label{NotWorkMirr1A}
\scalebox{0.9}
{\begin{tikzpicture}[font=\scriptsize,baseline]
\begin{scope}[auto,%
  every node/.style={draw, minimum size=0.5cm}, node distance=0.8cm];
\node[circle] (U2N1) at (1.5,0) {$2N$};
\node[draw=none, right=of U2N1] (dots) {\tiny$\cdots$};
\node[circle,  right=of dots] (U2N2)  {$2N$};
\node[circle, above left =of U2N1] (UN1)  {$N$};
\node[circle, below left =of U2N1] (UN2) {$N$};
\node[circle ,left =of UN1] (UN1S) {$N$};
\node[circle, right =of U2N2] (UNT1) {$2N-1$};
\node[draw=none, right=of UNT1] (dots1) {\tiny$\cdots$};
\node[circle, right=of dots1] (UNT2) {$1$};
\draw[draw=red,solid, snake it, thick,-] (UN1S)--(UN1);
\node[draw=none] at (-0.3,1.6) {{\red$T(U(N))$}};
\node[rectangle, left =of UN1S] (f1)  {$1$};
\node[rectangle, above =of U2N2] (f2)  {$1$};
\end{scope}
\draw (U2N1)--(dots)--(U2N2);
\draw (UN1)--(U2N1)
(UN2)--(U2N1);
\draw (f1)--(UN1S);
\draw (f2)--(U2N2);
\draw (U2N2)--(UNT1)--(dots1)--(UNT2);
\draw [decorate, decoration={brace, mirror}](1.3,-0.6)--(5,-0.6) node[black,midway,below, yshift=-0.1cm] {$n-2N-1~\text{nodes}$};
\end{tikzpicture}}
\ee
The Higgs and Coulomb branch dimensions read
\be	
\begin{split}
	\text{dim}_\mathbb {H} \,\cC_{(\ref{NotWorkMirr1A})} &=N+2N(n-2N-1)+\sum_{i=1}^{2N-1}i=2Nn-2N^2-2N~, \\
	\text{dim}_\mathbb {H} \,\cH_{(\ref{NotWorkMirr1A})} &=N+(N^2-N)+2N^2+2N^2+4N^2(n-2N-2)\\
	& \qquad +2N+\sum_{i=1}^{2N-1}i(i+1)-N^2-N^2-N^2\\
	& \qquad -4N^2(n-2N-1) -\sum_{i=1}^{2N-1}i(i+1) \\
	&= N~.
\end{split}
\ee
We see that these are inconsistent with mirror symmetry, if we assume that the gauge symmetry is completely broken and that the circular nodes that are connected by a $T$-link do not contribute to the Coulomb branch.  We see that these assumptions are violated or \eref{NotWorkMirr1A} is not a mirror theory of \eref{NotWorkEx1A} if we insert the $S$-fold next to the orientifold plane.

A similar issue also happens if we take $n_1=1$ and $n_2=n-1$ (with $n-1 \geq 2N$).  In which case, the putative mirror theory looks like
\be\label{NotWorkMirr1B}
\scalebox{0.9}
{\begin{tikzpicture}[font=\scriptsize,baseline]
\begin{scope}[auto,%
  every node/.style={draw, minimum size=0.5cm}, node distance=0.8cm];
\node[circle] (U2N1) at (1.5,0) {$2N$};
\node[draw=none, right=of U2N1] (dots) {\tiny$\cdots$};
\node[circle,  right=of dots] (U2N2)  {$2N$};
\node[circle, above left =of U2N1] (UN1)  {$N$};
\node[circle, below left =of U2N1] (UN2) {$N$};
\node[circle ,left =of UN1] (UN1S) {$N$};
\node[circle, right =of U2N2] (UNT1) {$2N-1$};
\node[draw=none, right=of UNT1] (dots1) {\tiny$\cdots$};
\node[circle, right=of dots1] (UNT2) {$1$};
\node[circle ,left =of UN2] (UNS) {$N$};
\draw[draw=red,solid, snake it, thick,-] (UN1S)--(UN1);
\draw[draw=red,solid, snake it, thick,-] (UNS)--(UN2);
\node[draw=none] at (-0.3,1.6) {{\red$T(U(N))$}};
\node[rectangle, left =of UN1S] (f1)  {$1$};
\node[rectangle, above =of U2N2] (f2)  {$1$};
\end{scope}
\draw (U2N1)--(dots)--(U2N2);
\draw (UN1)--(U2N1)
(UN2)--(U2N1);
\draw (f1)--(UN1S);
\draw (f2)--(U2N2);
\draw (U2N2)--(UNT1)--(dots1)--(UNT2);
\draw [decorate, decoration={brace, mirror}](1.3,-0.6)--(5,-0.6) node[black,midway,below, yshift=-0.1cm] {$n-2N-1~\text{nodes}$};
\end{tikzpicture}}
\ee
Upon computing the Higgs branch of this theory, the lower left part contributes a factor:
\be
\scalebox{0.8}{
\begin{tikzpicture}[baseline,font=\scriptsize]
\begin{scope}[auto,%
  every node/.style={draw, minimum size=0.8cm}, node distance=0.6cm];
\node[draw, circle, black] (node1) at (0,1.5-1.5) {$N$};
\end{scope}
\draw[draw=red,solid, snake it, thick,-] (node1)--(2,0);
\node[draw=none] at (2,0) {\large \red $\times$};
\end{tikzpicture}}
\ee
Assuming that the gauge symmetry is completely broken, we obtain a negative Higgs branch dimension:
\be
\frac{1}{2}(N^2-N)-N^2 = -\frac{1}{2}N(N+1)~.
\ee
This, again, confirms the statement that under the aforementioned assumptions, the $S$-fold cannot be inserted ``too close'' to the orientifold plane ($n_1 \geq 2$).  In other words, in order for the $S$-fold to co-exist with an orientifold fiveplane, it must be ``shielded'' by a sufficient number of fivebranes.

\subsubsection{$S$-folding the $USp(2N) \times U(2N) \times USp(2N)$ gauge theory}
Let us consider the following theory:
\be \label{USpUSfold}
\begin{tikzpicture}[baseline,font=\scriptsize]
\begin{scope}[auto,%
  every node/.style={draw, minimum size=1cm}, node distance=0.6cm];
\node[circle, blue] (USp1) at (0, 0) {$2N$};
\node[circle, right=of  USp1] (U2)  {$2N$};
\node[circle] (U3) at (4.2,0) {$2N$};
\draw[draw=red,solid, snake it, thick,-] (U2)--(U3);
\node[draw=none] at (2.9,0.4) {\red{$T(U(2N))$}};
\node[circle, blue, right=of U3] (USp2) {$2N$};
\node[rectangle, red, below=of USp1] (n1) {$2n_1$};
\node[rectangle, below=of U2]  (fl) {$F_1$};
\node[rectangle, below=of U3]  (fl1) {$F_2$};
\node[rectangle, red,  below=of USp2] (n2){$2n_2$};
\end{scope}
 \draw  (USp1)--(U2);
  \draw  (USp1)--(n1);
  \draw  (U2)--(fl);
  \draw  (U3)--(fl1);
   \draw  (USp2)--(n2);
   \draw  (U3)--(USp2); 
\end{tikzpicture}
\ee
The brane construction for this is given by \eref{braneUSpUUUSp}, with $m=1$ and with an $S$-fold inserted in the interval labelled by $f_1$.  The $S$-fold partitions $f_1$ D5 branes into $F_1$ and $F_2$ D5 branes on the left and on the right of the $S$-fold, respectively.  The dimension of the Higgs branch of this theory reads
\be
\begin{split}
	\text{dim}_\mathbb{H} \,\cH_{(\ref{USpUSfold})} &=2N n_1 +4N^2+2N F_1 + (4N^2-2N)+2N F_2+4N^2\\
	& \qquad +2N n_2 - N(2N+1)-4N^2-4N^2- N(2N+1) \\
	&=2N(F_1+F_2+n_1+n_2-2)~,
\end{split}
\ee
and, for the Coulomb branch, we find
\be
	\text{dim}_\mathbb{H} \,\cC_{(\ref{USpUSfold})}=2N.
\ee
We remark that it is not possible to insert an $S$-fold in the interval labelled by $n_1$ in the diagram \eref{braneUSpUUUSp}.  The reason is that such a brane interval corresponds to the gauge group $USp(2N)$, and not $USp'(2N)$.  We do not have the notion of a $T(USp(2N))$ link since $USp(2N)$ is not invariant under the $S$-duality.  This supports the point we made earlier that the $S$-fold cannot be inserted ``too close'' to the orientifold plane; it must be ``shielded'' by a sufficient numbers of fivebranes.

In order to obtain the mirror configuration, we can insert an $S$-fold anywhere between two D5-branes denoted by the black dots in \eref{mirrUSpUUUUSp}. (Recall that $m=1$ in this case.)  In terms of the quiver, this means that we can put the $T$-link anywhere in between the two $(2N)$-nodes attached by one flavour.  For example, for $N=1$, $n_1=n_2=3$, $F_1=1$ and $F_2=0$, the mirror theory is
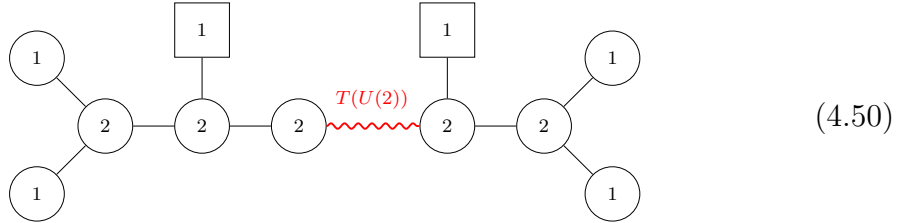
\be\label{USpUSfoldMirr}
\scalebox{0.9}{
\begin{tikzpicture}[baseline,font=\scriptsize]
\begin{scope}[auto,%
every node/.style={draw, minimum size=0.8cm}, node distance=0.6cm];
\node[circle] (U2N1) at (0, 0) {$2$};
\node[circle, above left=of U2N1] (UN1) {$1$};
\node[circle, below left=of U2N1] (UN2) {$1$};
\node[circle, right=of U2N1] (U2N2) {$2$};
\node[circle, right=of U2N2] (U2N3) {$2$};
\node[circle] (U2N4) at (5,0) {$2$};
\node[circle, right=of U2N4] (U2N5) {$2$};
\node[rectangle, above=of U2N2] (f1) {$1$};
\node[rectangle, above=of U2N4] (f2) {$1$};
\draw[draw=red,solid, snake it, thick,-] (U2N3)--(U2N4);
\node[draw=none] at (3.9,0.4) {\red{$T(U(2))$}};
\node[circle, above right=of U2N5] (UN3) {$1$};
\node[circle, below right=of U2N5] (UN4) {$1$};
\end{scope}
  \draw (UN1)--(U2N1)--(U2N2)--(U2N3);
  \draw (UN2)--(U2N1);
    \draw (U2N2)--(f1);
    \draw (U2N4)--(f2);   \draw (U2N4)--(U2N5);
    \draw (UN3)--(U2N5)--(UN4);
\end{tikzpicture}}
\ee
In order to compute the dimensions of Higgs and Coulomb branches of the mirror theory we can simply start with the corresponding non $S$-folded theory and observe that inserting a $T$-link implies the following:
\begin{itemize}
	\item{For the Higgs branch, we need to add the dimension of the $T(U(2N))$ link, that in this case gives 
	$4N^2-2N$ and subtract the gauging of the extra $U(2N)$, hence we subtract $4N^2$; in total we find that 
	\be
	\begin{split}
		\text{dim}_\mathbb{H} \,\cH_{\text{mirr of}\,\, \eref{USpUSfold}} &=\text{dim}_\mathbb{H}\, \cH_{\eref{mirrUSpUUUSp}}	+(4N^2-2N)-4N^2 \\
		&=\text{dim}_\mathbb{H}\, \cH_{\eref{mirrUSpUUUSp}}-2N \\
		&= (N+2N+N)-2N = 2N~.
	\end{split}
	\ee}
	\item{For the Coulomb branch, the result of inserting an $S$-fold is to add one gauge node and then consider 
	that the ones connected by the $T$-link are frozen, so in total we have
	\be
	\begin{split}
		\text{dim}_\mathbb{H}\, \cC_{\text{mirr of}\,\, \eref{USpUSfold}} &=\text{dim}_\mathbb{H}\, \cC_{\eref{mirrUSpUUUSp}}-2N \\
		&= 2N(F_1+F_2+n_1+n_2-2)~,~ \text{with $f_1=F_1+F_2$}~.
	\end{split}
	\ee}
\end{itemize}
These are in agreement with mirror symmetry.

In the above example of $N=1$, $n_1=n_2=3$, $f_1=1$ and $f_2=0$, one can compute the Hilbert series for (\ref{USpUSfold}) and its mirror (\ref{USpUSfoldMirr}).  The unrefined results are
\be 
\begin{split}
	H[\cH_{(\ref{USpUSfold})}] &= H[\cC_{(\ref{USpUSfoldMirr})}]\\
	&=\text{PE}\,[16 t^2 + 12 t^3 - 15 t^4 - 40 t^5 + 19 	t^6+ \dots ]\times \\
	& \qquad  \text{PE}\,[15 t^2 - 16 t^4 + 35 t^6 + \dots ]  ~,
\end{split}
\ee
and
\be
\begin{split}
	H[\cC_{(\ref{USpUSfold})}]&=H[\cH_{(\ref{USpUSfoldMirr})}] \\
	&=H[\cC_{\text{$USp(2)$ with 5 flv}}]^2=\text{PE}\,[t^4 + t^6 + t^8+\dots]^2~.
\end{split}
\ee
The above results deserve some explanations.  In \eref{USpUSfoldMirr}, the Coulomb branch symmetry can be seen from the after taking the two $U(2)$ gauge groups connected by the $T$-link to be two separate flavour symmetries. The left part gives an $SU(4) \times U(1)$ symmetry due to the fact that the balanced nodes form an $A_3$ Dynkin diagram and that there is one overbalanced node (namely, the $U(2)$ node that is attached to one flavour). The right part gives an $SU(4)$ symmetry due to the fact that the balanced nodes form an $A_3$ Dynkin diagram \cite{Gaiotto:2008ak}.  The Coulomb branch of (\ref{USpUSfold}) is identified with a product of two copies of the Coulomb branch of $USp(2)$ gauge theory with $5$ flavours due to the following reason. The nodes connected by the $T$-link do not contribute to the Coulomb branch and therefore each of the left and the right parts contains the $USp(2)$ gauge theory with $2N+n_1 = 2+3=5$ flavours.

\section{Models with an orientifold threeplane} \label{sec:O3}
\subsection{The cases without an $S$-fold}
In this subsection, we summarise brane constructions for the elliptic models with alternating orthogonal and symplectic gauge groups, in the absence of the $S$-fold.  Such brane configurations and their $S$-duals were studied extensively in \cite{Feng:2000eq} (see also \cite{Cabrera:2017njm} for a related discussion).  For brevity of the discussion, we shall not go through the detail on how to obtain the $S$-dual configurations but simply state the results.  The following quiver diagrams and their brane configurations will turn out to be useful for the discussion in the subsequent subsections.

\paragraph{\it The $SO(2N)\times USp(2N)$ gauge theory with two bifundamentals and $n$ flavours for $USp(2N)$ and its mirror.}  Their quivers are
\be \label{SO2NxUSp2N}
\scalebox{0.8}{
\begin{tikzpicture}[baseline]
\tikzstyle{every node}=[font=\footnotesize]
\node[draw, circle, red] (node1) at (0,2) {$2N$};
\node[draw, circle, blue] (node3) at (0,0) {$2N$};
\node[draw, rectangle, red] (sqnode) at (0,-1.5) {$2n$};
\draw[draw=black,solid,thick,-]  (node1) to [bend left] (node3) ; 
\draw[draw=black,solid,thick,-]  (node1) to [bend right] (node3) ; 
\draw[draw=black,solid,thick,-]  (node3)--(sqnode) ; 
\end{tikzpicture}}
\qquad \qquad\qquad \qquad
\scalebox{0.6}{
\begin{tikzpicture}[baseline, scale=0.6,font=\scriptsize]
\begin{scope}[auto,%
  every node/.style={draw, minimum size=1.2cm}];
\def \n {7}
\def \radius {4.2cm}
\def \margin {14} 
\draw[-, >=latex] ({360/\n * (1 - 3)+\margin}:\radius);
arc ({360/\n * (1 - 3)+\margin}:{360/\n * (1-2)-\margin}:\radius);
\node[draw=none] at ({360/\n * (2 - 2)}:\radius) {{\large $\vdots$}};
\draw[-, >=latex] ({360/\n * (2 - 3)+\margin}:\radius);
arc ({360/\n * (2 - 3)+\margin}:{360/\n * (2-2)-\margin}:\radius);
\draw[-, >=latex] ({360/\n * (3 - 3)+\margin}:\radius);
arc ({360/\n * (3 - 3)+\margin}:{360/\n * (3-2)-\margin}:\radius);
\draw[-, >=latex] ({360/\n * (4 - 3)+\margin}:\radius);
arc ({360/\n * (5 - 3)+\margin}:{360/\n * (5-2)-\margin}:\radius);
\draw[-, >=latex] ({360/\n * (5 - 3)+\margin}:\radius);
arc ({360/\n * (5 - 3)+\margin}:{360/\n * (5-2)-\margin}:\radius);
\node[draw, circle, blue] (n2) at ({360/\n * (0 - 2)}:\radius) {$2N$};
\node[draw, circle, red] at ({360/\n * (1 - 2)}:\radius) {$2N+1$};
\node[draw, circle,blue] at ({360/\n * (3 - 2)}:\radius) {$2N'$};
\node[draw, circle,red] at ({360/\n * (4 - 2)}:\radius) {$2N+1$};
\node[draw, circle,blue ] (n1) at ({360/\n * (5 - 2)}:\radius) {$2N$};
\foreach \s in {1,...,5}
{	
	\draw[-, >=latex] ({360/\n * (\s - 3)+\margin}:\radius) 
	arc ({360/\n * (\s - 3)+\margin}:{360/\n * (\s-2)-\margin}:\radius);
}
\draw[-, >=latex] ({360/\n * (0 - 3)+\margin}:\radius) 
arc ({360/\n * (0 - 3)+\margin}:{360/\n * (0-2)-\margin}:\radius);
\node[draw, circle,red] at ({360/\n * (6 - 2)}:\radius) {$2N$};
\node[draw, circle,red] at ({360/\n * (-1 - 2)}:\radius) {$2N$};
\draw[-, >=latex] ({360/\n * (-1 - 3)+\margin}:\radius) 
arc ({360/\n * (-1 - 3)+\margin}:{360/\n * (-1-2)-\margin}:\radius);
\node[draw=none] at (0,-11) {\large $\substack{\text{One red $(2N)$ node + two blue $(2N)$ nodes} \\~\\ \text{ with a half-flavour each, and alternating} \\~\\ \text{$(n-2)$ blue $(2N')$ nodes with no flavour} \\~ \\ \text{+ $(n-1)$ red $(2N+1)$ nodes with no flavour}}$};
\node[draw, rectangle ,red] (f1) at (-6.8,1.8) {$1$};
\node[draw, rectangle ,red] (f2) at (-1,-7.5) {$1$};
\draw[-,solid] (f1) to (n1);
\draw[-,solid] (f2) to (n2);
\end{scope}
\end{tikzpicture}}
\ee
Their brane configurations are, respectively, given by \cite[Fig. 23]{Feng:2000eq}:
\be \label{braneDC}
\scalebox{0.85}{
\begin{tikzpicture}[baseline]
\tikzstyle{every node}=[font=\footnotesize, node distance=0.45cm]
\tikzset{decoration={snake,amplitude=.4mm,segment length=2mm,
                       post length=0mm,pre length=0mm}}
\draw[blue,thick] (0,0) circle (1.5cm);
\draw[black,thick] (0,1) -- (0,2) node at (0,2.2) {$\frac{1}{2}\text{NS5}$};
\draw[black,thick] (0,-1) -- (0,-2) node at (0, -2.2) {};
\node[draw=none, red] at (-1.8,0) {$\mathbf{-}$};
\def \n {6}
\def \radius {1.2cm}
\def \margin {0} 
\node[draw=none, blue, thick] at (-1, 0) {{$2N$}}; 
\node[draw=none, blue, thick] (g3) at (1, 0) {{$2N$}}; 
\node[draw=none] (g1) at (0.75, 1.3) {$\bullet$};
\node[draw=none, blue] (g1) at (0.4, 1.7)  {$\mathbf{+}$};
\node[draw=none] (g1) at (1.3, 0.75) {$\bullet$};
\node[draw=none, blue] (g1) at (1.25, 1.25)  {$\mathbf{\tilde +}$};
\node[draw=none, blue] (g1) at (1.7, 0.45)  {$\mathbf{+}$};
\node[draw=none] at (1.7,0) {{\Large $\mathbf{\vdots}$}};
\node[draw=none] (g1) at (0.75, -1.3) {$\bullet$} node at (1,-1.6)  {$\frac{1}{2}\text{D5}$};
\node[draw=none, blue] (g1) at (0.4, -1.7)  {$\mathbf{+}$};
\end{tikzpicture}}
\qquad \qquad\qquad \qquad
\scalebox{0.85}{
\begin{tikzpicture}[baseline]
\tikzstyle{every node}=[font=\footnotesize]
\tikzset{decoration={snake,amplitude=.4mm,segment length=2mm,
                       post length=0mm,pre length=0mm}}
\draw[black,thick] (0,1) -- (0,2) node at (0,2.2) {$\frac{1}{2}\text{NS5}$};    
\draw[blue,thick] (0,0) circle (1.5cm);
\def \n {6}
\def \radius {1.2cm}
\def \margin {0} 
\foreach \s in {1,...,10}
{
	\node[draw=none] (\s) at ({360/\n * (\s - 2)+30}:{\radius-10}) {};
}
\node[draw=none, circle] (last) at ({360/3 * (3 - 1)+30}:{\radius-10}) {};
\node[draw=none,  below right= of 1] (f1) {};
\node[draw=none, above right= of 2] (f2) {};
\node[draw=none, above = of 3] (f3) {};
\node[draw=none, above left= of 4] (f4) {};
\node[draw=none,  below left= of 5] (f5) {};
\node[draw=none,  below = of last] (f6) {};
\node[draw=none] at (1.8,0) {{\Large $\mathbf{\vdots}$}};
\node[draw=none] at (0,-2.4) {};
\draw[-, >=latex,black, thick] (0.7, 0.7) to (1.56, 1.56);
\draw[-, >=latex,black, thick] (0.97, 0.26) to (2.13, 0.57);
\draw[-, >=latex,black, thick] (0.7, -0.7) to (1.56, -1.56);
\draw[-, >=latex,black, thick] (2.12, -0.57) to (0.98, -0.25);
\node[draw=none] (g1) at (0.6, 1.35) {$\bullet$};
\node[draw=none] (g2) at (0.6, -1.40) {$\bullet$};
\node[draw=none, blue, thick] (g3) at (0.45, 1.10) {\tiny{$2N$}}; 
\node[draw=none, blue, thick] (g5) at (0.8, 0.55) {\tiny{$2N+1$}}; 
\node[draw=none, blue, thick] (g5) at (0.8, -0.55) {\tiny{$2N+1$}}; 

\node[draw=none, blue, thick] (g8) at (0.5, -1.1) {\tiny{$2N$}}; 
\node[draw=none, blue, thick] (g10) at (-1.2, 0) {\tiny{$2N$}}; 
\node[draw=none, red] at (-1.8, 0)  {$\mathbf{-}$};
\node[draw=none, blue] at (0.35, 1.65)  {$\mathbf{+}$};
\node[draw=none, blue] at (0.95, 1.45)  {$\mathbf{\tilde +}$};
\node[draw=none, red] at (1.5, 0.85)  {$\mathbf{\tilde -}$};
\node[draw=none, blue] at (0.35, -1.65)  {$\mathbf{+}$};
\node[draw=none, blue] at (0.95, -1.45)  {$\mathbf{\tilde +}$};
\node[draw=none, red] at (1.5, -0.85)  {$\mathbf{\tilde -}$};
\draw[-, >=latex,black, thick] (last) to (f6);
\end{tikzpicture}}
\ee
where in the left diagram we have $n$ half-D5 branes, and in the right diagram we have $n$ half-NS5 branes.  Here and subsequently, we denote in blue the number of half-D3 branes at each interval between two succesive half-NS5 branes.  Note that one may also add flavours (say, $m$ flavours, or equivalently a blue rectangular node with label $2m$) to the $SO(2N)$ gauge group in the left diagram of \eref{SO2NxUSp2N}, the resulting mirror quiver can be obtained from the right diagram of \eref{SO2NxUSp2N} by simply replacing the $(2N)$ red node by a series of alternating $m+1$ red $(2N)$ nodes and $m$ blue $(2N)$ nodes:
\be
\scalebox{0.7}{
\begin{tikzpicture}[baseline,font=\scriptsize]
\begin{scope}[auto,%
  every node/.style={draw, minimum size=0.8cm}, node distance=0.6cm];
\node[draw=none] (emp1) at (-2,0) {};
\node[draw=none] (emp2) at (2,0) {};
\node[draw, circle, red] (node1) at (0,0) {$2N$};
\node[draw, rectangle, blue] (f1) at (0,-1.5) {$2m$};
\end{scope}
\draw (node1)--(emp1);
\draw (node1)--(emp2);
\draw (node1)--(f1);
\end{tikzpicture}}
\quad \longrightarrow \quad
\scalebox{0.7}{
\begin{tikzpicture}[baseline,font=\scriptsize]
\begin{scope}[auto,%
  every node/.style={draw, minimum size=0.8cm}, node distance=0.6cm];
\node[draw=none] (emp1) at (-1.5,0) {};
\node[draw, circle, red] (n1) at (0,0) {$2N$};
\node[draw, circle, blue] (n2) at (1.5,0) {$2N$};
\node[draw, circle, red] (n3) at (3,0) {$2N$};
\node[draw, circle, blue] (n35) at (4.5,0) {$2N$};
\node[draw=none] (dots) at (6,0) {$\cdots$};
\node[draw, circle, red] (n4) at (7.5,0) {$2N$};
\node[draw=none] (emp2) at (9,0) {};
\end{scope}
\draw (emp1)--(n1)--(n2)--(n3)--(n35)--(dots)--(n4)--(emp2);
\draw [decorate, decoration={brace,mirror}](-0.2,-0.5)--(7.6,-0.5) node[black,midway,yshift=-0.5cm] {\scriptsize $(m+1)$ red nodes \& $m$ blue nodes};
\end{tikzpicture}}
\ee

\paragraph{\it The $USp'(2N)\times SO(2N+1)$ gauge theory with two bifundamentals and $n$ flavours for $SO(2N+1)$ and its mirror.}  Their quivers are
\be \label{CNxBN}
\scalebox{0.8}{
\begin{tikzpicture}[baseline]
\tikzstyle{every node}=[font=\footnotesize]
\node[draw, circle, blue] (node1) at (0,2) {$2N'$};
\node[draw, circle, red] (node3) at (0,0) {\scriptsize $2N+1$};
\node[draw, rectangle, blue] (sqnode) at (0,-1.5) {$2n$};
\draw[draw=black,solid,thick,-]  (node1) to [bend left] (node3) ; 
\draw[draw=black,solid,thick,-]  (node1) to [bend right] (node3) ; 
\draw[draw=black,solid,thick,-]  (node3)--(sqnode) ; 
\end{tikzpicture}}
\qquad \qquad \qquad \qquad \qquad 
\scalebox{0.7}{
\begin{tikzpicture}[baseline, scale=0.6,font=\scriptsize]
\begin{scope}[auto,%
  every node/.style={draw, minimum size=1.2cm}];
\def \n {7}
\def \radius {4.2cm}
\def \margin {14} 
\draw[-, >=latex] ({360/\n * (1 - 3)+\margin}:\radius);
arc ({360/\n * (1 - 3)+\margin}:{360/\n * (1-2)-\margin}:\radius);
\node[draw=none] at ({360/\n * (2 - 2)}:\radius) {{\large $\vdots$}};
\draw[-, >=latex] ({360/\n * (2 - 3)+\margin}:\radius);
arc ({360/\n * (2 - 3)+\margin}:{360/\n * (2-2)-\margin}:\radius);
\draw[-, >=latex] ({360/\n * (3 - 3)+\margin}:\radius);
arc ({360/\n * (3 - 3)+\margin}:{360/\n * (3-2)-\margin}:\radius);
\draw[-, >=latex] ({360/\n * (4 - 3)+\margin}:\radius);
arc ({360/\n * (5 - 3)+\margin}:{360/\n * (5-2)-\margin}:\radius);
\draw[-, >=latex] ({360/\n * (5 - 3)+\margin}:\radius);
arc ({360/\n * (5 - 3)+\margin}:{360/\n * (5-2)-\margin}:\radius);
\node[draw, circle, red] at ({360/\n * (0 - 2)}:\radius) {$2N+2$};
\node[draw, circle, blue] at ({360/\n * (1 - 2)}:\radius) {$2N$};
\node[draw, circle,red] at ({360/\n * (3 - 2)}:\radius) {$2N+2$};
\node[draw, circle,blue] at ({360/\n * (4 - 2)}:\radius) {$2N$};
\node[draw, circle,red ] at ({360/\n * (5 - 2)}:\radius) {$2N+2$};
\foreach \s in {1,...,6}
{	
	\draw[-, >=latex] ({360/\n * (\s - 3)+\margin}:\radius) 
	arc ({360/\n * (\s - 3)+\margin}:{360/\n * (\s-2)-\margin}:\radius);
}
\draw[-, >=latex] ({360/\n * (0 - 3)+\margin}:\radius) 
arc ({360/\n * (0 - 3)+\margin}:{360/\n * (0-2)-\margin}:\radius);
\node[draw, circle,blue] (n1) at ({360/\n * (6 - 2)}:\radius) {$2N'$};
\node[draw, circle,blue] (n2) at ({360/\n * (-1 - 2)}:\radius) {$2N'$};
arc ({360/\n * (-1 - 3)+\margin}:{360/\n * (-1-2)-\margin}:\radius);
\node[draw=none] at (0,-9) {\large $\substack{\text{$n$ red circular nodes + $(n-1)$ blue} \\~ \\ \text{usual circular nodes + 2 blue nodes} \\~ \\ \text{connected by $T(USp'(2N))$}}$};
\node[draw, rectangle ,red, below=of n2] (f2) {$2$};
\draw[-,solid] (f2) to (n1);
\end{scope}
\end{tikzpicture}}
\ee
The corresponding brane configurations are respectively given by \cite[Fig. 29]{Feng:2000eq}:
\be \label{braneCNxBN}
\begin{tikzpicture}[baseline]
\tikzstyle{every node}=[font=\footnotesize, node distance=0.45cm]
\tikzset{decoration={snake,amplitude=.4mm,segment length=2mm,
                       post length=0mm,pre length=0mm}}
\draw[blue,thick] (0,0) circle (1.5cm);
\draw[black,thick] (0,1) -- (0,2) node at (0,2.2) {$\frac{1}{2}\text{NS5}$};
\draw[black,thick] (0,-1) -- (0,-2) node at (0, -2.2) {};
\node[draw=none, blue] at (-1.8,0) {$\mathbf{\tilde +}$};
\def \n {6}
\def \radius {1.2cm}
\def \margin {0} 
\node[draw=none, blue, thick] at (-0.9, 0) {\scriptsize {$2N$}}; 
\node[draw=none, blue, thick] (g3) at (0.9, 0) {\scriptsize {$2N+1$}}; 
\node[draw=none] (g1) at (0.75, 1.3) {$\bullet$};
\node[draw=none, red] (g1) at (0.4, 1.7)  {$\mathbf{\tilde -}$};
\node[draw=none] (g1) at (1.3, 0.75) {$\bullet$};
\node[draw=none, red] (g1) at (1.25, 1.25)  {$\mathbf{-}$};
\node[draw=none, red] (g1) at (1.7, 0.45)  {$\mathbf{\tilde -}$};
\node[draw=none] at (1.7,0) {{\Large $\mathbf{\vdots}$}};
\node[draw=none] (g1) at (0.75, -1.3) {$\bullet$} node at (1,-1.7) {\scriptsize $\frac{1}{2}\text{D5}$};
\node[draw=none, red] (g1) at (0.4, -1.7)  {$\mathbf{\tilde -}$};
\end{tikzpicture}
\qquad \qquad \qquad \qquad
\begin{tikzpicture}[baseline]
\tikzstyle{every node}=[font=\footnotesize]
\tikzset{decoration={snake,amplitude=.4mm,segment length=2mm,
                       post length=0mm,pre length=0mm}}
\draw[black,thick] (0,1) -- (0,2) node at (0,2.2) {$\frac{1}{2}\text{NS5}$};    
\draw[blue,thick] (0,0) circle (1.5cm);
\def \n {6}
\def \radius {1.2cm}
\def \margin {0} 
\foreach \s in {1,...,10}
{
	\node[draw=none] (\s) at ({360/\n * (\s - 2)+30}:{\radius-10}) {};
}
\node[draw=none, circle] (last) at ({360/3 * (3 - 1)+30}:{\radius-10}) {};
\node[draw=none,  below right= of 1] (f1) {};
\node[draw=none, above right= of 2] (f2) {};
\node[draw=none, above = of 3] (f3) {};
\node[draw=none, above left= of 4] (f4) {};
\node[draw=none,  below left= of 5] (f5) {};
\node[draw=none,  below = of last] (f6) {};
\node[draw=none] at (1.8,0) {{\Large $\mathbf{\vdots}$}};
\node[draw=none] at (0,-2.4) {};
\draw[-, >=latex,black, thick] (0.7, 0.7) to (1.56, 1.56);
\draw[-, >=latex,black, thick] (0.97, 0.26) to (2.13, 0.57);
\draw[-, >=latex,black, thick] (0.7, -0.7) to (1.56, -1.56);
\draw[-, >=latex,black, thick] (2.12, -0.57) to (0.98, -0.25);
\node[draw=none] (g1) at (-1.45, 0.4) {$\bullet$};
\node[draw=none] (g2) at (-1.45, -0.4) {$\bullet$};
\node[draw=none, blue] at (-1.3, 1.2)  {$\mathbf{+}$};
\node[draw=none, blue] at (-1.3, -1.2)  {$\mathbf{+}$};
\node[draw=none, blue, thick] (g3) at (0.45, 1.10) {\tiny{$2N+2$}}; 
\node[draw=none, blue, thick] (g5) at (1, 0.55) {\tiny{$2N$}}; 
\node[draw=none, blue, thick] (g5) at (1, -0.55) {\tiny{$2N$}}; 

\node[draw=none, blue, thick] (g8) at (0.5, -1.1) {\tiny{$2N+2$}}; 
\node[draw=none, blue, thick] (g10) at (-1.2, 0) {\tiny{$2N$}}; 
\node[draw=none, blue] at (-1.8, 0)  {$\mathbf{\tilde +}$};
\node[draw=none, red] at (0.65, 1.65)  {$\mathbf{-}$};
\node[draw=none, blue] at (1.5, 0.85)  {$\mathbf{+}$};
\node[draw=none, red] at (0.65, -1.65)  {$\mathbf{-}$};
\node[draw=none, blue] at (1.5, -0.85)  {$\mathbf{+}$};
\draw[-, >=latex,black, thick] (last) to (f6);
\end{tikzpicture}
\ee
where in the left diagrams there are $2n$ half-D5 branes, and on the right diagram there are $2n$ half NS5 branes.  One may also add flavours (say, $m$ flavours or equivalently a red square node with label $2m$) to the $USp'(2m)$ gauge group in the left diagram of \eref{CNxBN}, the resulting mirror quiver can be obtained from the right diagram of \eref{CNxBN} by making the following replacement:
\be
\scalebox{0.8}{
\begin{tikzpicture}[baseline,font=\scriptsize]
\begin{scope}[auto,%
  every node/.style={draw, minimum size=0.8cm}, node distance=0.6cm];
\node[draw=none] (emp1) at (-2,0) {};
\node[draw=none] (emp2) at (2,0) {};
\node[draw, circle, blue] (node1) at (0,0) {$2N'$};
\node[draw, rectangle, red] (f1) at (0,-1.5) {$2m$};
\end{scope}
\draw (node1)--(emp1);
\draw (node1)--(emp2);
\draw (node1)--(f1);
\end{tikzpicture}}
\quad \longrightarrow \quad
\scalebox{0.7}{
\begin{tikzpicture}[baseline,font=\scriptsize]
\begin{scope}[auto,%
  every node/.style={draw, minimum size=0.8cm}, node distance=0.6cm];
\node[draw=none] (emp1) at (-1.5,0) {};
\node[draw, circle, blue] (n1) at (0,0) {$2N$};
\node[draw, circle, red] (n2) at (1.5,0) {$2N$};
\node[draw, circle, blue] (n3) at (3,0) {$2N$};
\node[draw, circle, red] (n35) at (4.5,0) {$2N$};
\node[draw=none] (dots) at (6,0) {$\cdots$};
\node[draw, circle, blue] (n4) at (7.5,0) {$2N$};
\node[draw=none] (emp2) at (9,0) {};
\node[draw, rectangle, red] (f1) at (0,-1.5) {$1$};
\node[draw, rectangle, red] (f4) at (7.5,-1.5) {$1$};
\end{scope}
\draw (emp1)--(n1)--(n2)--(n3)--(n35)--(dots)--(n4)--(emp2);
\draw (f1)--(n1);
\draw (f4)--(n4);
\draw [decorate, decoration={brace}](1.0,0.5)--(6.5,0.5) node[black,midway,yshift=0.5cm] {\scriptsize $m$ red nodes \& $(m-1)$ blue nodes};
\end{tikzpicture}}
\ee

\subsection{Quiver with a $T(SO(2N))$ loop}
We start by examining the following brane configuration and the corresponding quiver:
\be \label{SO2Nhead}
\scalebox{0.95}{
\begin{tikzpicture}[baseline]
\tikzstyle{every node}=[font=\footnotesize, node distance=0.45cm]
\tikzset{decoration={snake,amplitude=.4mm,segment length=2mm,
                       post length=0mm,pre length=0mm}}
\draw[blue,thick] (0,0) circle (1.5cm);
\draw[draw=red,snake it,thick,-] (0,1) -- (0,2) ;
\node[draw=none] (g1) at (0.7, -1.33) {$\bullet$};
\node[draw=none] (g1) at (-0.7, -1.33) {$\bullet$};
\node[draw=none] (g1) at (1.33, -0.7) {$\bullet$};
\node[draw=none] (g1) at (-1.33, -0.7) {$\bullet$};
\node[draw=none] (g1) at (-1.8, -0.7)  {$\tfrac{1}{2}\text{D5}$};
\node[draw=none, blue] (g1) at (1, 0)  {$2N$};
\node[draw=none] at (0,-1.7) {{\Large $\mathbf{\dots}$}};
\def \n {6}
\def \radius {1.2cm}
\def \margin {0} 
\node[draw=none, red] (g1) at (-0.6, 1.7)  {$\mathbf{-}$};
\node[draw=none, red] at (0.6, 1.7)  {$\mathbf{-}$};
\node[draw=none, red] at (1.3, -1.2)  {$\mathbf{\tilde -}$};
\node[draw=none, red] at (-1.3, -1.2)  {$\mathbf{\tilde -}$};
\end{tikzpicture}}
\qquad \qquad \qquad\qquad
\begin{tikzpicture}[baseline]
\tikzstyle{every node}=[font=\footnotesize]
\node[draw, circle, red] (node1) at (0,1) {$2N$};
\draw[red,thick] (node1) edge [out=45,in=135,loop,looseness=5, snake it]  (node1);
\node[draw=none] at (0,2.4) {{\red $T(SO(2N))$}};
\node[draw, rectangle,blue] (sqnode) at (0,-1) {$2n$};
\draw (node1)--(sqnode);
\end{tikzpicture}
\ee
where in the left diagram the red wriggly denotes the $S$-fold and there are $2n$ half D5 branes.  In order to obtain the mirror theory, we apply $S$-duality to the above brane system.  The result is
\be \label{mirrSOhead}
\begin{tikzpicture}[baseline]
\tikzstyle{every node}=[font=\footnotesize]
\tikzset{decoration={snake,amplitude=.4mm,segment length=2mm,
                       post length=0mm,pre length=0mm}}
  
\draw[draw=red,snake it,thick,-] (0,1) -- (0,2) ;
\draw[blue,thick] (0,0) circle (1.5cm);\def \n {6}
\def \radius {1.2cm}
\def \margin {0} 
\foreach \s in {1,...,10}
{
	\node[draw=none] (\s) at ({360/\n * (\s - 2)+30}:{\radius-10}) {};
}
\node[draw=none, circle] (last) at ({360/3 * (3 - 1)+30}:{\radius-10}) {};
\node[draw=none,  below right= of 1] (f1) {};
\node[draw=none, above right= of 2] (f2) {};
\node[draw=none, above = of 3] (f3) {};
\node[draw=none, above left= of 4] (f4) {};
\node[draw=none,  below left= of 5] (f5) {};
\node[draw=none,  below = of last] (f6) {};
\node[draw=none] at (0,-1.7) {{\Large $\mathbf{\dots}$}};
\draw[-, >=latex,black, thick] (0.7, -0.7) to (1.56, -1.56);
\draw[-, >=latex,black, thick] (2.12, -0.57) to (0.98, -0.25);
\draw[-, >=latex,black, thick] (-0.7, -0.7) to (-1.56, -1.56);
\draw[-, >=latex,black, thick] (-2.12, -0.57) to (-0.98, -0.25);
\node[draw=none, blue, thick] (g10) at (1, 0.5) {\scriptsize {$2N$}}; 
\node[draw=none, blue, thick] at (-1, 0.5) {\scriptsize {$2N$}}; 
\node[draw=none, blue, thick] at (-1, -0.55) {\scriptsize {$2N$}}; 
\node[draw=none, blue, thick] at (1, -0.55) {\scriptsize {$2N$}}; 
\node[draw=none, red] at (0.65, 1.65)  {$\mathbf{-}$};
\node[draw=none, red] at (-0.65, 1.65)  {$\mathbf{-}$};\node[draw=none, blue] at (1.55, -0.85)  {$\mathbf{+}$};
\node[draw=none, blue] at (-1.55, -0.85)  {$\mathbf{+}$};
\node[draw=none] (g1) at (-2, -0.2)  {$\tfrac{1}{2}\text{NS5}$};
\end{tikzpicture}
\qquad \qquad 
\scalebox{0.8}{
\begin{tikzpicture}[baseline, font=\footnotesize]
\begin{scope}[auto,%
  every node/.style={draw, minimum size=1cm}];
\def \n {8}
\def \radius {2.5cm}
\def \margin {11} 
\draw[-, >=latex] ({360/\n * (1 - 3)+\margin}:\radius);
arc ({360/\n * (1 - 3)+\margin}:{360/\n * (1-2)-\margin}:\radius);
\node[draw=none] at ({360/\n * (2 - 2)}:\radius) {{\large $\vdots$}};
\draw[-, >=latex] ({360/\n * (2 - 3)+\margin}:\radius);
arc ({360/\n * (2 - 3)+\margin}:{360/\n * (2-2)-\margin}:\radius);
\draw[-, >=latex] ({360/\n * (3 - 3)+\margin}:\radius);
arc ({360/\n * (3 - 3)+\margin}:{360/\n * (3-2)-\margin}:\radius);
\draw[-, >=latex] ({360/\n * (4 - 3)+\margin}:\radius);
arc ({360/\n * (5 - 3)+\margin}:{360/\n * (5-2)-\margin}:\radius);
\draw[-, >=latex] ({360/\n * (5 - 3)+\margin}:\radius);
arc ({360/\n * (5 - 3)+\margin}:{360/\n * (5-2)-\margin}:\radius);
\node[draw, circle, blue] (n2) at ({360/\n * (0 - 2)}:\radius) {$2N$};
\node[draw, circle, red] at ({360/\n * (1 - 2)}:\radius) {$2N$};
\node[draw, circle,blue] at ({360/\n * (3 - 2)}:\radius) {$2N$};
\node[draw, circle,red] at ({360/\n * (4 - 2)}:\radius) {$2N$};
\node[draw, circle,blue] (n1) at ({360/\n * (5 - 2)}:\radius) {$2N$};
\node[draw, circle,red] at ({360/\n * (6 - 2)}:\radius) {$2N$};
\foreach \s in {1,...,6}
{	
	\draw[-, >=latex] ({360/\n * (\s - 3)+\margin}:\radius) 
	arc ({360/\n * (\s - 3)+\margin}:{360/\n * (\s-2)-\margin}:\radius);
}
\draw[-, >=latex] ({360/\n * (0 - 3)+\margin}:\radius) 
arc ({360/\n * (0 - 3)+\margin}:{360/\n * (0-2)-\margin}:\radius);
\node[draw, circle,red] at ({360/\n * (-1 - 2)}:\radius) {$2N$};
\draw[-, >=latex,red,thick,snake it] ({360/\n * (-1 - 3)+\margin}:\radius) 
arc ({360/\n * (-1 - 3)+\margin}:{360/\n * (-1-2)-\margin}:\radius);
\node[draw=none] at (0,-4) {\large $\substack{\text{$n$ blue circular nodes + $(n-1)$ red} \\~ \\ \text{usual circular nodes + 2 red nodes} \\~ \\ \text{connected by $T(SO(2N))$}}$};
\node[draw=none] at (-3.2,-1.2) {{\red $T(SO(2N))$}};
\end{scope}
\end{tikzpicture}}
\ee
where in the left diagram there are $2n$ half-NS5 branes.

In the absence of the $S$-fold, quivers \eref{SO2Nhead} and \eref{mirrSOhead} reduce to conventional Lagrangian theories that are related to each other by mirror symmetry.  In particular, \eref{SO2Nhead} reduces to a theory of free $4Nn$ half-hypermultiplets, namely
\be 
\begin{tikzpicture}[baseline]
\tikzstyle{every node}=[font=\footnotesize]
\node[draw, rectangle, red] (node1) at (-1,0) {$2N$};
\node[draw, rectangle,blue] (sqnode) at (1,0) {$2n$};
\draw (node1)--(sqnode);
\end{tikzpicture}
\ee
and quiver \eref{mirrSOhead} reduces to
\be \label{mirrSOheadx}
\scalebox{0.75}{
\begin{tikzpicture}[baseline, font=\footnotesize]
\begin{scope}[auto,%
  every node/.style={draw, minimum size=1cm}];
\def \n {7}
\def \radius {2.5cm}
\def \margin {11} 
\draw[-, >=latex] ({360/\n * (1 - 3)+\margin}:\radius);
arc ({360/\n * (1 - 3)+\margin}:{360/\n * (1-2)-\margin}:\radius);
\node[draw=none] at ({360/\n * (2 - 2)}:\radius) {{\large $\vdots$}};
\draw[-, >=latex] ({360/\n * (2 - 3)+\margin}:\radius);
arc ({360/\n * (2 - 3)+\margin}:{360/\n * (2-2)-\margin}:\radius);
\draw[-, >=latex] ({360/\n * (3 - 3)+\margin}:\radius);
arc ({360/\n * (3 - 3)+\margin}:{360/\n * (3-2)-\margin}:\radius);
\draw[-, >=latex] ({360/\n * (4 - 3)+\margin}:\radius);
arc ({360/\n * (5 - 3)+\margin}:{360/\n * (5-2)-\margin}:\radius);
\draw[-, >=latex] ({360/\n * (5 - 3)+\margin}:\radius);
arc ({360/\n * (5 - 3)+\margin}:{360/\n * (5-2)-\margin}:\radius);
\node[draw, circle, blue] (n2) at ({360/\n * (0 - 2)}:\radius) {$2N$};
\node[draw, circle, red] at ({360/\n * (1 - 2)}:\radius) {$2N$};
\node[draw, circle,blue] at ({360/\n * (3 - 2)}:\radius) {$2N$};
\node[draw, circle,red] at ({360/\n * (4 - 2)}:\radius) {$2N$};
\node[draw, circle,blue] (n1) at ({360/\n * (5 - 2)}:\radius) {$2N$};
\node[draw, circle,red] at ({360/\n * (6 - 2)}:\radius) {$2N$};
\foreach \s in {1,...,6}
{	
	\draw[-, >=latex] ({360/\n * (\s - 3)+\margin}:\radius) 
	arc ({360/\n * (\s - 3)+\margin}:{360/\n * (\s-2)-\margin}:\radius);
}
\draw[-, >=latex] ({360/\n * (0 - 3)+\margin}:\radius) 
arc ({360/\n * (0 - 3)+\margin}:{360/\n * (0-2)-\margin}:\radius);
\node[draw=none] at (6,-1) {$2n$ alternating red/blue circular nodes};
\end{scope}
\end{tikzpicture}}
\ee
where the two $SO(2N)$ gauge groups that were connected by $T(SO(2N))$ merged into a single $SO(2N)$ circular node.
It can be checked that the Higgs branch dimension of \eref{mirrSOheadx} is indeed zero:
\be \label{HiggsmirrSOheadx}
(2n)(2N^2) - n\left[ \frac{1}{2} (2N)(2N-1)\right] - n \left[ \frac{1}{2}(2N)(2N+1)\right] =0~,
\ee
and the quaternionic dimension of the Coulomb branch of \eref{mirrSOheadx} is $2Nn$.
These are in agreement with mirror symmetry.

\subsection*{Theory \eref{SO2Nhead}}
The Higgs branch of this theory is given by the hyperK\"ahler quotient:
\be
\CH_{\eref{SO2Nhead}}  = \frac{\CN_{so(2N)} \times \CN_{so(2N)} \times \CH\left( [S/O(2N)]-[USp(2n)] \right)}{S/O(2N)}~.
\ee
where the notation $S/O$ means that we may take the gauge group to be $SO(2N)$ or $O(2N)$.
The dimension of this space is
\be
\dim_\BH \, \CH_{\eref{SO2Nhead}} = \left[ \frac{1}{2} (2N)(2N-1) -N \right] +2N n -  \frac{1}{2} (2N)(2N-1) = (2n-1)N~.
\ee
Since the circular nodes that are connected by $T(SO(2N))$ do not contribute to the Coulomb branch, it follows that the Coulomb branch of \eref{SO2Nhead} is trivial:
\be
\dim_\BH \, \CC_{\eref{SO2Nhead}} = 0~.
\ee
Let us now discuss certain interesting special cases below.

\subsubsection*{\it The Higgs branch of \eref{SO2Nhead} for $N=1,\, 2$}

For $N=1$, since $\CN_{so(2)}$ is trivial, it follows that $\CH_{\eref{SO2Nhead}}$ is the Higgs branch of the 3d $\CN=4$ $S/O(2)$ gauge theory with $n$ flavours.  If the gauge group is taken to be $O(2)$, $\CH_{\eref{SO2Nhead}}$ is isomorphic to the closure of the minimal nilpotent orbit of $usp(2n)$.  On the other hand, if the gauge group is taken to be $SO(2)$, $\CH_{\eref{SO2Nhead}}$ turns out to be isomorphic to the closure of the minimal nilpotent orbit of $su(2n)$. The reason is because the generators of the moduli space with $SU(2)_R$-spin 1 are mesons and baryons; they transform in the representation $[2,0,\ldots,0]+[0,1,0,\ldots,0]$ of $usp(2n)$.  This representation combines into the adjoint representation $[1,0,\ldots,0,1]$ of $su(2n)$.

For $N=2$, let us denote the fundamental half-hypermultiplets by $Q^i_a$ with $i,j,k,l=1,\ldots, 2n$ and $a,b,c,d=1,2,3,4$, and the generators of $\CN_{so(4)}$ by a rank-two antisymmetric tensor $X_{ab}$.  We find that for the $O(4)$ gauge group, the generators of the Higgs branch are as follows: 
\bi
\item The mesons $M^{ij} = Q^i_a Q^j_b \delta^{ab}$, with $SU(2)_R$-spin $1$, transforming in the adjoint representation $[2,0,\ldots,0]$ of $usp(2n)$.
\item The combinations $Q^i_a Q^j_b X_{ab}$, with $SU(2)_R$-spin $2$, transforming in the adjoint representation $[0,1,0,\ldots,0]$ of $usp(2n)$.
\ei
For the $SO(4)$ gauge group, we have, in addition to the above, the following generators of the Higgs branch:
\bi
\item The baryons $B^{ijkl}= \epsilon^{abcd} Q^i_a Q^j_b Q^k_c Q^l_d$, with $SU(2)_R$-spin $2$, transforming in the adjoint representation $[0,0,0,1,0,\ldots,0]+[0,1,0,\ldots,0]$ of $usp(2n)$.
\item The combinations $ \epsilon^{abcd} Q^i_a Q^j_b X_{cd}$, with $SU(2)_R$-spin $2$, transforming in the adjoint representation $[0,1,0,\ldots,0]$ of $usp(2n)$. 
\item The $USp(2n)$ flavour singlet $\epsilon^{abcd} X_{ab} X_{cd}$, with $SU(2)_R$-spin $2$.
\ei

\subsubsection*{\it The Higgs branch of \eref{SO2Nhead} for $n=1$}
In this case, it does not matter whether we take the gauge group to be $SO(2N)$ or $O(2N)$, the Higgs branch is the same.  The corresponding Hilbert series is
\be \label{HSSO2Nheadn1}
H[\CH_{\eref{SO2Nhead}}|_{n=1}] = \PE \left[ \chi^{su(2)}_{[2]}(x) \sum_{j=0}^{N-1} t^{4j+2} - \sum_{l=N}^{2N-1} t^{4l} \right]~.
\ee
Indeed, for $N=n=1$, we recover the nilpotent cone of $su(2)$, which is isomorphic to $\BC^2/\BZ_2$.

\subsection*{Theory \eref{mirrSOhead}}
Since the nodes that are connected by $T(SO(2N))$ do not contribute to the Coulomb branch, it follows that the dimension of the Coulomb branch is
\be
\CC_{\eref{mirrSOhead}} = (2n-1)N~.
\ee
Note, however, that quiver \eref{mirrSOhead} is a ``bad'' theory in the sense of \cite{Gaiotto:2008ak}, due to the fact that each $USp(2N)$ gauge group has $2N$ flavours.  Nevertheless, we shall analyse the case of $n=1$ and general $N$ in detail below.  In which case, we shall see that the result is consistent with mirror symmetry.

The computation of the Higgs branch dimension of \eref{mirrSOhead} indicates that the gauge symmetry is not completely broken at a generic point of the Higgs branch.  Indeed, if we assume (incorrectly) that the gauge symmetry is completely broken, we would obtain the $\dim_\BH \CH_{\eref{mirrSOhead}}$ to be
\be
\eref{HiggsmirrSOheadx} + \left[\frac{1}{2}(2N)(2N-1)-N \right] - \frac{1}{2}(2N)(2N-1) = -N~.
\ee
We conjecture that the $SO(2N) \times SO(2N)$ gauge group connected by $T(SO(2N))$ is broken to $SO(2)^N$, whose dimension is $N$.  This statement can be checked explicitly in the case of $N=1$, where $T(SO(2))$ is trivial.  Taking into account such an unbroken symmetry, we obtain $\dim_\BH \CH_{\eref{mirrSOhead}}=0$, which is in agreement with the Coulomb branch of \eref{SO2Nhead}.

\subsubsection*{\it The special case of $n=1$}

In this case, the Coulomb branch of \eref{mirrSOhead} is equal to that of the $USp(2N)$ gauge theory with $2N$ flavours.  As pointed out in \cite{Assel:2018exy}, the most singular locus of the Coulomb branch consists of two points, related by a $\BZ_2$ global symmetry.  The infrared theory at any of these two points is an interacting SCFT, which we denote by $\CT_N$.

For $n=N=1$, the corresponding singularity is of an $A_1$ type \cite{Seiberg:1996nz}, and the corresponding SCFT is $\CT_2 = T(SU(2))$ whose Higgs/Coulomb branch is a nilpotent cone of $su(2)$; this is indeed in agreement with the Higgs branch of \eref{SO2Nhead}.  The situation here is the same as that described on Page 30 of \cite{Assel:2018exy}, namely mirror symmetry is realized locally at each of the two singular points.  The Higgs branch of \eref{SO2Nhead} has one component, whereas the the Coulomb branch of \eref{mirrSOhead} (for $N=n=1$) splits into two components, each of which is isomorphic to the former.

For $n=1$ and $N>1$, the mirror theory of $\CT_N$ is described by the following quiver \cite{Ferlito:2016grh}:
\be \label{Dshapedthing}
\scalebox{0.8}{
\begin{tikzpicture}[font=\scriptsize,baseline]
\begin{scope}[auto,%
  every node/.style={draw, minimum size=1cm}, node distance=1cm];
\node[circle] (U1L1) at (0, -1) {$N-1$};
\node[circle] (U1L2) at (0, 1) {$N$};
\node[rectangle] (flv) at (-1.5, 1) {$2$};
\node[circle] (U2L) at (1.5, 0) {$2N-2$};
\node[circle] (U2R) at (3, 0) {$2N-3$};
\node[draw=none] (U1R1) at (4.5, 0) {$\ldots$};
\node[circle] (U1R2) at (6, 0) {$1$};
\end{scope}
\draw (U1L1)--(U2L)
(flv)--(U1L2)--(U2L)
(U2L)--(U2R)
(U1R1)--(U2R)
(U1R1)--(U1R2);
\end{tikzpicture}}
\ee
By mirror symmetry, the Coulomb branch of $\CT_N$ is equal to the Higgs branch of \eref{Dshapedthing}, whose Hilbert series is given by \cite[(D.11)]{Assel:2018exy}:
\be
H[\CC_{\CT_N}] (t, x)= H[\CH_{\eref{Dshapedthing}}] (t,x) = \PE \left[ \chi^{su(2)}_{[2]}(x) \sum_{j=0}^{N-1} t^{4j+2} - \sum_{l=N}^{2N-1} t^{4l} \right]~.
\ee
This is perfectly in agreement with \eref{HSSO2Nheadn1}.

\subsection{Quivers with a $T(SO(2N))$ link or a $T(USp'(2N))$ link}
Let us insert an $S$-fold in the brane interval marked by red minus sign ($\red{-}$) in each brane set-up in \eref{braneDC}.  This leads to the presence of $T(SO(2N))$ link in the corresponding quiver diagram. In particular, the insertion of an $S$-fold in the left diagram of \eref{braneDC} leads to the following configuration:
\be \label{triangleDDC}
\scalebox{0.85}{
\begin{tikzpicture}[baseline]
\tikzstyle{every node}=[font=\footnotesize, node distance=0.45cm]
\tikzset{decoration={snake,amplitude=.4mm,segment length=2mm,
                       post length=0mm,pre length=0mm}}
\draw[blue,thick] (0,0) circle (1.5cm);
\draw[black,thick] (0,1) -- (0,2) node at (0,2.2) {$\frac{1}{2}\text{NS5}$};
\draw[black,thick] (0,-1) -- (0,-2) node at (0, -2.2) {};
\node[draw=none, red] at (-1.8,0.5) {$\mathbf{-}$};
\node[draw=none, red] at (-1.8,-0.5) {$\mathbf{-}$};
\draw[draw=red,solid,thick,snake it]  (-2,0)--(-0.5,0) ; 
\def \n {6}
\def \radius {1.2cm}
\def \margin {0} 
\node[draw=none, blue, thick] at (-1, 0.5) {{$2N$}}; 
\node[draw=none, blue, thick] at (-1, -0.5) {{$2N$}}; 
\node[draw=none, blue, thick] (g3) at (1, 0) {{$2N$}}; 
\node[draw=none] (g1) at (0.75, 1.3) {$\bullet$};
\node[draw=none, blue] (g1) at (0.4, 1.7)  {$\mathbf{+}$};
\node[draw=none] (g1) at (1.3, 0.75) {$\bullet$};
\node[draw=none, blue] (g1) at (1.25, 1.25)  {$\mathbf{\tilde +}$};
\node[draw=none, blue] (g1) at (1.7, 0.45)  {$\mathbf{+}$};
\node[draw=none] at (1.7,0) {{\Large $\mathbf{\vdots}$}};
\node[draw=none] (g1) at (0.75, -1.3) {$\bullet$} node at (1,-1.6)  {$\frac{1}{2}\text{D5}$};
\node[draw=none, blue] (g1) at (0.4, -1.7)  {$\mathbf{+}$};
\end{tikzpicture}}
\qquad \qquad \qquad
\scalebox{0.8}{
\begin{tikzpicture}[baseline]
\tikzstyle{every node}=[font=\footnotesize]
\node[draw, circle, red] (node1) at (-2,2) {$2N$};
\node[draw, circle, red] (node2) at (2,2) {$2N$};
\node[draw, circle, blue] (node3) at (0,0) {$2N$};
\node[draw, rectangle, red] (sqnode) at (0,-1.5) {$2n$};
\draw[draw=black,solid,thick,-]  (node2)--(node3) ; 
\draw[draw=black,solid,thick,-]  (node1)--(node3) ; 
\draw[draw=red,snake it,thick,-]  (node1)--(node2) node at (0,2.3) {{\red $T(SO(2N))$}} ; 
\draw[draw=black,solid,thick,-]  (node3)--(sqnode) ; 
\end{tikzpicture}}
\ee

The mirror theory can be obtained from the $S$-dual configuration of the above, or simply inserting an $S$-fold to the left interval of the right diagram in \eref{braneDC}.  The result is
\be \label{mirrtriangleDDC}
\scalebox{0.9}{
\begin{tikzpicture}[baseline]
\tikzstyle{every node}=[font=\footnotesize]
\tikzset{decoration={snake,amplitude=.4mm,segment length=2mm,
                       post length=0mm,pre length=0mm}}
\draw[black,thick] (0,1) -- (0,2) node at (0,2.2) {$\frac{1}{2}\text{NS5}$};    
\draw[blue,thick] (0,0) circle (1.5cm);
\def \n {6}
\def \radius {1.2cm}
\def \margin {0} 
\foreach \s in {1,...,10}
{
	\node[draw=none] (\s) at ({360/\n * (\s - 2)+30}:{\radius-10}) {};
}
\node[draw=none, circle] (last) at ({360/3 * (3 - 1)+30}:{\radius-10}) {};
\node[draw=none,  below right= of 1] (f1) {};
\node[draw=none, above right= of 2] (f2) {};
\node[draw=none, above = of 3] (f3) {};
\node[draw=none, above left= of 4] (f4) {};
\node[draw=none,  below left= of 5] (f5) {};
\node[draw=none,  below = of last] (f6) {};
\node[draw=none] at (1.8,0) {{\Large $\mathbf{\vdots}$}};
\node[draw=none] at (0,-2.4) {};
\draw[-, >=latex,black, thick] (0.7, 0.7) to (1.56, 1.56);
\draw[-, >=latex,black, thick] (0.97, 0.26) to (2.13, 0.57);
\draw[-, >=latex,black, thick] (0.7, -0.7) to (1.56, -1.56);
\draw[-, >=latex,black, thick] (2.12, -0.57) to (0.98, -0.25);
\node[draw=none] (g1) at (0.6, 1.35) {$\bullet$};
\node[draw=none] (g2) at (0.6, -1.40) {$\bullet$};
\node[draw=none, blue, thick] (g3) at (0.45, 1.10) {\tiny{$2N$}}; 
\node[draw=none, blue, thick] (g5) at (0.8, 0.55) {\tiny{$2N+1$}}; 
\node[draw=none, blue, thick] (g5) at (0.8, -0.55) {\tiny{$2N+1$}}; 
\node[draw=none, blue, thick] (g8) at (0.5, -1.1) {\tiny{$2N$}}; 
\draw[draw=red,solid,thick,snake it]  (-2,0)--(-0.5,0) ; 
\node[draw=none, blue, thick] (g10) at (-1.2, 0.3) {\tiny{$2N$}}; 
\node[draw=none, blue, thick] (g10) at (-1.2, -0.3) {\tiny{$2N$}}; 
\node[draw=none, red] at (-1.8, 0.3)  {$\mathbf{-}$};
\node[draw=none, red] at (-1.8, -0.3)  {$\mathbf{-}$};
\node[draw=none, blue] at (0.35, 1.65)  {$\mathbf{+}$};
\node[draw=none, blue] at (0.95, 1.45)  {$\mathbf{\tilde +}$};
\node[draw=none, red] at (1.5, 0.85)  {$\mathbf{\tilde -}$};
\node[draw=none, blue] at (0.35, -1.65)  {$\mathbf{+}$};
\node[draw=none, blue] at (0.95, -1.45)  {$\mathbf{\tilde +}$};
\node[draw=none, red] at (1.5, -0.85)  {$\mathbf{\tilde -}$};
\draw[-, >=latex,black, thick] (last) to (f6);
\end{tikzpicture}}
\qquad \qquad\qquad
\scalebox{0.65}{
\begin{tikzpicture}[baseline, scale=0.6,font=\scriptsize]
\begin{scope}[auto,%
  every node/.style={draw, minimum size=1.2cm}];
\def \n {8}
\def \radius {4.2cm}
\def \margin {14} 
\draw[-, >=latex] ({360/\n * (1 - 3)+\margin}:\radius);
arc ({360/\n * (1 - 3)+\margin}:{360/\n * (1-2)-\margin}:\radius);
\node[draw=none] at ({360/\n * (2 - 2)}:\radius) {{\large $\vdots$}};
\draw[-, >=latex] ({360/\n * (2 - 3)+\margin}:\radius);
arc ({360/\n * (2 - 3)+\margin}:{360/\n * (2-2)-\margin}:\radius);
\draw[-, >=latex] ({360/\n * (3 - 3)+\margin}:\radius);
arc ({360/\n * (3 - 3)+\margin}:{360/\n * (3-2)-\margin}:\radius);
\draw[-, >=latex] ({360/\n * (4 - 3)+\margin}:\radius);
arc ({360/\n * (5 - 3)+\margin}:{360/\n * (5-2)-\margin}:\radius);
\draw[-, >=latex] ({360/\n * (5 - 3)+\margin}:\radius);
arc ({360/\n * (5 - 3)+\margin}:{360/\n * (5-2)-\margin}:\radius);
\node[draw, circle, blue] (n2) at ({360/\n * (0 - 2)}:\radius) {$2N$};
\node[draw, circle, red] at ({360/\n * (1 - 2)}:\radius) {$2N+1$};
\node[draw, circle,blue] at ({360/\n * (3 - 2)}:\radius) {$2N'$};
\node[draw, circle,red] at ({360/\n * (4 - 2)}:\radius) {$2N+1$};
\node[draw, circle,blue ] (n1) at ({360/\n * (5 - 2)}:\radius) {$2N$};
\foreach \s in {1,...,6}
{	
	\draw[-, >=latex] ({360/\n * (\s - 3)+\margin}:\radius) 
	arc ({360/\n * (\s - 3)+\margin}:{360/\n * (\s-2)-\margin}:\radius);
}
\draw[-, >=latex] ({360/\n * (0 - 3)+\margin}:\radius) 
arc ({360/\n * (0 - 3)+\margin}:{360/\n * (0-2)-\margin}:\radius);
\node[draw, circle,red] at ({360/\n * (6 - 2)}:\radius) {$2N$};
\node[draw, circle,red] at ({360/\n * (-1 - 2)}:\radius) {$2N$};
\draw[-, >=latex,red,thick,snake it] ({360/\n * (-1 - 3)+\margin}:\radius) 
arc ({360/\n * (-1 - 3)+\margin}:{360/\n * (-1-2)-\margin}:\radius);
\node[draw=none] at (0,-10) {\large $\substack{\text{$n$ blue circular nodes + $(n-1)$ red} \\~ \\ \text{usual circular nodes + 2 red nodes} \\~ \\ \text{connected by $T(SO(2N))$}}$};
\node[draw=none] at (-5.7,-1.5) {{\red $T(SO(2N))$}};
\node[draw, rectangle ,red] (f1) at (-6.5,3) {$1$};
\node[draw, rectangle ,red] (f2) at (0,-7.5) {$1$};
\draw[-,solid] (f1) to (n1);
\draw[-,solid] (f2) to (n2);
\end{scope}
\end{tikzpicture}}
\ee
where the number of half-NS5 branes is $2n$.  Note that for $n=1$, the theory is self-mirror.  


\subsubsection*{Theory \eref{triangleDDC}}
The Higgs branch of \eref{triangleDDC} is described by the hyperK\"ahler quotient
\be \label{HKtriangleDDC1}
\resizebox{.91\textwidth}{!}{$
\begin{aligned}
\CH_{\eref{triangleDDC}} &=  \Big( \CN_{so(2N)}  \times \CH([SO(2N)]-[USp(2N)]) \times \CN_{so(2N)} \times \CH(SO(2N)]-[USp(2N)])  \times \\
& \qquad \CH([USp(2N)]-[SO(2n)]) \Big) /\left( SO(2N) \times SO(2N) \times USp(2N) \right) \\
&= \frac{\CN_{usp(2N)} \times \CN_{usp(2N)} \times  \CH([USp(2N)]-[SO(2n)])}{USp(2N)}~,
\end{aligned} $}
\ee
where we have used \eref{specialusp2n} to obtain the last line.   We remark that both red circular nodes can be chosen to be either $SO(2N)$ or $O(2N)$ and the results for both options are the same, thanks to the equality between \eref{specialusp2n} and \eref{specialusp2no}.  Moreover, the hyperK\"ahler quotient in the last line of \eref{HKtriangleDDC1} suggests the equality between \eref{HKtriangleDDC1} and the Higgs branch of the following theory:
\be \label{Uspheadso2n}
\scalebox{0.95}{
\begin{tikzpicture}[baseline]
\tikzstyle{every node}=[font=\footnotesize, node distance=0.45cm]
\tikzset{decoration={snake,amplitude=.4mm,segment length=2mm,
                       post length=0mm,pre length=0mm}}
\draw[blue,thick] (0,0) circle (1.5cm);
\draw[draw=red,snake it,thick,-] (0,1) -- (0,2) ;
\node[draw=none] (g1) at (0.7, -1.33) {$\bullet$};
\node[draw=none] (g1) at (-0.7, -1.33) {$\bullet$};
\node[draw=none] (g1) at (1.33, -0.7) {$\bullet$};
\node[draw=none] (g1) at (-1.33, -0.7) {$\bullet$};
\node[draw=none] (g1) at (-1.8, -0.7)  {$\tfrac{1}{2}\text{D5}$};
\node[draw=none, blue] (g1) at (1, 0)  {$2N$};
\node[draw=none, blue] (g1) at (-1, 0)  {$2N$};
\node[draw=none] at (0,-1.7) {{\Large $\mathbf{\dots}$}};
\def \n {6}
\def \radius {1.2cm}
\def \margin {0} 
\node[draw=none, blue] (g1) at (-0.6, 1.7)  {$\mathbf{\tilde{+}}$};
\node[draw=none, blue] at (0.6, 1.7)  {$\mathbf{\tilde{+}}$};
\node[draw=none, red] at (1.3, -1.2)  {$\mathbf{+}$};
\node[draw=none, red] at (-1.3, -1.2)  {$\mathbf{+}$};
\end{tikzpicture}}
\qquad \qquad \qquad
\begin{tikzpicture}[baseline]
\tikzstyle{every node}=[font=\footnotesize]
\node[draw, circle, blue] (node1) at (0,1) {$2N'$};
\draw[red,thick] (node1) edge [out=45,in=135,loop,looseness=5, snake it]  (node1);
\node[draw=none] at (0,2.4) {{\red $T(USp'(2N))$}};
\node[draw, rectangle,red] (sqnode) at (0,-1) {$2n$};
\draw (node1)--(sqnode);
\end{tikzpicture}
\ee
where the blue circular node is a $USp'(2N)$ gauge group.  In other words, we have the following equality of the Higgs branch between two different gauge theories:
\be
\CH_{\eref{triangleDDC}} = \CH_{\eref{Uspheadso2n}} ~.
\ee
The quaternionic dimension of \eref{HKtriangleDDC1} is 
\be
\begin{split}
\dim_\BH \CH_{\eref{triangleDDC}} &= \left[ \frac{1}{2}(2N)(2N-1) - N \right] +2(4N^2) +2 N n  \\
& \qquad - \left[ 2 \times \frac{1}{2}(2N)(2N-1) \right] - \frac{1}{2}(2N)(2N+1) \\
&= (2n-1)N~.
\end{split}
\ee

Since the nodes that are connected by $T(SO(N))$ does not contribute to the Coulomb branch of the theory, the Coulomb branch of \eref{triangleDDC} is isomorphic to the Coulomb branch of the 3d $\CN=4$ $USp(2N)$ gauge theory with $2N+n$ flavours, whose Hilbert series is given by \cite[(5.14)]{Cremonesi:2013lqa}. Its quaternionic dimension is
\be
\begin{split}
\dim_\BH \CC_{\eref{triangleDDC}} &=  N~.
\end{split}
\ee

\paragraph{Example: $n=1$.} The theory is self-mirror.  One can check that the Hilbert series of the quotient \eref{HKtriangleDDC} is indeed equal to the Coulomb branch of $USp(2N)$ gauge theory with $2N+1$ flavours \cite[(5.14)]{Cremonesi:2013lqa}, which is
\be \label{USp2Nw2Np1}
\PE \left [ \sum_{j=1}^{2N} t^{2j} + \sum_{j=1}^{N} t^{4j} - \sum_{j=1}^{N} t^{4j+4N} \right]~.
\ee
Note that for $N=n=1$, we have $\BC^2/\BZ_4$, as expected from the Coulomb branch of $USp(2)$ with 3 flavours.

There is another way to check that theory \eref{triangleDDC} for $n=1$ (and a general $N$) is self-mirror.  We can easily compute a mirror theory of \eref{Uspheadso2n}, with $n=1$, by applying $S$-duality to the brane system; see \eref{mirrUsphead}. The result is
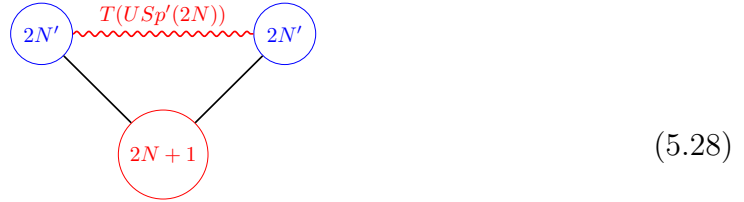
\be \label{BNquiv}
\scalebox{0.8}{
\begin{tikzpicture}[baseline]
\tikzstyle{every node}=[font=\footnotesize]
\node[draw, circle, blue] (node1) at (-2,2) {$2N'$};
\node[draw, circle, blue] (node2) at (2,2) {$2N'$};
\node[draw, circle, red] (node3) at (0,0) {$2N+1$};
\draw[draw=black,solid,thick,-]  (node2)--(node3) ; 
\draw[draw=black,solid,thick,-]  (node1)--(node3) ; 
\draw[draw=red,snake it,thick,-]  (node1)--(node2) node at (0,2.3) {{\red $T(USp'(2N))$}} ; 
\end{tikzpicture}}
\ee
The Coulomb branch of this theory is isomorphic to that of 3d $\CN=4$ $SO(2N+1)$ gauge theory with $2N$ flavours, whose Hilbert series is given in \cite[(5.18)]{Cremonesi:2013lqa}.  However, as pointed out in that reference, this turns out to be isomorphic to the Coulomb branch of the $USp(2N)$ gauge theory with $2N+1$ flavours, whose Hilbert series is given by \eref{USp2Nw2Np1}.  We thus establish the self-duality of \eref{triangleDDC} for $n=1$.

\subsubsection*{Theory \eref{mirrtriangleDDC}}
The Higgs branch dimension of \eref{mirrtriangleDDC} is
\be
\begin{split}
\dim_{\BH} \, \CH_{\eref{mirrtriangleDDC}} &= (2)(2N^2)+(2n-2)N(2N+1)+ \left[\frac{1}{2}(2N)(2N-1)-N \right]  \\
& \quad + N+N - n\left[ \frac{1}{2}(2N)(2N+1) \right] - 2\left[ \frac{1}{2}(2N)(2N-1) \right] \\
& \quad - (n-1)\left[ \frac{1}{2}(2N+1)(2N) \right]  \\
&= N~.
\end{split}
\ee
The Coulomb branch dimension of \eref{mirrtriangleDDC} is equal to the total rank of the gauge groups that are not connected by $T(SO(2N))$:
\be
\dim_{\BH}\, \CC_{\eref{mirrtriangleDDC}} = (2n-1)N~.
\ee
These agree with the dimensions of the Coulomb and the Higgs branches of \eref{triangleDDC}.

Similarly to the previous discussion, the red circular nodes that are connected by $T(SO(2N))$ can be taken as $O(2N)$ or $SO(2N)$ without affecting the Higgs branch moduli space of \eref{mirrtriangleDDC}.  Moreover, we find that this applies to other red circular nodes in the quiver, namely the choice between $O(2N+1)$ and $SO(2N+1)$ does not change the Higgs branch of the theory.  This can be checked directly using the Hilbert series. 


It is worth pointing out that there is another gauge theory that gives the same Coulomb branch as \eref{triangleDDC}.  This is the mirror theory of \eref{Uspheadso2n} which is given by
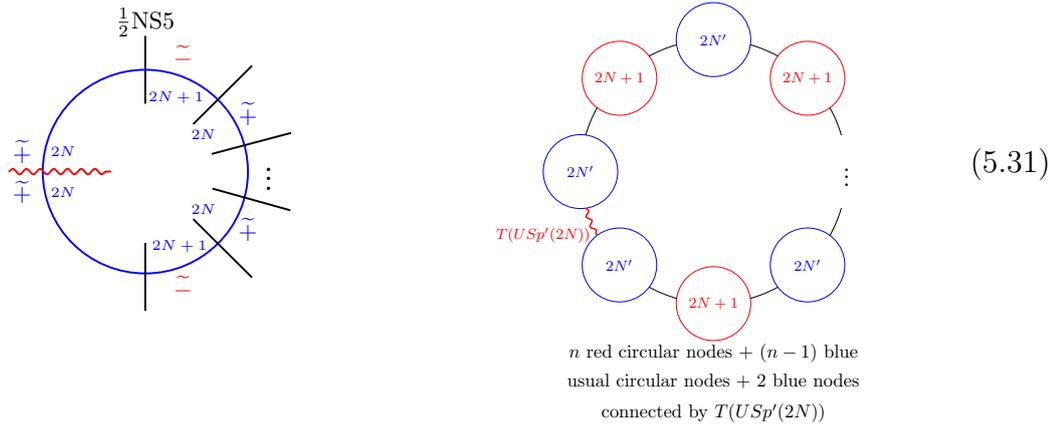
\be \label{mirrUsphead}
\scalebox{0.9}{
\begin{tikzpicture}[baseline]
\tikzstyle{every node}=[font=\footnotesize]
\tikzset{decoration={snake,amplitude=.4mm,segment length=2mm,
                       post length=0mm,pre length=0mm}}
\draw[black,thick] (0,1) -- (0,2) node at (0,2.2) {$\frac{1}{2}\text{NS5}$};    
\draw[blue,thick] (0,0) circle (1.5cm);
\def \n {6}
\def \radius {1.2cm}
\def \margin {0} 
\foreach \s in {1,...,10}
{
	\node[draw=none] (\s) at ({360/\n * (\s - 2)+30}:{\radius-10}) {};
}
\node[draw=none, circle] (last) at ({360/3 * (3 - 1)+30}:{\radius-10}) {};
\node[draw=none,  below right= of 1] (f1) {};
\node[draw=none, above right= of 2] (f2) {};
\node[draw=none, above = of 3] (f3) {};
\node[draw=none, above left= of 4] (f4) {};
\node[draw=none,  below left= of 5] (f5) {};
\node[draw=none,  below = of last] (f6) {};
\node[draw=none] at (1.8,0) {{\Large $\mathbf{\vdots}$}};
\node[draw=none] at (0,-2.4) {};
\draw[-, >=latex,black, thick] (0.7, 0.7) to (1.56, 1.56);
\draw[-, >=latex,black, thick] (0.97, 0.26) to (2.13, 0.57);
\draw[-, >=latex,black, thick] (0.7, -0.7) to (1.56, -1.56);
\draw[-, >=latex,black, thick] (2.12, -0.57) to (0.98, -0.25);
\node[draw=none, blue, thick] (g3) at (0.45, 1.10) {\tiny{$2N+1$}}; 
\node[draw=none, blue, thick] (g5) at (0.85, 0.55) {\tiny{$2N$}}; 
\node[draw=none, blue, thick] (g5) at (0.85, -0.55) {\tiny{$2N$}}; 
\node[draw=none, blue, thick] (g8) at (0.5, -1.1) {\tiny{$2N+1$}}; 
\draw[draw=red,solid,thick,snake it]  (-2,0)--(-0.5,0) ; 
\node[draw=none, blue, thick] (g10) at (-1.2, 0.3) {\tiny{$2N$}}; 
\node[draw=none, blue, thick] (g10) at (-1.2, -0.3) {\tiny{$2N$}}; 
\node[draw=none, blue] at (-1.8, 0.3)  {$\mathbf{\tilde{+}}$};
\node[draw=none, blue] at (-1.8, -0.3)  {$\mathbf{\tilde{+}}$};
\node[draw=none, red] at (0.55, 1.7)  {$\mathbf{\tilde{-}}$};
\node[draw=none, red] at (0.55, -1.7)  {$\mathbf{\tilde{-}}$};
\node[draw=none, blue] at (1.5, 0.85)  {$\mathbf{\tilde +}$};
\node[draw=none, blue] at (1.5, -0.85)  {$\mathbf{\tilde +}$};
\draw[-, >=latex,black, thick] (last) to (f6);
\end{tikzpicture}}
\qquad \qquad\qquad
\scalebox{0.7}{
\begin{tikzpicture}[baseline, font=\scriptsize]
\begin{scope}[auto,%
  every node/.style={draw, minimum size=1.4cm}];
\def \n {8}
\def \radius {2.5cm}
\def \margin {16} 
\draw[-, >=latex] ({360/\n * (1 - 3)+\margin}:\radius);
arc ({360/\n * (1 - 3)+\margin}:{360/\n * (1-2)-\margin}:\radius);
\node[draw=none] at ({360/\n * (2 - 2)}:\radius) {{\large $\vdots$}};
\draw[-, >=latex] ({360/\n * (2 - 3)+\margin}:\radius);
arc ({360/\n * (2 - 3)+\margin}:{360/\n * (2-2)-\margin}:\radius);
\draw[-, >=latex] ({360/\n * (3 - 3)+\margin}:\radius);
arc ({360/\n * (3 - 3)+\margin}:{360/\n * (3-2)-\margin}:\radius);
\draw[-, >=latex] ({360/\n * (4 - 3)+\margin}:\radius);
arc ({360/\n * (5 - 3)+\margin}:{360/\n * (5-2)-\margin}:\radius);
\draw[-, >=latex] ({360/\n * (5 - 3)+\margin}:\radius);
arc ({360/\n * (5 - 3)+\margin}:{360/\n * (5-2)-\margin}:\radius);
\node[draw, circle, red] (n2) at ({360/\n * (0 - 2)}:\radius) {$2N+1$};
\node[draw, circle, blue] at ({360/\n * (1 - 2)}:\radius) {$2N'$};
\node[draw, circle,red] at ({360/\n * (3 - 2)}:\radius) {$2N+1$};
\node[draw, circle,blue] at ({360/\n * (4 - 2)}:\radius) {$2N'$};
\node[draw, circle,red] (n1) at ({360/\n * (5 - 2)}:\radius) {$2N+1$};
\node[draw, circle,blue] at ({360/\n * (6 - 2)}:\radius) {$2N'$};
\foreach \s in {1,...,6}
{	
	\draw[-, >=latex] ({360/\n * (\s - 3)+\margin}:\radius) 
	arc ({360/\n * (\s - 3)+\margin}:{360/\n * (\s-2)-\margin}:\radius);
}
\draw[-, >=latex] ({360/\n * (0 - 3)+\margin}:\radius) 
arc ({360/\n * (0 - 3)+\margin}:{360/\n * (0-2)-\margin}:\radius);
\node[draw, circle,blue] at ({360/\n * (-1 - 2)}:\radius) {$2N'$};
\draw[-, >=latex,red,thick,snake it] ({360/\n * (-1 - 3)+\margin}:\radius) 
arc ({360/\n * (-1 - 3)+\margin}:{360/\n * (-1-2)-\margin}:\radius);
\node[draw=none] at (0,-4) {\large $\substack{\text{$n$ red circular nodes + $(n-1)$ blue} \\~ \\ \text{usual circular nodes + 2 blue nodes} \\~ \\ \text{connected by $T(USp'(2N))$}}$};
\node[draw=none] at (-3.2,-1.2) {{\red $T(USp'(2N))$}};
\end{scope}
\end{tikzpicture}}
\ee
where the number of half-NS5 branes is $2n$.
We expect that the Coulomb branch of \eref{mirrUsphead} has to be equal to the Coulomb branch of \eref{mirrtriangleDDC}.  This can be seen as follows.  Let us focus on \eref{mirrUsphead}. Note that the two blue circular nodes that are connected by $T(USp'(2N))$ do not contribute to the Coulomb branch computation, so we can take them to be two flavour nodes that are not connected.  As pointed out below \cite[(5.18)]{Cremonesi:2013lqa}, the Coulomb branch of the $SO(2N+1)$ gauge theory with $2N$ flavours is the same as that of Coulomb branch of the $USp(2N)$ gauge theory with $2N+1$ flavours.  We can apply this fact to every node in quiver \eref{mirrUsphead} and see that the resulting quiver has the same Coulomb branch as that of \eref{mirrtriangleDDC}.

\subsection{More quivers with a $T(USp'(2N))$ link}
Let us insert an $S$-fold in the interval labelled by $\blue{\tilde{+}}$ in each of the diagram in \eref{braneCNxBN}.   Doing so in the left diagram yields the following theory:
\be \label{triangleCCD}
\begin{tikzpicture}[baseline]
\tikzstyle{every node}=[font=\footnotesize, node distance=0.45cm]
\tikzset{decoration={snake,amplitude=.4mm,segment length=2mm,
                       post length=0mm,pre length=0mm}}
\draw[blue,thick] (0,0) circle (1.5cm);
\draw[draw=red,solid,thick,snake it]  (-2,0)--(-0.5,0) ; 
\draw[black,thick] (0,1) -- (0,2) node at (0,2.2) {$\frac{1}{2}\text{NS5}$};
\draw[black,thick] (0,-1) -- (0,-2) node at (0, -2.2) {};
\node[draw=none, blue] at (-1.8,0.4) {$\mathbf{\tilde +}$};
\node[draw=none, blue] at (-1.8,-0.4) {$\mathbf{\tilde +}$};
\def \n {6}
\def \radius {1.2cm}
\def \margin {0} 
\node[draw=none, blue, thick] at (-0.9, 0.5) {\scriptsize {$2N$}}; 
\node[draw=none, blue, thick] at (-0.9,-0.5) {\scriptsize {$2N$}}; 
\node[draw=none, blue, thick] (g3) at (0.9, 0) {\scriptsize {$2N+1$}}; 
\node[draw=none] (g1) at (0.75, 1.3) {$\bullet$};
\node[draw=none, red] (g1) at (0.4, 1.7)  {$\mathbf{\tilde -}$};
\node[draw=none] (g1) at (1.3, 0.75) {$\bullet$};
\node[draw=none, red] (g1) at (1.25, 1.25)  {$\mathbf{-}$};
\node[draw=none, red] (g1) at (1.7, 0.45)  {$\mathbf{\tilde -}$};
\node[draw=none] at (1.7,0) {{\Large $\mathbf{\vdots}$}};
\node[draw=none] (g1) at (0.75, -1.3) {$\bullet$} node at (1,-1.7) {\scriptsize $\frac{1}{2}\text{D5}$};
\node[draw=none, red] (g1) at (0.4, -1.7)  {$\mathbf{\tilde -}$};
\end{tikzpicture}
\qquad\qquad\qquad 
\scalebox{0.8}{
\begin{tikzpicture}[baseline]
\tikzstyle{every node}=[font=\footnotesize]
\node[draw, circle, blue] (node1) at (-2,2) {$2N'$};
\node[draw, circle, blue] (node2) at (2,2) {$2N'$};
\node[draw, circle, red] (node3) at (0,0) {$2N+1$};
\node[draw, rectangle, blue] (sqnode) at (0,-1.5) {$2n$};
\draw[draw=black,solid,thick,-]  (node2)--(node3) ; 
\draw[draw=black,solid,thick,-]  (node1)--(node3) ; 
\draw[draw=red,snake it,thick,-]  (node1)--(node2) node at (0,2.3) {{\red $T(USp'(2N))$}} ; 
\draw[draw=black,solid,thick,-]  (node3)--(sqnode) ; 
\end{tikzpicture}}
\ee
On the other hand, inserting an $S$-fold to the right diagram yields the mirror configuration:
\be \label{mirrtriangleCCD}
\begin{tikzpicture}[baseline]
\tikzstyle{every node}=[font=\footnotesize]
\tikzset{decoration={snake,amplitude=.4mm,segment length=2mm,
                       post length=0mm,pre length=0mm}}
\draw[black,thick] (0,1) -- (0,2) node at (0,2.2) {$\frac{1}{2}\text{NS5}$};    
\draw[blue,thick] (0,0) circle (1.5cm);
\draw[draw=red,solid,thick,snake it]  (-2,0)--(-0.5,0) ; 
\def \n {6}
\def \radius {1.2cm}
\def \margin {0} 
\foreach \s in {1,...,10}
{
	\node[draw=none] (\s) at ({360/\n * (\s - 2)+30}:{\radius-10}) {};
}
\node[draw=none, circle] (last) at ({360/3 * (3 - 1)+30}:{\radius-10}) {};
\node[draw=none,  below right= of 1] (f1) {};
\node[draw=none, above right= of 2] (f2) {};
\node[draw=none, above = of 3] (f3) {};
\node[draw=none, above left= of 4] (f4) {};
\node[draw=none,  below left= of 5] (f5) {};
\node[draw=none,  below = of last] (f6) {};
\node[draw=none] at (1.8,0) {{\Large $\mathbf{\vdots}$}};
\node[draw=none] at (0,-2.4) {};
\draw[-, >=latex,black, thick] (0.7, 0.7) to (1.56, 1.56);
\draw[-, >=latex,black, thick] (0.97, 0.26) to (2.13, 0.57);
\draw[-, >=latex,black, thick] (0.7, -0.7) to (1.56, -1.56);
\draw[-, >=latex,black, thick] (2.12, -0.57) to (0.98, -0.25);
\node[draw=none] (g1) at (-1.45, 0.4) {$\bullet$};
\node[draw=none] (g2) at (-1.45, -0.4) {$\bullet$};
\node[draw=none, blue] at (-1.3, 1.2)  {$\mathbf{+}$};
\node[draw=none, blue] at (-1.3, -1.2)  {$\mathbf{+}$};
\node[draw=none, blue, thick] (g3) at (0.45, 1.10) {\tiny{$2N+2$}}; 
\node[draw=none, blue, thick] (g5) at (1, 0.55) {\tiny{$2N$}}; 
\node[draw=none, blue, thick] (g5) at (1, -0.55) {\tiny{$2N$}}; 
\node[draw=none, blue, thick] (g8) at (0.5, -1.1) {\tiny{$2N+2$}}; 
\node[draw=none, blue, thick] (g10) at (-0.8, 0.8) {\tiny{$2N$}}; 
\node[draw=none, blue, thick] (g10) at (-0.8, -0.8) {\tiny{$2N$}}; 
\node[draw=none, blue] at (-1.8, 0.3)  {$\mathbf{\tilde +}$};
\node[draw=none, blue] at (-1.8, -0.3)  {$\mathbf{\tilde +}$};
\node[draw=none, red] at (0.65, 1.65)  {$\mathbf{-}$};
\node[draw=none, blue] at (1.5, 0.85)  {$\mathbf{+}$};
\node[draw=none, red] at (0.65, -1.65)  {$\mathbf{-}$};
\node[draw=none, blue] at (1.5, -0.85)  {$\mathbf{+}$};
\draw[-, >=latex,black, thick] (last) to (f6);
\end{tikzpicture}
\qquad \qquad \qquad
\scalebox{0.7}{
\begin{tikzpicture}[baseline, scale=0.6,font=\scriptsize]
\begin{scope}[auto,%
  every node/.style={draw, minimum size=1.2cm}];
\def \n {8}
\def \radius {4.2cm}
\def \margin {14} 
\draw[-, >=latex] ({360/\n * (1 - 3)+\margin}:\radius);
arc ({360/\n * (1 - 3)+\margin}:{360/\n * (1-2)-\margin}:\radius);
\node[draw=none] at ({360/\n * (2 - 2)}:\radius) {{\large $\vdots$}};
\draw[-, >=latex] ({360/\n * (2 - 3)+\margin}:\radius);
arc ({360/\n * (2 - 3)+\margin}:{360/\n * (2-2)-\margin}:\radius);
\draw[-, >=latex] ({360/\n * (3 - 3)+\margin}:\radius);
arc ({360/\n * (3 - 3)+\margin}:{360/\n * (3-2)-\margin}:\radius);
\draw[-, >=latex] ({360/\n * (4 - 3)+\margin}:\radius);
arc ({360/\n * (5 - 3)+\margin}:{360/\n * (5-2)-\margin}:\radius);
\draw[-, >=latex] ({360/\n * (5 - 3)+\margin}:\radius);
arc ({360/\n * (5 - 3)+\margin}:{360/\n * (5-2)-\margin}:\radius);
\node[draw, circle, red] at ({360/\n * (0 - 2)}:\radius) {$2N+2$};
\node[draw, circle, blue] at ({360/\n * (1 - 2)}:\radius) {$2N$};
\node[draw, circle,red] at ({360/\n * (3 - 2)}:\radius) {$2N+2$};
\node[draw, circle,blue] at ({360/\n * (4 - 2)}:\radius) {$2N$};
\node[draw, circle,red ] at ({360/\n * (5 - 2)}:\radius) {$2N+2$};
\foreach \s in {1,...,6}
{	
	\draw[-, >=latex] ({360/\n * (\s - 3)+\margin}:\radius) 
	arc ({360/\n * (\s - 3)+\margin}:{360/\n * (\s-2)-\margin}:\radius);
}
\draw[-, >=latex] ({360/\n * (0 - 3)+\margin}:\radius) 
arc ({360/\n * (0 - 3)+\margin}:{360/\n * (0-2)-\margin}:\radius);
\node[draw, circle,blue] (n1) at ({360/\n * (6 - 2)}:\radius) {$2N'$};
\node[draw, circle,blue] (n2) at ({360/\n * (-1 - 2)}:\radius) {$2N'$};
\draw[-, >=latex,red,thick,snake it] ({360/\n * (-1 - 3)+\margin}:\radius) 
arc ({360/\n * (-1 - 3)+\margin}:{360/\n * (-1-2)-\margin}:\radius);
\node[draw=none] at (0,-10) {\large $\substack{\text{$n$ red circular nodes + $(n-1)$ blue} \\~ \\ \text{usual circular nodes + 2 blue nodes} \\~ \\ \text{connected by $T(USp'(2N))$}}$};
\node[draw=none] at (-5.7,-2.1) {{\red $T(USp'(2N))$}};
\node[draw, rectangle ,red, left=of n1] (f1) {$1$};
\node[draw, rectangle ,red, below=of n2] (f2) {$1$};
\draw[-,solid] (f1) to (n1);
\draw[-,solid] (f2) to (n2);
\end{scope}
\end{tikzpicture}}
\ee


\subsubsection*{Theory \eref{triangleCCD}}
The Higgs branch of \eref{triangleCCD} is described by the hyperK\"ahler quotient
\be \label{HKtriangleDDC}
\resizebox{.91\textwidth}{!}{$
\begin{aligned}
\CH_{\eref{triangleCCD}} &= \frac{\CN_{so(2N+1)} \times \CN_{so(2N+1)} \times  \CH([SO(2N+1)]-[USp(2n)])}{SO(2N+1)}~,
\end{aligned} $}
\ee
where we have used \eref{HKGUSp2NHO} and \eref{clOson} (with $n=2N+1$).  The dimension of this is
\be
\begin{split}
\dim_\BH \CH_{\eref{triangleCCD}} &=\left[ \frac{1}{2} (2N+1)(2N)-N \right] +(2N+1) n -\frac{1}{2}(2N+1)(2N) \\
 &= 2 nN+n -N~.
 \end{split}
\ee
A special case of $N=n=1$ is particularly simple.  The corresponding Higgs branch is a complete intersection with the Hilbert series
\be \label{HiggsNn1DDC}
H[ \CH_{\eref{triangleDDC}}|_{N=n=1} ](t; x) = \PE \left[ \chi^{su(2)}_{[2]}(x) t^2 +  \chi^{su(2)}_{[1]}(x) t^3 -t^8 \right]~.
\ee

The Coulomb branch of \eref{triangleCCD} is isomorphic to that of the 3d $\CN=4$ $SO(2N+1)$ gauge theory with $2N+n$ flavours, whose Hilbert series is given by \cite[(5.18)]{Cremonesi:2013lqa}.   Note that this is also equal to that of the Coulomb branch of the $USp(2N)$ gauge theory with $2N+n+1$ flavours.


\subsubsection*{Theory \eref{mirrtriangleCCD}}
The quaternionic dimension of the Coulomb branch of this theory is 
\be
\dim_{\BH} \CC_{\eref{mirrtriangleCCD}} = n(N+1)+(n-1)N = 2n N +n -N~.
\ee
This matches with the dimension of the Higgs branch of \eref{triangleCCD}. It should be noted that \eref{mirrtriangleCCD} is a ``bad'' theory in the sense of \cite{Gaiotto:2008ak}, due to the fact that each $SO(2N+2)$ gauge group effectively has $2N$ flavours.  Nevertheless, we shall analyse the case of $N=n=1$ below.

Let us now turn to the Higgs branch.  In the absence of the $S$-fold, it was pointed out below \cite[(7.1)]{Feng:2000eq} that the gauge symmetry is not completely broken at a generic point on the Higgs branch, but is broken to $n$ copies of $SO(2)$.  We conjecture that this still holds for \eref{mirrtriangleCCD}.  Indeed, if we assume that this is true, we obtain the quaternionic dimension of the Higgs branch to be
\be
\begin{split}
&\dim_{\BH} \CH_{\eref{mirrtriangleCCD}}  \\
&= \left[\frac{1}{2}(2N)(2N+1)-N \right] +N +N +(2n)N(2N+2) \\
& - (n) \left[ \frac{1}{2}(2N+2)(2N+1) \right] - (n-1+2) \left[ \frac{1}{2}(2N)(2N+1) \right] + {\blue n} \\
&= N~,
\end{split}
\ee 
where ${\blue n}$ in the second line is there due to the unbroken symmetry $SO(2)^n$ at a generic point of the Higgs branch.   This is in agreement with the dimension of the Coulomb branch of \eref{triangleCCD}, and is indeed consistent with mirror symmetry.

\subsubsection*{\it The special case of $N=n=1$}
In this case, the Coulomb branch of \eref{mirrtriangleCCD} is equal to that of the $SO(4)$ gauge theory with $2$ flavours (which has a $USp(4)$ flavour symmetry).  Although the latter is a bad theory, there is a mirror theory which has a ``good'' Lagrangian description.  The latter is denoted by $T^{(2,1^2)}(USp(4))$, whose quiver description is (see \cite[Table 2]{Cremonesi:2014uva}):
\be \label{altmirrN1n1}
\scalebox{0.8}{
\begin{tikzpicture}[baseline]
\tikzstyle{every node}=[font=\scriptsize, minimum size=0.5cm]
\node[draw, circle, red] (n1) at (-2,0) {$2$};
\node[draw, circle, blue] (n2) at (0,0) {$2$};
\node[draw, circle, red] (n3) at (2,0) {$3$};
\node[draw, rectangle, red] (f1) at (0,1.5) {$1$};
\node[draw, rectangle, blue] (f2) at (2,1.5) {$2$};
\draw[draw=black,solid,thick,-]  (n1)--(n2)--(n3) ; 
\draw[draw=black,solid,thick,-]  (f1)--(n2) ; 
\draw[draw=black,solid,thick,-]  (f2)--(n3) ; 
\end{tikzpicture}}
\ee
where each red circular node should be taken as an $SO$ gauge group.  The Higgs branch Hilbert series of \eref{altmirrN1n1} is indeed in agreement with \eref{HiggsNn1DDC}, consistent with mirror symmetry.


\section{Models with the exceptional group $G_2$} \label{sec:G2}
\subsection{Self-mirror models with a $T(G_2)$ link}
In this section, we turn to models with a $T(G_2)$ link connecting between two $G_2$ gauge groups.  We do not have the Type IIB brane construction for such theories.  Nevertheless, it is still possible to make some interesting statements regarding the moduli space.  We consider the following quiver:
\be \label{G2USp4}
\scalebox{0.75}{
\begin{tikzpicture}[baseline, scale=0.6,font=\scriptsize]
\begin{scope}[auto,%
  every node/.style={draw, minimum size=1.2cm}];
\def \n {8}
\def \radius {4.2cm}
\def \margin {14} 
\draw[-, >=latex] ({360/\n * (1 - 3)+\margin}:\radius);
arc ({360/\n * (1 - 3)+\margin}:{360/\n * (1-2)-\margin}:\radius);
\node[draw=none] at ({360/\n * (2 - 2)}:\radius) {{\large $\vdots$}};
\draw[-, >=latex] ({360/\n * (2 - 3)+\margin}:\radius);
arc ({360/\n * (2 - 3)+\margin}:{360/\n * (2-2)-\margin}:\radius);
\draw[-, >=latex] ({360/\n * (3 - 3)+\margin}:\radius);
arc ({360/\n * (3 - 3)+\margin}:{360/\n * (3-2)-\margin}:\radius);
\draw[-, >=latex] ({360/\n * (4 - 3)+\margin}:\radius);
arc ({360/\n * (5 - 3)+\margin}:{360/\n * (5-2)-\margin}:\radius);
\draw[-, >=latex] ({360/\n * (5 - 3)+\margin}:\radius);
arc ({360/\n * (5 - 3)+\margin}:{360/\n * (5-2)-\margin}:\radius);
\node[draw, circle, blue] (n2) at ({360/\n * (0 - 2)}:\radius) {$4'$};
\node[draw, circle, black] at ({360/\n * (1 - 2)}:\radius) {$G_2$};
\node[draw, circle,blue] at ({360/\n * (3 - 2)}:\radius) {$4'$};
\node[draw, circle,black] at ({360/\n * (4 - 2)}:\radius) {$G_2$};
\node[draw, circle,blue ] (n1) at ({360/\n * (5 - 2)}:\radius) {$4'$};
\foreach \s in {1,...,6}
{	
	\draw[-, >=latex] ({360/\n * (\s - 3)+\margin}:\radius) 
	arc ({360/\n * (\s - 3)+\margin}:{360/\n * (\s-2)-\margin}:\radius);
}
\draw[-, >=latex] ({360/\n * (0 - 3)+\margin}:\radius) 
arc ({360/\n * (0 - 3)+\margin}:{360/\n * (0-2)-\margin}:\radius);
\node[draw, circle,black] at ({360/\n * (6 - 2)}:\radius) {$G_2$};
\node[draw, circle,black] at ({360/\n * (-1 - 2)}:\radius) {$G_2$};
\draw[-, >=latex,red,thick,snake it] ({360/\n * (-1 - 3)+\margin}:\radius) 
arc ({360/\n * (-1 - 3)+\margin}:{360/\n * (-1-2)-\margin}:\radius);
\node[draw=none] at (9.3,-1) {\large $\substack{\text{$n$ blue nodes + $(n-1)$ $G_2$ usual} \\~ \\ \text{circular nodes + 2 $G_2$ nodes} \\~ \\ \text{connected by $T(G_2)$}}$};
\node[draw=none] at (-5.1,-1.8) {{\red $T(G_2)$}};
\end{scope}
\end{tikzpicture}}
\ee
Note that every gauge group in the quiver has the same rank, in the same way as the preceding sections.  The Higgs branch dimension of this quiver is
\be 
\dim_{\BH}\, \CH_{\eref{G2USp4}} = (14-2)+\frac{1}{2}(2n)(4)(7)- 10n-14(n-1+2) = 2(2n-1)~.
\ee
On the other hand, the Coulomb branch dimension of this quiver is
\be
\dim_{\BH}\, \CC_{\eref{G2USp4}} = 2(2n-1)~.
\ee
Observe that the dimensions of the Higgs and Coulomb branches are equal.  Indeed, we claim that quiver \eref{G2USp4} if {\bf self-mirror}.   We shall consider some special cases and compute the Hilbert series to support this statement below.

In the absence of $S$-fold, the two $G_2$ gauge groups merge into a single gauge group and quiver \eref{G2USp4} reduces to
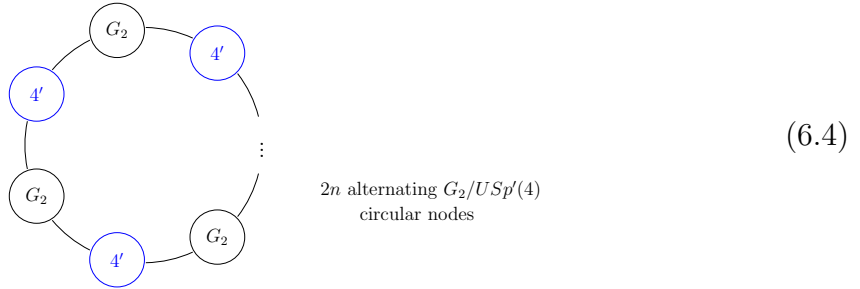
\be \label{altG2USp4}
\scalebox{0.6}{
\begin{tikzpicture}[baseline,font=\normalsize]
\begin{scope}[auto,%
  every node/.style={draw, minimum size=1.2cm}];
\def \n {7}
\def \radius {2.6cm}
\def \margin {14} 
\draw[-, >=latex] ({360/\n * (1 - 3)+\margin}:\radius);
arc ({360/\n * (1 - 3)+\margin}:{360/\n * (1-2)-\margin}:\radius);
\node[draw=none] at ({360/\n * (2 - 2)}:\radius) {{\large $\vdots$}};
\draw[-, >=latex] ({360/\n * (2 - 3)+\margin}:\radius);
arc ({360/\n * (2 - 3)+\margin}:{360/\n * (2-2)-\margin}:\radius);
\draw[-, >=latex] ({360/\n * (3 - 3)+\margin}:\radius);
arc ({360/\n * (3 - 3)+\margin}:{360/\n * (3-2)-\margin}:\radius);
\draw[-, >=latex] ({360/\n * (4 - 3)+\margin}:\radius);
arc ({360/\n * (5 - 3)+\margin}:{360/\n * (5-2)-\margin}:\radius);
\draw[-, >=latex] ({360/\n * (5 - 3)+\margin}:\radius);
arc ({360/\n * (5 - 3)+\margin}:{360/\n * (5-2)-\margin}:\radius);
\node[draw, circle, blue] (n2) at ({360/\n * (0 - 2)}:\radius) {$4'$};
\node[draw, circle, black] at ({360/\n * (1 - 2)}:\radius) {$G_2$};
\node[draw, circle,blue] at ({360/\n * (3 - 2)}:\radius) {$4'$};
\node[draw, circle,black] at ({360/\n * (4 - 2)}:\radius) {$G_2$};
\node[draw, circle,blue ] (n1) at ({360/\n * (5 - 2)}:\radius) {$4'$};
\foreach \s in {1,...,6}
{	
	\draw[-, >=latex] ({360/\n * (\s - 3)+\margin}:\radius) 
	arc ({360/\n * (\s - 3)+\margin}:{360/\n * (\s-2)-\margin}:\radius);
}
\draw[-, >=latex] ({360/\n * (0 - 3)+\margin}:\radius) 
arc ({360/\n * (0 - 3)+\margin}:{360/\n * (0-2)-\margin}:\radius);
\node[draw, circle,black] at ({360/\n * (6 - 2)}:\radius) {$G_2$};
\node[draw=none] at (6.3,-1) {$2n$ alternating $G_2$/$USp'(4)$};
\node[draw=none] at (6,-1.5) {circular nodes};
\end{scope}
\end{tikzpicture}}
\ee
It can also be checked that the Higgs and Coulomb branch dimensions of this quiver are equal:
\be
\dim_{\BH} \, \CH_{\eref{altG2USp4}} =\dim_{\BH} \, \CC_{\eref{altG2USp4}} =4n~.
\ee
Again, we claim that quiver \eref{altG2USp4} is also self-mirror.  Indeed, one can check using the Hilbert series (say for $n=1,\, 2$), in a similar way as that will be presented below, that the Higgs and Coulomb branches of \eref{altG2USp4} are equal.

Since we do not know the brane configurations for \eref{G2USp4} and \eref{altG2USp4}, we cannot definitely confirm if the gauge nodes labelled by $4$ is really $USp(4)$ or $USp'(4)$.  Nevertheless, we conjecture that such gauge nodes are $USp'(4)$, due to the fact that we can perform an ``$S$-folding'' and obtain another quiver which is self-dual.  The latter is depicted in \eref{G2USp4withprime} and will be discussed in detail in the next subsection.

\subsection*{\it The case of $n=1$}
In this case, \eref{G2USp4} reduces to the following quiver:
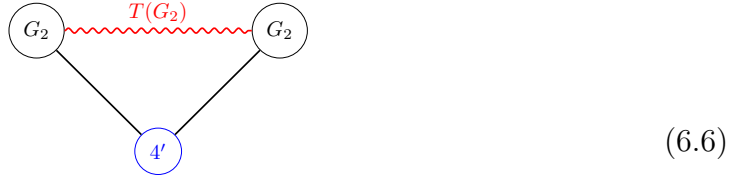
\be \label{triangleG2G2C}
\scalebox{0.8}{
\begin{tikzpicture}[baseline]
\tikzstyle{every node}=[font=\footnotesize]
\node[draw, circle, black] (node1) at (-2,2) {$G_2$};
\node[draw, circle, black] (node2) at (2,2) {$G_2$};
\node[draw, circle, blue] (node3) at (0,0) {$4'$};
\draw[draw=black,solid,thick,-]  (node2)--(node3) ; 
\draw[draw=black,solid,thick,-]  (node1)--(node3) ; 
\draw[draw=red,snake it,thick,-]  (node1)--(node2) node at (0,2.3) {{\red $T(G_2)$}} ; 
\end{tikzpicture}}
\ee
The Higgs branch Hilbert series can be computed as
\be \label{HSHiggstriangleG2G2C}
H[ \CH_{\eref{triangleG2G2C}}] (t) = \int {\mathrm d} \mu_{USp(4)}(\vec z)  \left \{ H[ \CH_{\eref{usp2ng2}} ](t; \vec z) \right \}^2 \PE\left[ -\chi^{C_2}_{[2,0]}(\vec z) t^2 \right]~,
\ee
where $\vec z = (z_1, z_2)$ and $H[ \CH_{\eref{usp2ng2}} ](t; \vec z)$ is given by \eref{HSusp2ng2}.  The integration yields
\be
H[ \CH_{\eref{triangleG2G2C}}] (t) = \PE \left[ t^4 + t^6 + 2 t^8 + t^{10} + t^{12} - t^{20} - t^{24} \right]~.
\ee
This is the Coulomb branch Hilbert series of 3d $\CN=4$ $USp(4)$ gauge theory with 7 flavours \cite[(5.14)]{Cremonesi:2013lqa}.    On the other hand, since the vector multiplet of  the $G_2$ gauge groups connected by $T(G_2)$ do not contribute to the Coulomb branch, the Coulomb branch of \eref{triangleG2G2C} is also isomorphic to the Coulomb branch of 3d $\CN=4$ $USp(4)$ gauge theory with 7 flavours.  

We see that the Higgs and the Coulomb branches of \eref{triangleG2G2C} are equal to each other.  We thus expect that theory \eref{triangleG2G2C} is self-mirror.

\subsection*{\it The case of $n=2$}
In this case, \eref{G2USp4} reduces to the following quiver:
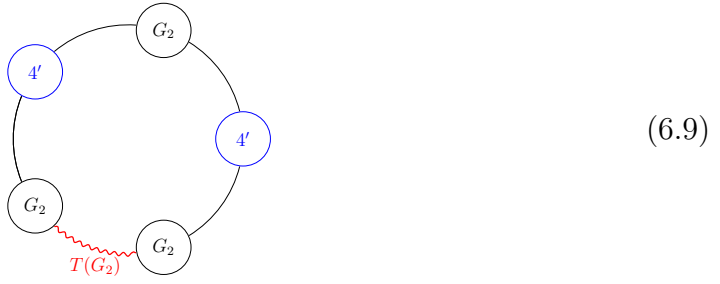
\be \label{G2USp4n2}
\scalebox{0.6}{
\begin{tikzpicture}[baseline, scale=0.6,font=\normalsize]
\begin{scope}[auto,%
  every node/.style={draw, minimum size=1.2cm}];
\def \n {5}
\def \radius {4.2cm}
\def \margin {14} 
\draw[-, >=latex] ({360/\n * (1 - 3)+\margin}:\radius);
arc ({360/\n * (1 - 3)+\margin}:{360/\n * (1-2)-\margin}:\radius);
\draw[-, >=latex] ({360/\n * (2 - 3)+\margin}:\radius);
arc ({360/\n * (2 - 3)+\margin}:{360/\n * (2-2)-\margin}:\radius);
\draw[-, >=latex] ({360/\n * (3 - 3)+\margin}:\radius);
arc ({360/\n * (3 - 3)+\margin}:{360/\n * (3-2)-\margin}:\radius);
\draw[-, >=latex] ({360/\n * (4 - 3)+\margin}:\radius);
arc ({360/\n * (4 - 3)+\margin}:{360/\n * (4-2)-\margin}:\radius);
\node[draw, circle, black] at ({360/\n * (1 - 2)}:\radius) {$G_2$};
\node[draw, circle,blue] at ({360/\n * (2 - 2)}:\radius) {$4'$};
\node[draw, circle,black] at ({360/\n * (3 - 2)}:\radius) {$G_2$};
\node[draw, circle,blue ]  at ({360/\n * (4 - 2)}:\radius) {$4'$};
\foreach \s in {2,...,5}
{	
	\draw[-, >=latex] ({360/\n * (\s - 3)+\margin}:\radius) 
	arc ({360/\n * (\s - 3)+\margin}:{360/\n * (\s-2)-\margin}:\radius);
}
\draw[-, >=latex] ({360/\n * (0 - 3)+\margin}:\radius) 
arc ({360/\n * (0 - 3)+\margin}:{360/\n * (0-2)-\margin}:\radius);
\node[draw, circle,black] at ({360/\n * (5 - 2)}:\radius) {$G_2$};
\draw[-, >=latex,red,thick,snake it] ({360/\n * (1 - 3)+\margin}:\radius) 
arc ({360/\n * (1 - 3)+\margin}:{360/\n * (1-2)-\margin}:\radius);
\node[draw=none] at (-1.2,-4.7) {{\red $T(G_2)$}};
\end{scope}
\end{tikzpicture}}
\ee
The Higgs branch Hilbert series can be computed similarly as before:
\be \label{HSHiggsG2USp4n2}
\begin{split}
H[ \CH_{\eref{G2USp4n2}}] (t) &= \int {\mathrm d} \mu_{USp(4)}(\vec u)\int {\mathrm d} \mu_{USp(4)}(\vec v) \int {\mathrm d} \mu_{G_2}(\vec w)  \times \\
& \quad H[ \CH_{\eref{usp2ng2}} ](t; \vec u) H[ \CH_{\eref{usp2ng2}} ](t; \vec v)  \PE \left[ \chi^{C_2}_{[1,0]}(\vec u) \chi^{G_2}_{[1,0]}(\vec w) + \vec u \leftrightarrow \vec v  \right]\\
& \quad \PE\left[ -\chi^{C_2}_{[2,0]}(\vec u) t^2-\chi^{C_2}_{[2,0]}(\vec v) t^2 - \chi^{G_2}_{[0,1]}(\vec w) t^2\right]~.
\end{split}
\ee 
The Coulomb branch Hilbert series can be computed as if the two $G_2$ symmetries that are connected by $T(G_2)$ becomes two separated flavour nodes:
\be \label{HSCoulG2USp4n2}
\begin{split}
H[\CC_{\eref{G2USp4n2}}] (t) &= \sum_{n_1, n_2 \geq 0} \,\, \sum_{a_1 \geq a_2 \geq 0} \,\, \sum_{b_1 \geq b_2 \geq 0} t^{2\Delta(\vec n, \vec a, \vec b)} P_{G_2}(t; \vec n) P_{C_2}(t; \vec a) P_{C_2}(t; \vec b)\\
\end{split}
\ee
where $\vec n=(n_1,n_2)$ are the fluxes of the $G_2$ gauge group, $\vec a=(a_1, a_2)$ and $\vec b = (b_1,b_2)$ are the fluxes for the two $USp(4)$ gauge groups.  Here $\Delta(\vec n, \vec a, \vec b)$ is the dimension of the monopole operator:
\be 
\begin{split}
\Delta(\vec n, \vec a, \vec b) &= \Delta^{\text{hyp}}_{G_2-C_2}( \vec 0, \vec a) + \Delta^{\text{hyp}}_{G_2-C_2}(\vec 0, \vec b)+  \Delta^{\text{hyp}}_{G_2-C_2}(\vec n, \vec a) + \Delta^{\text{hyp}}_{G_2-C_2}(\vec n, \vec b) \\
& \qquad - \Delta^{\text{vec}}_{G_2}(\vec n)- \Delta^{\text{vec}}_{C_2}(\vec a)- \Delta^{\text{vec}}_{C_2}(\vec b) \\
2\Delta^{\text{hyp}}_{G_2-C_2}(\vec n, \vec a) &=  \frac{1}{2} \sum_{\pm} \sum_{i=1}^2 \Big[ |n_1 \pm a_i | + |n_1+n_2 \pm a_i| + |2n_1+n_2 \pm a_i| +  \\
& \qquad \qquad \qquad \qquad + (n_1 \rightarrow -n_1, \, n_2 \rightarrow -n_2) + | \pm a_i| \Big] \\
\Delta^{\text{vec}}_{G_2}(\vec n) &= |n_1|+|n_2|+|n_1+n_2|+|2n_1 +n_2| +|3n_1+n_2|+|3n_1+2n_2| \\
\Delta^{\text{vec}}_{C_2}(\vec a) &= |2a_1|+|2a_2|+|a_1+a_2|+|a_1-a_2|~.
\end{split}
\ee
The dressing factors $P_{C_2}(t; \vec a)$ and $P_{G_2}(t; \vec n)$ are given by \cite[(A.8), (5.27)]{Cremonesi:2013lqa}:
\be
\begin{split}
P_{C_2}(t; a_1, a_2) &= \begin{cases} (1-t^2)^{-2} \quad & a_1 >a_2>0 \\ (1-t^2)^{-1}(1-t^4)^{-1} \quad & a_1 >a_2=0 \,\, \text{or} \,\, a_1=a_2>0 \\ (1-t^4)^{-1}(1-t^8)^{-1} \quad & a_1=a_2=0
\end{cases} \\
P_{G_2}(t; n_1, n_2) &= \begin{cases} (1-t^2)^{-2} \quad & n_1 >n_2>0 \\ (1-t^2)^{-1}(1-t^4)^{-1} \quad & n_1=0, n_2>0 \,\, \text{or} \,\, n_1>0, n_2=0 \\ (1-t^4)^{-1}(1-t^{12})^{-1} \quad & n_1=n_2=0
\end{cases}~.
\end{split}
\ee

Upon calculating the integrals and the summations, we check up to order $t^8$ that the Higgs branch and the Coulomb branch Hilbert series are equal to each other:
\be
H[ \CH_{\eref{G2USp4n2}}] (t) = H[\CC_{\eref{G2USp4n2}}] (t)  =  \PE \left[ 4 t^4 + 5 t^6 + 10 t^8 + \ldots \right]~.
\ee
This again supports our claim that \eref{G2USp4n2} is self-mirror.

\subsection{Self-mirror models with a $T(USp'(4))$ link}
We can obtain another variation of \eref{G2USp4} by simply $S$-folding one of the $USp'(4)$ gauge nodes in \eref{altG2USp4}.  The result is
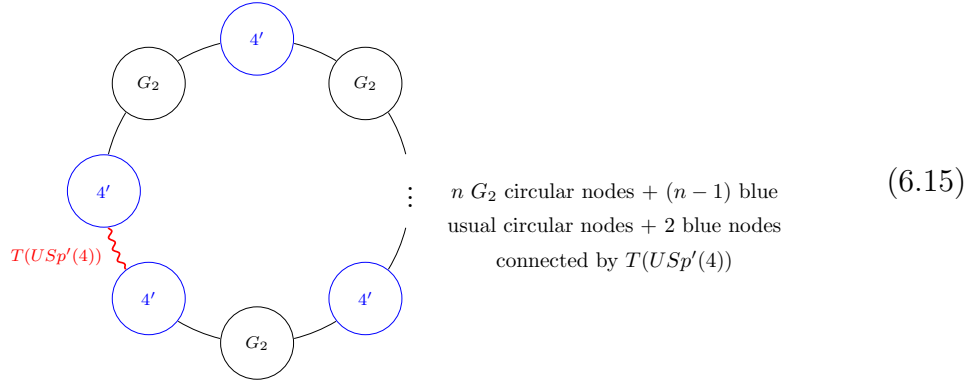
\be \label{G2USp4withprime}
\scalebox{0.8}{
\begin{tikzpicture}[baseline, scale=0.6,font=\scriptsize]
\begin{scope}[auto,%
  every node/.style={draw, minimum size=1.2cm}];
\def \n {8}
\def \radius {4.2cm}
\def \margin {14} 
\draw[-, >=latex] ({360/\n * (1 - 3)+\margin}:\radius);
arc ({360/\n * (1 - 3)+\margin}:{360/\n * (1-2)-\margin}:\radius);
\node[draw=none] at ({360/\n * (2 - 2)}:\radius) {{\large $\vdots$}};
\draw[-, >=latex] ({360/\n * (2 - 3)+\margin}:\radius);
arc ({360/\n * (2 - 3)+\margin}:{360/\n * (2-2)-\margin}:\radius);
\draw[-, >=latex] ({360/\n * (3 - 3)+\margin}:\radius);
arc ({360/\n * (3 - 3)+\margin}:{360/\n * (3-2)-\margin}:\radius);
\draw[-, >=latex] ({360/\n * (4 - 3)+\margin}:\radius);
arc ({360/\n * (5 - 3)+\margin}:{360/\n * (5-2)-\margin}:\radius);
\draw[-, >=latex] ({360/\n * (5 - 3)+\margin}:\radius);
arc ({360/\n * (5 - 3)+\margin}:{360/\n * (5-2)-\margin}:\radius);
\node[draw, circle, black] (n2) at ({360/\n * (0 - 2)}:\radius) {$G_2$};
\node[draw, circle, blue] at ({360/\n * (1 - 2)}:\radius) {$4'$};
\node[draw, circle,black] at ({360/\n * (3 - 2)}:\radius) {$G_2$};
\node[draw, circle,blue] at ({360/\n * (4 - 2)}:\radius) {$4'$};
\node[draw, circle,black ] (n1) at ({360/\n * (5 - 2)}:\radius) {$G_2$};
\foreach \s in {1,...,6}
{	
	\draw[-, >=latex] ({360/\n * (\s - 3)+\margin}:\radius) 
	arc ({360/\n * (\s - 3)+\margin}:{360/\n * (\s-2)-\margin}:\radius);
}
\draw[-, >=latex] ({360/\n * (0 - 3)+\margin}:\radius) 
arc ({360/\n * (0 - 3)+\margin}:{360/\n * (0-2)-\margin}:\radius);
\node[draw, circle,blue] at ({360/\n * (6 - 2)}:\radius) {$4'$};
\node[draw, circle,blue] at ({360/\n * (-1 - 2)}:\radius) {$4'$};
\draw[-, >=latex,red,thick,snake it] ({360/\n * (-1 - 3)+\margin}:\radius) 
arc ({360/\n * (-1 - 3)+\margin}:{360/\n * (-1-2)-\margin}:\radius);
\node[draw=none] at (9.8,-1) {\large $\substack{\text{$n$ $G_2$ circular nodes + $(n-1)$ blue} \\~ \\ \text{usual circular nodes + 2 blue nodes} \\~ \\ \text{connected by $T(USp'(4))$}}$};
\node[draw=none] at (-5.5,-1.8) {{\red $T(USp'(4))$}};
\end{scope}
\end{tikzpicture}}
\ee
The dimension of the Higgs branch is indeed equal to that of the Coulomb branch:
\be
\dim_\BH \, \CH_{\eref{G2USp4withprime}} = \dim_\BH \, \CC_{\eref{G2USp4withprime}}  = 2(2n-1)~.
\ee
We claim that \eref{G2USp4withprime} is also self-mirror for any $n \geq 1$.  One can indeed check, for example in the cases of $n=1,\,2$, that the Higgs and the Coulomb branch Hilbert series are equal, in the same way as presented in the preceding subsection.   As an example, for $n=1$, these are equal to the Coulomb branch Hilbert series of the $G_2$ gauge theory with 4 flavours \cite[(5.28)]{Cremonesi:2013lqa}:
\be
H[ \CH_{\eref{G2USp4withprime}}|_{n=1} ] = H[ \CC_{\eref{G2USp4withprime}}|_{n=1} ]  = \PE \left[ 2 t^4 + t^6 + t^8 + t^{10} + 2t^{12}+ \ldots \right]~.
\ee 

We finally remark that since we can perform an ``$S$-folding'' at any blue node, this confirms that each blue node labelled by $4$ is indeed $USp'(4)$.

\subsection{More mirror pairs by adding flavours}
In this subsection, we add fundamental flavours to the self-mirror models discussed earlier and obtain mirror pairs that are not self-dual.

\subsubsection{Models with a $T(G_2)$ link}
Let us start the discussion by adding $n$ flavours to the $USp'(4)$ gauge group in \eref{triangleG2G2C}.  This yields
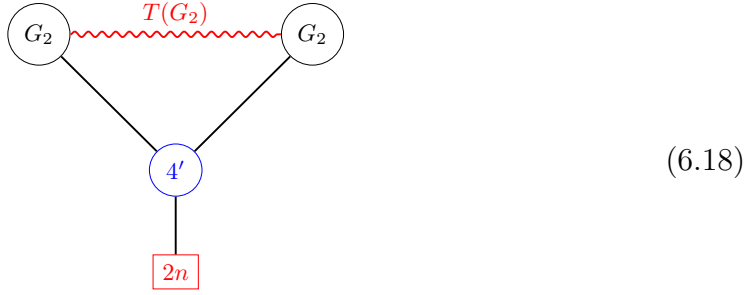
\be \label{flvtriangleG2G2C}
\scalebox{0.9}{
\begin{tikzpicture}[baseline]
\tikzstyle{every node}=[font=\footnotesize]
\node[draw, circle, black] (node1) at (-2,2) {$G_2$};
\node[draw, circle, black] (node2) at (2,2) {$G_2$};
\node[draw, circle, blue] (node3) at (0,0) {$4'$};
\node[draw, rectangle, red] (sqnode) at (0,-1.5) {$2n$};
\draw[draw=black,solid,thick,-]  (node2)--(node3) ; 
\draw[draw=black,solid,thick,-]  (node1)--(node3) ; 
\draw[draw=red,snake it,thick,-]  (node1)--(node2) node at (0,2.3) {{\red $T(G_2)$}} ; 
\draw[draw=black,solid,thick,-]  (node3)--(sqnode) ; 
\end{tikzpicture}}
\ee
where the flavour symmetry is $SO(2n)$.  The dimensions of the Higgs and Coulomb branches of this quiver are
\be
\begin{split}
\dim_\BH \, \CH_{\eref{flvtriangleG2G2C}} = 4n+2~, \qquad
\dim_\BH \, \CC_{\eref{flvtriangleG2G2C}} = 2~.
\end{split}
\ee 

We propose that \eref{flvtriangleG2G2C} is mirror dual to
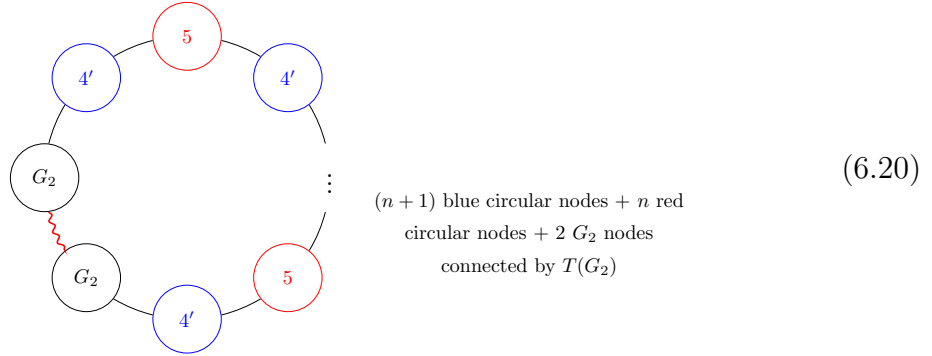
\be \label{mirrflvtriangleG2G2C}
\scalebox{0.75}{
\begin{tikzpicture}[baseline, font=\footnotesize]
\begin{scope}[auto,%
  every node/.style={draw, minimum size=1.2cm}];
\def \n {8}
\def \radius {2.5cm}
\def \margin {14} 
\draw[-, >=latex] ({360/\n * (1 - 3)+\margin}:\radius);
arc ({360/\n * (1 - 3)+\margin}:{360/\n * (1-2)-\margin}:\radius);
\node[draw=none] at ({360/\n * (2 - 2)}:\radius) {{\large $\vdots$}};
\draw[-, >=latex] ({360/\n * (2 - 3)+\margin}:\radius);
arc ({360/\n * (2 - 3)+\margin}:{360/\n * (2-2)-\margin}:\radius);
\draw[-, >=latex] ({360/\n * (3 - 3)+\margin}:\radius);
arc ({360/\n * (3 - 3)+\margin}:{360/\n * (3-2)-\margin}:\radius);
\draw[-, >=latex] ({360/\n * (4 - 3)+\margin}:\radius);
arc ({360/\n * (5 - 3)+\margin}:{360/\n * (5-2)-\margin}:\radius);
\draw[-, >=latex] ({360/\n * (5 - 3)+\margin}:\radius);
arc ({360/\n * (5 - 3)+\margin}:{360/\n * (5-2)-\margin}:\radius);
\node[draw, circle, blue] (n2) at ({360/\n * (0 - 2)}:\radius) {$4'$};
\node[draw, circle, red] at ({360/\n * (1 - 2)}:\radius) {$5$};
\node[draw, circle,blue] at ({360/\n * (3 - 2)}:\radius) {$4'$};
\node[draw, circle,red] at ({360/\n * (4 - 2)}:\radius) {$5$};
\node[draw, circle,blue] (n1) at ({360/\n * (5 - 2)}:\radius) {$4'$};
\node[draw, circle,black] at ({360/\n * (6 - 2)}:\radius) {$G_2$};
\foreach \s in {1,...,6}
{	
	\draw[-, >=latex] ({360/\n * (\s - 3)+\margin}:\radius) 
	arc ({360/\n * (\s - 3)+\margin}:{360/\n * (\s-2)-\margin}:\radius);
}
\draw[-, >=latex] ({360/\n * (0 - 3)+\margin}:\radius) 
arc ({360/\n * (0 - 3)+\margin}:{360/\n * (0-2)-\margin}:\radius);
\node[draw, circle,black] at ({360/\n * (-1 - 2)}:\radius) {$G_2$};
\draw[-, >=latex,red,thick,snake it] ({360/\n * (-1 - 3)+\margin}:\radius) 
arc ({360/\n * (-1 - 3)+\margin}:{360/\n * (-1-2)-\margin}:\radius);
\node[draw=none] at (6,-1) {\large $\substack{\text{$(n+1)$ blue circular nodes + $n$ red} \\~ \\ \text{circular nodes + 2 $G_2$ nodes} \\~ \\ \text{connected by $T(G_2)$}}$};
\end{scope}
\end{tikzpicture}}
\ee
The Higgs branch dimension of this model is
\be
\begin{aligned}
\dim_{\BH} \,\CH_{\eref{mirrflvtriangleG2G2C}} &= (14-2) +2 \left(\frac{1}{2} \times 7 \times 4 \right) + 10(2n) \\
& \qquad -14 -14-10(n+1)-10n \\
&= 2~.
\end{aligned}
\ee
and the Coulomb branch dimension of this is $\dim_{\BH} \,\CC_{\eref{mirrflvtriangleG2G2C}} = 2(2n+1)$.  This is consistent with mirror symmetry.  We shall soon match the Higgs/Coulomb branch Hilbert series of \eref{flvtriangleG2G2C} with the Coulomb/Higgs branch Hilbert series of \eref{mirrflvtriangleG2G2C} for $n=1$.  

Although we do not have a brane construction for \eref{mirrflvtriangleG2G2C} due to the presence of the $G_2$ gauge groups, the part of the quiver that contains alternating $USp'(4)$/$SO(5)$ gauge groups could be ``realised'' by a series of brane segments involving alternating $\Ott^+$/$\Ott^-$ across NS5 branes.   In other words, starting from \eref{flvtriangleG2G2C}, the mirror theory \eref{mirrflvtriangleG2G2C} can be obtained by making the following replacement:
\be \label{replace1}
\scalebox{0.8}{
\begin{tikzpicture}[baseline,font=\scriptsize]
\begin{scope}[auto,%
  every node/.style={draw, minimum size=0.8cm}, node distance=0.6cm];
\node[draw=none] (emp1) at (-1,1) {};
\node[draw=none] (emp2) at (-1,-1) {};
\node[draw, circle, blue] (node1) at (0,0) {$4'$};
\node[draw, rectangle,red] (sqnode) at (2,0) {$2n$};
\end{scope}
\draw (node1)--(sqnode);
\draw (node1)--(emp1);
\draw (node1)--(emp2);
\end{tikzpicture}}
\qquad \longrightarrow \qquad
\scalebox{0.7}{
\begin{tikzpicture}[baseline, font=\footnotesize]
\begin{scope}[auto,%
  every node/.style={draw, minimum size=1.2cm}];
\def \n {8}
\def \radius {2.5cm}
\def \margin {14} 
\draw[-, >=latex] ({360/\n * (1 - 3)+\margin}:\radius);
arc ({360/\n * (1 - 3)+\margin}:{360/\n * (1-2)-\margin}:\radius);
\node[draw=none] at ({360/\n * (2 - 2)}:\radius) {{\large $\vdots$}};
\draw[-, >=latex] ({360/\n * (2 - 3)+\margin}:\radius);
arc ({360/\n * (2 - 3)+\margin}:{360/\n * (2-2)-\margin}:\radius);
\draw[-, >=latex] ({360/\n * (3 - 3)+\margin}:\radius);
arc ({360/\n * (3 - 3)+\margin}:{360/\n * (3-2)-\margin}:\radius);
\draw[-, >=latex] ({360/\n * (4 - 3)+\margin}:\radius);
arc ({360/\n * (5 - 3)+\margin}:{360/\n * (5-2)-\margin}:\radius);
\draw[-, >=latex] ({360/\n * (5 - 3)+\margin}:\radius);
arc ({360/\n * (5 - 3)+\margin}:{360/\n * (5-2)-\margin}:\radius);
\node[draw, circle, blue] (n2) at ({360/\n * (0 - 2)}:\radius) {$4'$};
\node[draw, circle, red] at ({360/\n * (1 - 2)}:\radius) {$5$};
\node[draw, circle,blue] at ({360/\n * (3 - 2)}:\radius) {$4'$};
\node[draw, circle,red] at ({360/\n * (4 - 2)}:\radius) {$5$};
\node[draw, circle,blue] (n1) at ({360/\n * (5 - 2)}:\radius) {$4'$};
\foreach \s in {1,...,6}
{	
	\draw[-, >=latex] ({360/\n * (\s - 3)+\margin}:\radius) 
	arc ({360/\n * (\s - 3)+\margin}:{360/\n * (\s-2)-\margin}:\radius);
}
\draw[-, >=latex] ({360/\n * (0 - 3)+\margin}:\radius) 
arc ({360/\n * (0 - 3)+\margin}:{360/\n * (0-2)-\margin}:\radius);
\node[draw=none] at (5,-1) {\large $\substack{\text{$(n+1)$ blue circular nodes } \\~ \\ \text{+ $n$ red circular nodes}}$};
\end{scope}
\end{tikzpicture}}
\ee


In the absence of the $S$-fold, the two $G_2$ gauge groups that were connected by $T(G_2)$ merge into a single one.  We thus obtain the mirror pair between the following elliptic models:
\be \label{pairG2USp4}
\scalebox{0.9}{
\begin{tikzpicture}[baseline]
\tikzstyle{every node}=[font=\footnotesize]
\node[draw, circle, black] (node1) at (-2,1) {$G_2$};
\node[draw, circle, blue] (node3) at (-2,-1) {$4'$};
\node[draw, rectangle, red] (sqnode) at (-2,-2.5) {$2n$};
\draw[draw=black,solid,thick,-]  (node1) to [bend left] (node3) ;
\draw[draw=black,solid,thick,-]  (node1) to [bend right] (node3) ; 
\draw[draw=black,solid,thick,-]  (node3)--(sqnode) ; 
\end{tikzpicture}}
\qquad \qquad \longleftrightarrow \qquad \qquad
\scalebox{0.75}{
\begin{tikzpicture}[baseline, font=\footnotesize]
\begin{scope}[auto,%
  every node/.style={draw, minimum size=1.2cm}];
\def \n {7}
\def \radius {2.5cm}
\def \margin {14} 
\draw[-, >=latex] ({360/\n * (1 - 3)+\margin}:\radius);
arc ({360/\n * (1 - 3)+\margin}:{360/\n * (1-2)-\margin}:\radius);
\node[draw=none] at ({360/\n * (2 - 2)}:\radius) {{\large $\vdots$}};
\draw[-, >=latex] ({360/\n * (2 - 3)+\margin}:\radius);
arc ({360/\n * (2 - 3)+\margin}:{360/\n * (2-2)-\margin}:\radius);
\draw[-, >=latex] ({360/\n * (3 - 3)+\margin}:\radius);
arc ({360/\n * (3 - 3)+\margin}:{360/\n * (3-2)-\margin}:\radius);
\draw[-, >=latex] ({360/\n * (4 - 3)+\margin}:\radius);
arc ({360/\n * (5 - 3)+\margin}:{360/\n * (5-2)-\margin}:\radius);
\draw[-, >=latex] ({360/\n * (5 - 3)+\margin}:\radius);
arc ({360/\n * (5 - 3)+\margin}:{360/\n * (5-2)-\margin}:\radius);
\node[draw, circle, blue] (n2) at ({360/\n * (0 - 2)}:\radius) {$4'$};
\node[draw, circle, red] at ({360/\n * (1 - 2)}:\radius) {$5$};
\node[draw, circle,blue] at ({360/\n * (3 - 2)}:\radius) {$4'$};
\node[draw, circle,red] at ({360/\n * (4 - 2)}:\radius) {$5$};
\node[draw, circle,blue] (n1) at ({360/\n * (5 - 2)}:\radius) {$4'$};
\node[draw, circle,black] at ({360/\n * (6 - 2)}:\radius) {$G_2$};
\foreach \s in {1,...,6}
{	
	\draw[-, >=latex] ({360/\n * (\s - 3)+\margin}:\radius) 
	arc ({360/\n * (\s - 3)+\margin}:{360/\n * (\s-2)-\margin}:\radius);
}
\draw[-, >=latex] ({360/\n * (0 - 3)+\margin}:\radius) 
arc ({360/\n * (0 - 3)+\margin}:{360/\n * (0-2)-\margin}:\radius);
\node[draw=none] at (0,-4) {\large $\substack{\text{$(n+1)$ blue circular nodes +} \\~ \\ \text{$n$ red circular nodes + 1 $G_2$ node}}$};
\end{scope}
\end{tikzpicture}}
\ee

\subsubsection*{\it The case of $n=1$}
Let us first focus on \eref{flvtriangleG2G2C}.  The Higgs branch Hilbert series can be computed simply by putting the term $\PE[(x+x^{-1}) \chi^{C_2}_{[1,0]}(\vec z) t]$ in the integrand of \eref{HSHiggstriangleG2G2C}, where $x$ is the $SO(2)$ flavour fugacity.  Performing the integral, we obtain (after setting $x=1$):
\be \label{unrefn1}
H \left[ \CH_{\eref{flvtriangleG2G2C}}|_{n=1} \right] (t; x=1) = 1 + t^2 + 9 t^4 + 15 t^6 + 60 t^8 + 113 t^{10}+ \ldots~.
\ee
The Coulomb branch Hilbert series for \eref{flvtriangleG2G2C} is equal to that of the 3d $\CN=4$ $USp(4)$ gauge theory with $7+1=8$ flavours.  The latter is given by
\be \label{CoulUSp4w8flv}
H \left[ \CC_{\eref{flvtriangleG2G2C}}|_{n=1} \right] (t) =  \PE \left[t^4 + 2 t^8 + t^{10} + t^{12} + t^{14} - t^{24} - t^{28} \right]~.
\ee

Let us now turn to \eref{mirrflvtriangleG2G2C}.  The Higgs branch Hilbert series is given by \eref{HSHiggsG2USp4n2} with the following replacement:
\be
\int {\mathrm d} \mu_{G_2}(\vec w) \rightarrow \int {\mathrm d} \mu_{SO(5)}(\vec w)~, \quad \chi^{G_2}_{[1,0]}(\vec w) \rightarrow  \chi^{B_2}_{[1,0]}(\vec w)~, \quad \chi^{G_2}_{[0,1]}(\vec w) \rightarrow  \chi^{B_2}_{[0,2]}(\vec w)~.
\ee
We checked that the result of this agrees with \eref{CoulUSp4w8flv} up to order $t^{10}$.  The Coulomb branch Hilbert series of \eref{mirrflvtriangleG2G2C} can be obtained in a similar way from \eref{HSCoulG2USp4n2} with the following replacement:
\be
\begin{split}
\Delta^{\text{vec}}_{G_2}(\vec n) &\rightarrow \Delta^{\text{vec}}_{B_2}(\vec n) = |n_1|+|n_2|+|n_1+n_2|+|n_1-n_2| \\
\Delta^{\text{hyp}}_{G_2-C_2}(\vec n, \vec a \,\, \text{or} \, \,\vec b) &\rightarrow \Delta^{\text{hyp}}_{B_2-C_2}(\vec n, \vec a \,\,  \text{or} \,\, \vec b)\\
P_{G_2}(t; \vec n) &\rightarrow P_{C_2}(t;\vec n)
\end{split}
\ee
with 
\be
\Delta^{\text{hyp}}_{B_2-C_2}(\vec n, \vec a) = \frac{1}{2} \times \frac{1}{2} \sum_{s_1, s_2 =0}^1 \sum_{j=1}^2 \left[ |(-1)^{s_2} a_j|  +\sum_{i=1}^2  |(-1)^{s_1} n_i +(-1)^{s_2} a_j| \right]~.
\ee
Again, we checked that the result of this agrees with \eref{unrefn1} up to order $t^{10}$.

\subsubsection*{\it Generalisation of \eref{flvtriangleG2G2C} to a polygon with flavours added}
We can generalise \eref{flvtriangleG2G2C} to a polygon consisting of alternating $G_2$/$USp'(4)$ gauge groups, with $n$ flavours added to one of the $USp'(4)$ gauge group.  This is depicted below.
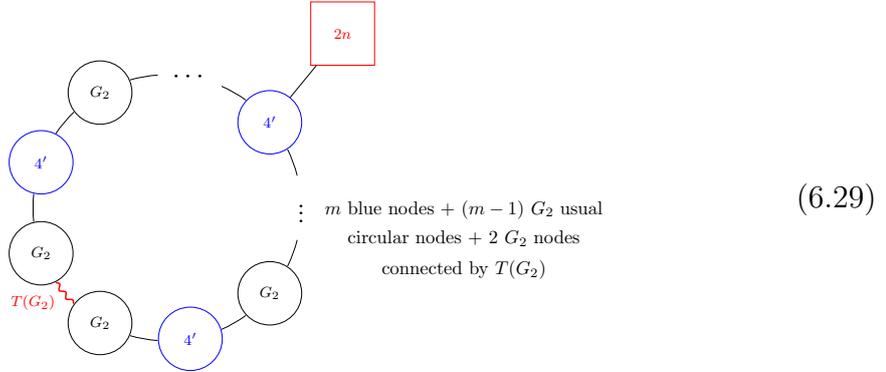
\be \label{flvpolyG2G2C}
\scalebox{0.7}{
\begin{tikzpicture}[baseline, scale=0.6,font=\scriptsize]
\begin{scope}[auto,%
  every node/.style={draw, minimum size=1.2cm}];
\def \n {9}
\def \radius {4.2cm}
\def \margin {14} 
\draw[-, >=latex] ({360/\n * (1 - 3)+\margin}:\radius);
arc ({360/\n * (1 - 3)+\margin}:{360/\n * (1-2)-\margin}:\radius);
\draw[-, >=latex] ({360/\n * (2 - 3)+\margin}:\radius);
arc ({360/\n * (2 - 3)+\margin}:{360/\n * (2-2)-\margin}:\radius);
\draw[-, >=latex] ({360/\n * (3 - 3)+\margin}:\radius);
arc ({360/\n * (3 - 3)+\margin}:{360/\n * (3-2)-\margin}:\radius);
\draw[-, >=latex] ({360/\n * (4 - 3)+\margin}:\radius);
arc ({360/\n * (5 - 3)+\margin}:{360/\n * (5-2)-\margin}:\radius);
\draw[-, >=latex] ({360/\n * (5 - 3)+\margin}:\radius);
arc ({360/\n * (5 - 3)+\margin}:{360/\n * (5-2)-\margin}:\radius);
\node[draw, circle, blue] (n2) at ({360/\n * (0 - 2)}:\radius) {$4'$};
\node[draw, circle, black] at ({360/\n * (1 - 2)}:\radius) {$G_2$};
\node[draw=none] at ({360/\n * (2 - 2)}:\radius) {{\large $\vdots$}};
\node[draw, circle,blue] (nn) at ({360/\n * (3 - 2)}:\radius) {$4'$};
\node[draw,rectangle,red] (ff) at (5.5,5.5) {$2n$};
\node[draw=none] at ({360/\n * (4 - 2)}:\radius) {\large $\cdots$};
\node[draw, circle,black] at ({360/\n * (5 - 2)}:\radius) {$G_2$};
\node[draw, circle,blue ] (n1) at ({360/\n * (6 - 2)}:\radius) {$4'$};
\foreach \s in {1,...,7}
{	
	\draw[-, >=latex] ({360/\n * (\s - 3)+\margin}:\radius) 
	arc ({360/\n * (\s - 3)+\margin}:{360/\n * (\s-2)-\margin}:\radius);
}
\draw[-, >=latex] ({360/\n * (0 - 3)+\margin}:\radius) 
arc ({360/\n * (0 - 3)+\margin}:{360/\n * (0-2)-\margin}:\radius);
\node[draw, circle,black] at ({360/\n * (7 - 2)}:\radius) {$G_2$};
\node[draw, circle,black] at ({360/\n * (-1 - 2)}:\radius) {$G_2$};
\draw[-, >=latex,red,thick,snake it] ({360/\n * (-1 - 3)+\margin}:\radius) 
arc ({360/\n * (-1 - 3)+\margin}:{360/\n * (-1-2)-\margin}:\radius);
\node[draw=none] at (9.3,-1) {\large $\substack{\text{$m$ blue nodes + $(m-1)$ $G_2$ usual} \\~ \\ \text{circular nodes + 2 $G_2$ nodes} \\~ \\ \text{connected by $T(G_2)$}}$};
\node[draw=none] at (-4.2,-3) {{\red $T(G_2)$}};
\draw[-,solid] (ff) to (nn);
\end{scope}
\end{tikzpicture}}
\ee
The mirror theory can simply be obtain by applying the replacement rule \eref{replace1}.  For example, we have the following mirror pair
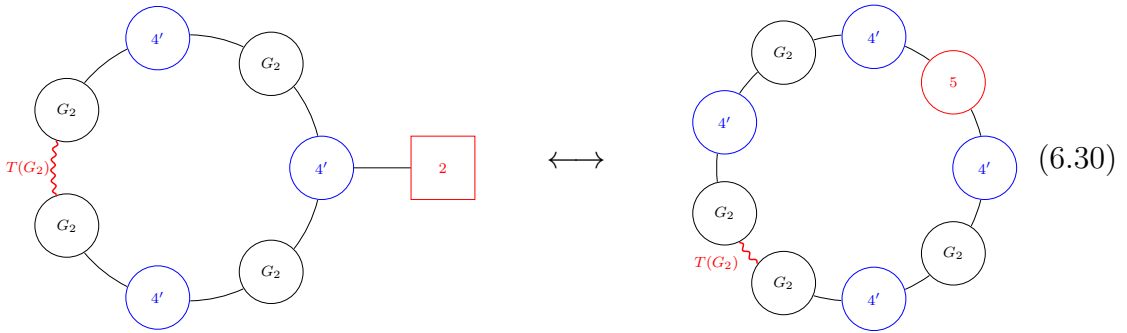
\be \label{exampleG2}
\scalebox{0.7}{
\begin{tikzpicture}[baseline, scale=0.6,font=\scriptsize]
\begin{scope}[auto,%
  every node/.style={draw, minimum size=1.2cm}];
\def \n {7}
\def \radius {4.2cm}
\def \margin {14} 
\draw[-, >=latex] ({360/\n * (1 - 3)+\margin}:\radius);
arc ({360/\n * (1 - 3)+\margin}:{360/\n * (1-2)-\margin}:\radius);
\draw[-, >=latex] ({360/\n * (2 - 3)+\margin}:\radius);
arc ({360/\n * (2 - 3)+\margin}:{360/\n * (2-2)-\margin}:\radius);
\draw[-, >=latex] ({360/\n * (3 - 3)+\margin}:\radius);
arc ({360/\n * (3 - 3)+\margin}:{360/\n * (3-2)-\margin}:\radius);
\draw[-, >=latex] ({360/\n * (4 - 3)+\margin}:\radius);
arc ({360/\n * (5 - 3)+\margin}:{360/\n * (5-2)-\margin}:\radius);
\draw[-, >=latex] ({360/\n * (5 - 3)+\margin}:\radius);
arc ({360/\n * (5 - 3)+\margin}:{360/\n * (5-2)-\margin}:\radius);
\node[draw, circle, blue] (n2) at ({360/\n * (0 - 2)}:\radius) {$4'$};
\node[draw, circle, black] at ({360/\n * (1 - 2)}:\radius) {$G_2$};
\node[draw, circle,blue] (nn) at ({360/\n * (2 - 2)}:\radius) {$4'$};
\node[draw,rectangle,red] (ff) at (8,0) {$2$};
\node[draw, circle,black] at ({360/\n * (3 - 2)}:\radius) {$G_2$};
\node[draw, circle,blue ]  at ({360/\n * (4 - 2)}:\radius) {$4'$};
\node[draw, circle,black ]  at ({360/\n * (5 - 2)}:\radius) {$G_2$};
\foreach \s in {1,...,5}
{	
	\draw[-, >=latex] ({360/\n * (\s - 3)+\margin}:\radius) 
	arc ({360/\n * (\s - 3)+\margin}:{360/\n * (\s-2)-\margin}:\radius);
}
\draw[-, >=latex] ({360/\n * (0 - 3)+\margin}:\radius) 
arc ({360/\n * (0 - 3)+\margin}:{360/\n * (0-2)-\margin}:\radius);
\node[draw, circle,black] at ({360/\n * (-1 - 2)}:\radius) {$G_2$};
\draw[-, >=latex,red,thick,snake it] ({360/\n * (-1 - 3)+\margin}:\radius) 
arc ({360/\n * (-1 - 3)+\margin}:{360/\n * (-1-2)-\margin}:\radius);
\node[draw=none] at (-5,0) {{\red $T(G_2)$}};
\draw[-,solid] (ff) to (nn);
\end{scope}
\end{tikzpicture}}
\qquad  \longleftrightarrow \qquad
\scalebox{0.7}{
\begin{tikzpicture}[baseline, scale=0.6,font=\scriptsize]
\begin{scope}[auto,%
  every node/.style={draw, minimum size=1.2cm}];
\def \n {9}
\def \radius {4.2cm}
\def \margin {14} 
\draw[-, >=latex] ({360/\n * (1 - 3)+\margin}:\radius);
arc ({360/\n * (1 - 3)+\margin}:{360/\n * (1-2)-\margin}:\radius);
\draw[-, >=latex] ({360/\n * (2 - 3)+\margin}:\radius);
arc ({360/\n * (2 - 3)+\margin}:{360/\n * (2-2)-\margin}:\radius);
\draw[-, >=latex] ({360/\n * (3 - 3)+\margin}:\radius);
arc ({360/\n * (3 - 3)+\margin}:{360/\n * (3-2)-\margin}:\radius);
\draw[-, >=latex] ({360/\n * (4 - 3)+\margin}:\radius);
arc ({360/\n * (5 - 3)+\margin}:{360/\n * (5-2)-\margin}:\radius);
\draw[-, >=latex] ({360/\n * (5 - 3)+\margin}:\radius);
arc ({360/\n * (5 - 3)+\margin}:{360/\n * (5-2)-\margin}:\radius);
\node[draw, circle, blue] (n2) at ({360/\n * (0 - 2)}:\radius) {$4'$};
\node[draw, circle, black] at ({360/\n * (1 - 2)}:\radius) {$G_2$};
\node[draw, circle, blue] at ({360/\n * (2 - 2)}:\radius) {$4'$};
\node[draw, circle, red] (nn) at ({360/\n * (3 - 2)}:\radius) {$5$};
\node[draw, circle, blue] at ({360/\n * (4 - 2)}:\radius) {$4'$};
\node[draw, circle,black] at ({360/\n * (5 - 2)}:\radius) {$G_2$};
\node[draw, circle,blue ] (n1) at ({360/\n * (6 - 2)}:\radius) {$4'$};
\foreach \s in {1,...,7}
{	
	\draw[-, >=latex] ({360/\n * (\s - 3)+\margin}:\radius) 
	arc ({360/\n * (\s - 3)+\margin}:{360/\n * (\s-2)-\margin}:\radius);
}
\draw[-, >=latex] ({360/\n * (0 - 3)+\margin}:\radius) 
arc ({360/\n * (0 - 3)+\margin}:{360/\n * (0-2)-\margin}:\radius);
\node[draw, circle,black] at ({360/\n * (7 - 2)}:\radius) {$G_2$};
\node[draw, circle,black] at ({360/\n * (-1 - 2)}:\radius) {$G_2$};
\draw[-, >=latex,red,thick,snake it] ({360/\n * (-1 - 3)+\margin}:\radius) 
arc ({360/\n * (-1 - 3)+\margin}:{360/\n * (-1-2)-\margin}:\radius);
\node[draw=none] at (-4.2,-3) {{\red $T(G_2)$}};
\end{scope}
\end{tikzpicture}}
\ee

As emphasised before, as a by-product, one may obtain a mirror pair between the usual field theories, without an $S$-fold, by simply merging the two $G_2$ nodes that are connected by $T(G_2)$.  The replacement rule described in \eref{replace1} still applies.  As an example, \eref{exampleG2} becomes
\be
\scalebox{0.7}{
\begin{tikzpicture}[baseline, scale=0.6,font=\scriptsize]
\begin{scope}[auto,%
  every node/.style={draw, minimum size=1.2cm}];
\def \n {6}
\def \radius {4.2cm}
\def \margin {14} 
\draw[-, >=latex] ({360/\n * (1 - 3)+\margin}:\radius);
arc ({360/\n * (1 - 3)+\margin}:{360/\n * (1-2)-\margin}:\radius);
\draw[-, >=latex] ({360/\n * (2 - 3)+\margin}:\radius);
arc ({360/\n * (2 - 3)+\margin}:{360/\n * (2-2)-\margin}:\radius);
\draw[-, >=latex] ({360/\n * (3 - 3)+\margin}:\radius);
arc ({360/\n * (3 - 3)+\margin}:{360/\n * (3-2)-\margin}:\radius);
\draw[-, >=latex] ({360/\n * (4 - 3)+\margin}:\radius);
arc ({360/\n * (5 - 3)+\margin}:{360/\n * (5-2)-\margin}:\radius);
\draw[-, >=latex] ({360/\n * (5 - 3)+\margin}:\radius);
arc ({360/\n * (5 - 3)+\margin}:{360/\n * (5-2)-\margin}:\radius);
\node[draw, circle, blue] (n2) at ({360/\n * (0 - 2)}:\radius) {$4'$};
\node[draw, circle, black] at ({360/\n * (1 - 2)}:\radius) {$G_2$};
\node[draw, circle,blue] (nn) at ({360/\n * (2 - 2)}:\radius) {$4'$};
\node[draw,rectangle,red] (ff) at (8,0) {$2$};
\node[draw, circle,black] at ({360/\n * (3 - 2)}:\radius) {$G_2$};
\node[draw, circle,blue ]  at ({360/\n * (4 - 2)}:\radius) {$4'$};
\node[draw, circle,black ]  at ({360/\n * (5 - 2)}:\radius) {$G_2$};
\foreach \s in {1,...,5}
{	
	\draw[-, >=latex] ({360/\n * (\s - 3)+\margin}:\radius) 
	arc ({360/\n * (\s - 3)+\margin}:{360/\n * (\s-2)-\margin}:\radius);
}
\draw[-, >=latex] ({360/\n * (0 - 3)+\margin}:\radius) 
arc ({360/\n * (0 - 3)+\margin}:{360/\n * (0-2)-\margin}:\radius);
\node[draw, circle,black] at ({360/\n * (-1 - 2)}:\radius) {$G_2$};
\draw[-,solid] (ff) to (nn);
\end{scope}
\end{tikzpicture}}
\qquad  \longleftrightarrow \qquad
\scalebox{0.7}{
\begin{tikzpicture}[baseline, scale=0.6,font=\scriptsize]
\begin{scope}[auto,%
  every node/.style={draw, minimum size=1.2cm}];
\def \n {8}
\def \radius {4.2cm}
\def \margin {14} 
\draw[-, >=latex] ({360/\n * (1 - 3)+\margin}:\radius);
arc ({360/\n * (1 - 3)+\margin}:{360/\n * (1-2)-\margin}:\radius);
\draw[-, >=latex] ({360/\n * (2 - 3)+\margin}:\radius);
arc ({360/\n * (2 - 3)+\margin}:{360/\n * (2-2)-\margin}:\radius);
\draw[-, >=latex] ({360/\n * (3 - 3)+\margin}:\radius);
arc ({360/\n * (3 - 3)+\margin}:{360/\n * (3-2)-\margin}:\radius);
\draw[-, >=latex] ({360/\n * (4 - 3)+\margin}:\radius);
arc ({360/\n * (5 - 3)+\margin}:{360/\n * (5-2)-\margin}:\radius);
\draw[-, >=latex] ({360/\n * (5 - 3)+\margin}:\radius);
arc ({360/\n * (5 - 3)+\margin}:{360/\n * (5-2)-\margin}:\radius);
\node[draw, circle, blue] (n2) at ({360/\n * (0 - 2)}:\radius) {$4'$};
\node[draw, circle, black] at ({360/\n * (1 - 2)}:\radius) {$G_2$};
\node[draw, circle, blue] at ({360/\n * (2 - 2)}:\radius) {$4'$};
\node[draw, circle, red] (nn) at ({360/\n * (3 - 2)}:\radius) {$5$};
\node[draw, circle, blue] at ({360/\n * (4 - 2)}:\radius) {$4'$};
\node[draw, circle,black] at ({360/\n * (5 - 2)}:\radius) {$G_2$};
\node[draw, circle,blue ] (n1) at ({360/\n * (6 - 2)}:\radius) {$4'$};
\foreach \s in {1,...,7}
{	
	\draw[-, >=latex] ({360/\n * (\s - 3)+\margin}:\radius) 
	arc ({360/\n * (\s - 3)+\margin}:{360/\n * (\s-2)-\margin}:\radius);
}
\draw[-, >=latex] ({360/\n * (0 - 3)+\margin}:\radius) 
arc ({360/\n * (0 - 3)+\margin}:{360/\n * (0-2)-\margin}:\radius);
\node[draw, circle,black] at ({360/\n * (7 - 2)}:\radius) {$G_2$};
\node[draw, circle,black] at ({360/\n * (-1 - 2)}:\radius) {$G_2$};
\end{scope}
\end{tikzpicture}}
\ee

\subsubsection{Models with a $T(USp'(4))$ link}
Instead of $S$-folding the $G_2$ node as in \eref{flvpolyG2G2C}, we can $S$-fold the $USp'(4)$ node and obtain
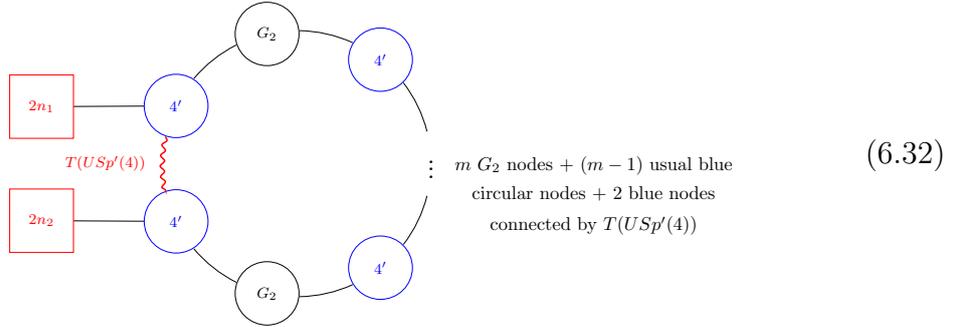
\be \label{twoflvUSp4p}
\scalebox{0.7}{
\begin{tikzpicture}[baseline, scale=0.6,font=\scriptsize]
\begin{scope}[auto,%
  every node/.style={draw, minimum size=1.2cm}];
\def \n {7}
\def \radius {4.2cm}
\def \margin {14} 
\draw[-, >=latex] ({360/\n * (1 - 3)+\margin}:\radius);
arc ({360/\n * (1 - 3)+\margin}:{360/\n * (1-2)-\margin}:\radius);
\draw[-, >=latex] ({360/\n * (2 - 3)+\margin}:\radius);
arc ({360/\n * (2 - 3)+\margin}:{360/\n * (2-2)-\margin}:\radius);
\draw[-, >=latex] ({360/\n * (3 - 3)+\margin}:\radius);
arc ({360/\n * (3 - 3)+\margin}:{360/\n * (3-2)-\margin}:\radius);
\draw[-, >=latex] ({360/\n * (4 - 3)+\margin}:\radius);
arc ({360/\n * (5 - 3)+\margin}:{360/\n * (5-2)-\margin}:\radius);
\draw[-, >=latex] ({360/\n * (5 - 3)+\margin}:\radius);
arc ({360/\n * (5 - 3)+\margin}:{360/\n * (5-2)-\margin}:\radius);
\node[draw, circle, black] (n2) at ({360/\n * (0 - 2)}:\radius) {$G_2$};
\node[draw, circle, blue] at ({360/\n * (1 - 2)}:\radius) {$4'$};
\node[draw= none,black] (nn) at ({360/\n * (2 - 2)}:\radius) {\Large $\vdots$};
\node[draw,rectangle,red] (f1) at (-8,1.8) {$2n_1$};
\node[draw,rectangle,red] (f2) at (-8,-1.8) {$2n_2$};
\node[draw, circle,blue] at ({360/\n * (3 - 2)}:\radius) {$4'$};
\node[draw, circle,black]  at ({360/\n * (4 - 2)}:\radius) {$G_2$};
\node[draw, circle,blue]  (n1) at ({360/\n * (5 - 2)}:\radius) {$4'$};
\foreach \s in {1,...,5}
{	
	\draw[-, >=latex] ({360/\n * (\s - 3)+\margin}:\radius) 
	arc ({360/\n * (\s - 3)+\margin}:{360/\n * (\s-2)-\margin}:\radius);
}
\draw[-, >=latex] ({360/\n * (0 - 3)+\margin}:\radius) 
arc ({360/\n * (0 - 3)+\margin}:{360/\n * (0-2)-\margin}:\radius);
\node[draw, circle,blue] (n2) at ({360/\n * (-1 - 2)}:\radius) {$4'$};
\draw[-, >=latex,red,thick,snake it] ({360/\n * (-1 - 3)+\margin}:\radius) 
arc ({360/\n * (-1 - 3)+\margin}:{360/\n * (-1-2)-\margin}:\radius);
\node[draw=none] at (9.3,-1) {\large $\substack{\text{$m$ $G_2$ nodes + $(m-1)$ usual blue} \\~ \\ \text{circular nodes + 2 blue nodes} \\~ \\ \text{connected by $T(USp'(4))$}}$};
\node[draw=none] at (-6,0) {{\red $T(USp'(4))$}};
\draw[-,solid] (f1) to (n1);
\draw[-,solid] (f2) to (n2);
\end{scope}
\end{tikzpicture}}
\ee
The mirror theory of this quiver can be obtained by applying the replacement rule \eref{replace1}, with one of the external legs on each side being a $T$-link.   As an example, we have the following mirror pair:
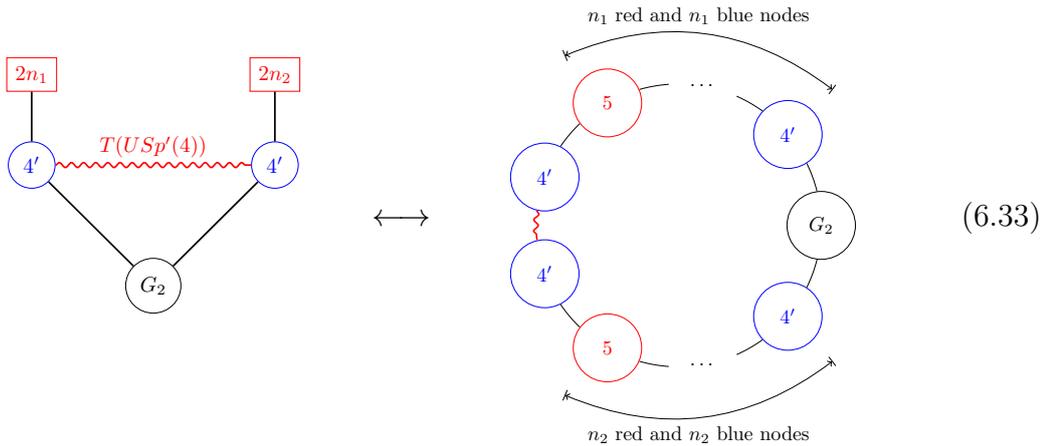
\be \label{quiv44G2}
\scalebox{0.8}{
\begin{tikzpicture}[baseline]
\tikzstyle{every node}=[font=\footnotesize]
\node[draw, circle, blue] (node1) at (-2,2-1) {$4'$};
\node[draw, circle, blue] (node2) at (2,2-1) {$4'$};
\node[draw, circle, black] (node3) at (0,0-1) {$G_2$};
\node[draw, rectangle, red] (sqnode1) at (-2,3.5-1) {$2n_1$};
\node[draw, rectangle, red] (sqnode2) at (2,3.5-1) {$2n_2$};
\draw[draw=black,solid,thick,-]  (node2)--(node3) ; 
\draw[draw=black,solid,thick,-]  (node1)--(node3) ; 
\draw[draw=red,snake it,thick,-]  (node1)--(node2) node at (0,2.3-1) {{\red $T(USp'(4))$}} ; 
\draw[draw=black,solid,thick,-]  (node1)--(sqnode1) ; 
\draw[draw=black,solid,thick,-]  (node2)--(sqnode2) ; 
\end{tikzpicture}}
\qquad \longleftrightarrow \qquad
\scalebox{0.75}{
\begin{tikzpicture}[baseline, font=\footnotesize]
\begin{scope}[auto,%
  every node/.style={draw, minimum size=1.2cm}];
\def \n {9}
\def \radius {2.5cm}
\def \margin {14} 
\draw[-, >=latex] ({360/\n * (1 - 3)+\margin}:\radius);
arc ({360/\n * (1 - 3)+\margin}:{360/\n * (1-2)-\margin}:\radius);
\node[draw, circle, black] at ({360/\n * (2 - 2)}:\radius) {$G_2$};
\draw[-, >=latex] ({360/\n * (2 - 3)+\margin}:\radius);
arc ({360/\n * (2 - 3)+\margin}:{360/\n * (2-2)-\margin}:\radius);
\draw[-, >=latex] ({360/\n * (3 - 3)+\margin}:\radius);
arc ({360/\n * (3 - 3)+\margin}:{360/\n * (3-2)-\margin}:\radius);
\draw[-, >=latex] ({360/\n * (4 - 3)+\margin}:\radius);
arc ({360/\n * (5 - 3)+\margin}:{360/\n * (5-2)-\margin}:\radius);
\draw[-, >=latex] ({360/\n * (5 - 3)+\margin}:\radius);
arc ({360/\n * (5 - 3)+\margin}:{360/\n * (5-2)-\margin}:\radius);
\node[draw, circle, blue] (n1)at ({360/\n * (1 - 2)}:\radius) {$4'$};
\node[draw, circle, blue] (n2) at ({360/\n * (3 - 2)}:\radius) {$4'$};
\node[draw=none] at ({360/\n * (4 - 2)}:\radius) {$\cdots$};
\node[draw, circle,red]  at ({360/\n * (5 - 2)}:\radius) {$5$};
\node[draw, circle,blue] (b2) at ({360/\n * (6 - 2)}:\radius) {$4'$};
\node[draw, circle, red] (b1) at ({360/\n * (8 - 2)}:\radius) {$5$};
\node[draw=none] at ({360/\n * (9 - 2)}:\radius) {$\cdots$};
\foreach \s in {1,2,3,4,5,6,8,9}
{	
	\draw[-, >=latex] ({360/\n * (\s - 3)+\margin}:\radius) 
	arc ({360/\n * (\s - 3)+\margin}:{360/\n * (\s-2)-\margin}:\radius);
}
\node[draw, circle,blue] at ({360/\n * (7 - 2)}:\radius) {$4'$};
\draw[-, >=latex,red,thick,snake it] ({360/\n * (7 - 3)+\margin}:\radius) 
arc ({360/\n * (7 - 3)+\margin}:{360/\n * (7-2)-\margin}:\radius);
\end{scope}
\draw[draw=black,|<->|]  (-2,3) to[bend left] (2.7,2.4) node at (0.35,3.7) {$n_1$ red and $n_1$ blue nodes} ;
\draw[draw=black,|<->|]  (-2,-3) to[bend right] (2.7,-2.4) node at (0.35,-3.7) {$n_2$ red and $n_2$ blue nodes} ;
\end{tikzpicture}}
\ee

Yet another generalisation one can possibly consider is to add flavour to any of the $4'$-node that is not connected by the $T$-link in \eref{twoflvUSp4p}.  The mirror theory can simply be obtained, again, by applying the replacement rule given by \eref{replace1}.

As emphasised before, as a by-product, one may obtain a mirror pair between the usual field theories, without an $S$-fold, by simply merging the two $USp'(4)$ nodes that are connected by $T(USp'(4))$.  

\section{Conclusions and Perspectives}
We propose new classes of three dimensional $S$-fold CFTs and study their moduli spaces in detail.  These generalise and extend the previous results of \cite{Assel:2018vtq, Garozzo:2018kra}, whose central role was played by quivers that contain a $T(U(N))$ theory connecting two unitary gauge groups.  In this paper, we explore the possibility of replacing $T(U(N))$ by a more general $T(G)$ theory, where $G$ is self-dual under the $S$-duality.  In particular, we investigate the cases where $G$ is taken be $SO(2N)$, $USp'(2N)$ and $G_2$.  

For $G=SO(2N)$ and $USp'(2N)$, we propose that the quiver can be realised from an insertion of an $S$-fold into a brane configuration that involves D3 branes on top of orientifold threeplanes, possibly with NS5 and D5 branes \cite{Feng:2000eq}.  In which case, the $S$-fold needs to be inserted in an interval of the D3 brane where the orientifold is of the $\Ot^-$ type or the $\tilde{\Ot}^+$ type for $G=SO(2N)$ or $USp'(2N)$, respectively.  The resulting quiver contains alternating orthogonal and symplectic gauge groups, along with a $T(G)$ theory connecting two gauge groups $G$ in the quiver.   Moreover, we also obtain the mirror theory by performing $S$-duality on the brane system.  Under the action of latter, the $\O3^-$ and $\tilde{\Ot}^+$ planes, as well as the $S$-fold remain invariant.  Hence the resulting mirror theory can be obtained from the $S$-dual configuration discussed in \cite{Feng:2000eq}, with an $S$-fold inserted in the position corresponding to the original set-up.

As observed in our previous paper \cite{Garozzo:2018kra}, the $U(N)$ gauge nodes, with zero Chern--Simons levels, that are connected by $T(U(N))$ link do not contribute to Coulomb branch of the quiver.  We dub this phenomenon the ``freezing rule''.  This has been successfully tested by study the moduli spaces of various quiver theories and their mirrors.  The results turned out be nicely consistent with mirror symmetry, namely the Higgs and Coulomb branches of the original theory get exchanged with the Coulomb and Higgs branches of the mirror theory.  In this paper, we perform a similar consistency checks.  We find that the freezing rule still holds for the quiver with $T(SO(2N))$ and $T(USp'(2N))$ and the results are consistent with mirror symmetry.  Such consistency supports the statement that the $S$-fold can be present in the background of $\Ot^-$ and $\tilde{\Ot}^+$ planes.

Following the same logic, we also investigate the presence of the $S$-fold in the background of orientifold fiveplanes.  In particular, we examine the insertion of the $S$-fold into the brane configurations involving orientifold fiveplanes, studied in \cite{Hanany:1999sj}.  The corresponding quiver contains several interesting features such as the presence of the antisymmetric hypermultiplet, along with the $T(U(N))$ link connecting two unitary gauge groups.  The mirror configuration consists of an $\ON$ plane that gives rise to a bifurcation in the mirror quiver \cite{Kapustin:1998fa}, with the $S$-fold inserted in the position corresponding to the original set-up.   An important result that we discover for this class of theories is that, in order for the freezing rule to hold and for the moduli spaces of the mirror pair to be consistent with mirror symmetry, the $S$-fold must not be inserted ``too close'' to the orientifold plane; there must be a sufficient number of NS5 branes that separate the $S$-fold from the orientifold plane.  This suggests that the NS5 branes provide a certain ``screening effect'' or ``shielding effect'' in the combination of the orientifold plane and the $S$-fold.  We hope to understand this phenomenon better in the future.

Finally, we propose a novel class of circular quivers that contains the exceptional $G_2$ gauge groups.  In particular, the quiver contains alternating $G_2$ and $USp'(4)$ gauge groups, possibly with flavours under $USp'(4)$.   Although the Type IIB brane configuration for this class of theories is not known and the $S$-fold supergravity solution for the exceptional group is not available, we propose that it is possible to ``insert an S-fold'' into the $G_2$ and/or $USp'(4)$ gauge groups in the aforementioned quivers.  This results in the presence of the $T(G_2)$ link connecting two $G_2$ gauge groups, and/or the $T(USp'(4))$ link connecting two $USp'(4)$ gauge groups.  Furthermore, we propose the mirror theory which is also a circular quiver, consisting of the $G_2$, $USp'(4)$ and possibly $SO(5)$ gauge groups if the original theory has the flavours under $USp'(4)$.  To the best of our knowledge, such mirror pairs are new and have never been studied in the literature before.  We check, using the Hilbert series, that moduli spaces of such pairs satisfy the freezing rule and are consistent with mirror symmetry.  This, again, serves as strong evidence for the existence of an $S$-fold of the $G_2$ type.

%
%
The results in this paper leads to a number of open questions. First of all, we restricts ourself to models with equal-rank gauge nodes; this avoids problem arising from non-complete Higgsing of the gauge symmetries. It would be interesting to generalise all the result to the unequal-rank cases. This amounts to consider $S$-fold configurations with fractional branes.

Secondly, so far we have taken the Chern--Simons levels of all gauge groups connected by the $T$-link to be zero.  It would be interesting to study the moduli spaces as well as the duality between theories with non-zero Chern--Simons levels.

Finally, it would be interesting to generalise our result on the $G_2$ group to other exceptional groups, including $F_4$ and $E_{6,7,8}$, which are also invariant under the $S$-duality.  It is natural to expect that the $S$-fold associated with such groups should exist and, in that case, it should be possible to find quivers as well as their mirror theories that describe such $S$-fold CFTs.  Moreover, it would be nice to find a string or an F-theoretic construction for such theories.  This would certainly lead to a deeper understanding of such CFTs.

\acknowledgments
We would like to thank Antonio Amariti, Benjamin Assel, Constantin Bachas, Craig Lawrie, Alessandro Tomasiello and Alberto Zaffaroni for useful discussions.

\appendix
\section{Models with an $\Of^+$ plane} \label{app:O5plus}
In this appendix, we analyse models with $\text{O5}^+$ plane.  In particular, we focus on a theory with one symmetric hypermultiplet and its mirror theory.  One of the important features is that the mirror theory does not admit a conventional Lagrangian description.  Nevertheless, we can represent it by a quiver diagram with a ``multiple-lace'', in the same sense of the Dynkin diagram of the $C_N$ algebra \cite{Hanany:2001iy, Cremonesi:2014xha}.   As pointed out in \cite{Cremonesi:2014xha}, it is possible to compute the Coulomb branch Hilbert series of such a mirror theory with the multiple-lace, and equate the result with the Higgs branch Hilbert series of the original theory with one symmetric hypermultiplet.  

The point of this appendix is to demonstrate that one may insert an $S$-fold into the brane system of the original theory and the corresponding mirror configuration, and still obtain a consistent mirror theory.  Again, one can compute the Coulomb branch Hilbert series of the latter and match it with the Higgs branch Hilbert series of the former.

\subsubsection*{\it Models without an $S$-fold}
We start by looking at the following theory:
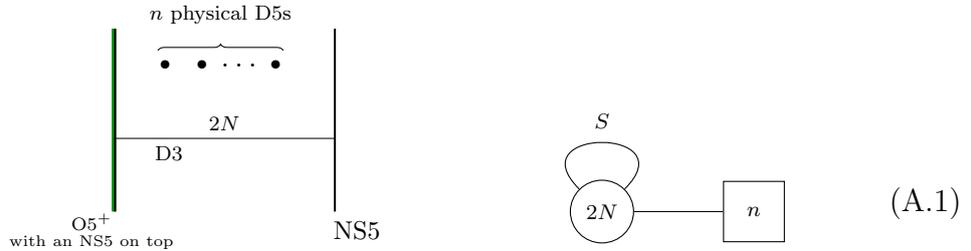
\be \label{O5brane}
\scalebox{0.97}{
\begin{tikzpicture} [baseline, scale=1, transform shape]
\draw [ultra thick,black!40!green] (0,0)--(0,2.5) node[black,midway, xshift =-0.3cm, yshift=-1.5cm] {\footnotesize $\overset{\text{O5}^+}{{\tiny \text{with an NS5 on top}}}$}; 
\draw [thick,black] (0.02,0)--(0.02,2.5);
\node[midway, color=black, xshift=0.7cm, yshift=2cm] {\scriptsize $\bullet$};
\node[midway, color=black, xshift=1.2cm, yshift=2cm] {\scriptsize $\bullet$};
\node[midway, color=black, xshift=1.7cm, yshift=2cm] {$~\ldots~$};
\node[midway, color=black, xshift=2.2cm, yshift=2cm] {\scriptsize $\bullet$};
\draw [decorate, decoration={brace}](0.6,2.2)--(2.3,2.2) node[black,midway,yshift=0.5cm] {\scriptsize $n~\text{physical D5s}$};
\draw [thick,black] (3,0)--(3,2.5) node[black,midway, xshift =0.3cm, yshift=-1.5cm] {\footnotesize NS5};
\draw (0,1)--(3,1) node[black,midway, yshift=0.2cm] {\scriptsize $2N$} node[black,near start, yshift=-0.2cm] {\scriptsize D3};
\end{tikzpicture}}
\qquad \qquad 
\begin{tikzpicture}[baseline,font=\scriptsize]
\begin{scope}[auto,%
  every node/.style={draw, minimum size=0.8cm}, node distance=0.6cm];
\node[draw, circle, black] (node1) at (0,0) {$2N$};
\draw[black] (node1) edge [out=45,in=135,loop,looseness=5]  (node1) node[draw=none] at (0,1.2) {$S$} ;
\node[draw, rectangle,black] (sqnode) at (2,0) {$n$};
\end{scope}
\draw (node1)--(sqnode);
\end{tikzpicture}
\ee
The presence of the $\text{O5}^+$ plane gives rise to a rank-two symmetric hypermultiplet.
The Higgs and Coulomb branch dimensions for theory are as follows
\be
\begin{split}
\text{dim}_\mathbb{H}\, \cC_{(\ref{O5brane})}&=2N~, \\
\text{dim}_\mathbb{H}\, \cH_{(\ref{O5brane})}&=\frac{1}{2} \, 2N (2N+1)+2N n -4N^2=2Nn-2N^2+N~.
\end{split}
\ee
Applying $S$-duality to the brane system (\ref{O5brane}) we get 

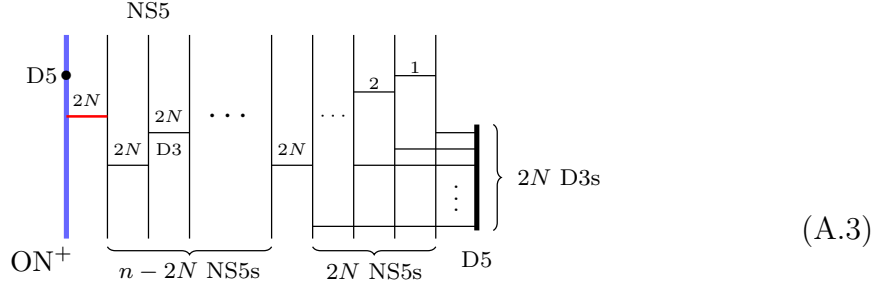
\be
\scalebox{1.2}{
\begin{tikzpicture} [baseline=0, scale=0.9, transform shape]
\draw [ultra thick, blue!60] (0,0)--(0,2.5) node[black,midway, xshift =-0.3cm, yshift=-1.5cm] {\footnotesize $\text{ON}^+$}; 
\node[midway, color=black, xshift=0cm, yshift=2cm] {\scriptsize $\bullet$} node[midway, color=black, xshift=-0.3cm, yshift=2cm] {\scriptsize D5};
\draw (0.5,0)--(0.5,2.5); \draw (1,0)--(1,2.5); \draw (1.5,0)--(1.5,2.5); \draw (2.5,0)--(2.5,2.5); \draw (3,0)--(3,2.5); 
\draw (3.5,0)--(3.5,2.5); \draw (4,0)--(4,2.5); \draw (4.5,0)--(4.5,2.5); 
%
%
\draw[thick,red] (0,1.5)--(0.5,1.5) node[black,midway, yshift=0.2cm] {\tiny $2N$};
\draw (0.5,0.9)--(1,0.9) node[black,midway, yshift=0.2cm] {\tiny $2N$};
\draw (1,1.3)--(1.5,1.3) node[black,midway, yshift=0.2cm] {\tiny $2N$} node[black,midway, yshift=-0.2cm] {\tiny D3};
\draw (2.5,0.9)--(3,0.9) node[black,midway, yshift=0.2cm] {\tiny $2N$};
\draw (4,2)--(4.5,2) node[black,midway, yshift=0.1cm] {\tiny $1$};
\draw (3.5,1.8)--(4,1.8) node[black,midway, yshift=0.1cm] {\tiny $2$};
\draw node at (1,2.8) {\scriptsize NS5};
\draw (5,1.3)--(4.5,1.3) ;
\draw (5,1.1)--(4,1.1) ;
\draw (5,0.9)--(3.5,0.9);
\draw node at (4.75,0.6) {\scriptsize $\vdots$ };
\draw (5,0.15)--(3,0.15);
\draw node at (2,1.5) {$\cdots$ };
\draw node at (3.25,1.5) {\tiny $\mathbf{\cdots}$ };
\draw [ultra thick, black] (5,0.1)--(5,1.4) node[black,midway, yshift=-1cm] {\scriptsize D5}; 
\draw [decorate, decoration={brace, mirror}](3,-0.1)--(4.5,-0.1) node[black,midway,yshift=-0.3cm] {\scriptsize $2N~\text{NS5s}$};
\draw [decorate, decoration={brace, mirror}](0.5,-0.1)--(2.5,-0.1) node[black,midway,yshift=-0.3cm] {\scriptsize $n-2N~\text{NS5s}$};
\draw [decorate, decoration={brace, mirror}](5.2,0.1)--(5.2,1.4) node[black,midway,xshift=0.8cm] {\scriptsize $2N~\text{D3s}$};
\end{tikzpicture}}
\ee
and, after moving the rightmost D5 brane into the brane interval, we arrive at
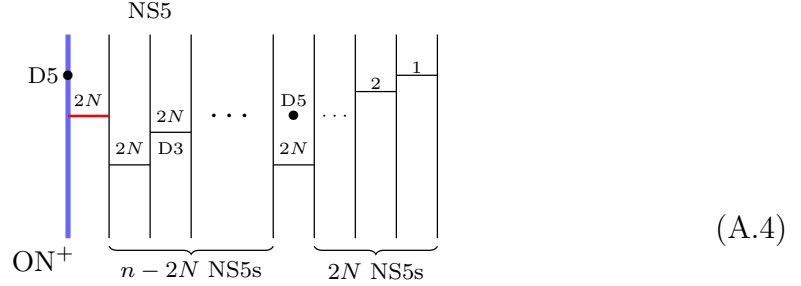
\be 
\scalebox{1.2}{
\begin{tikzpicture} [baseline=0, scale=0.9, transform shape]
\draw [ultra thick, blue!60] (0,0)--(0,2.5) node[black,midway, xshift =-0.3cm, yshift=-1.5cm] {\footnotesize $\ON^+$}; 
\node[midway, color=black, xshift=0cm, yshift=2cm] {\scriptsize $\bullet$} node[midway, color=black, xshift=-0.3cm, yshift=2cm] {\scriptsize D5};
\draw (0.5,0)--(0.5,2.5); \draw (1,0)--(1,2.5); \draw (1.5,0)--(1.5,2.5); \draw (2.5,0)--(2.5,2.5); \draw (3,0)--(3,2.5); 
\draw (3.5,0)--(3.5,2.5); \draw (4,0)--(4,2.5); \draw (4.5,0)--(4.5,2.5); 
%
%
\draw[thick,red] (0,1.5)--(0.5,1.5) node[black,midway, yshift=0.2cm] {\tiny $2N$};
\draw (0.5,0.9)--(1,0.9) node[black,midway, yshift=0.2cm] {\tiny $2N$};
\draw (1,1.3)--(1.5,1.3) node[black,midway, yshift=0.2cm] {\tiny $2N$} node[black,midway, yshift=-0.2cm] {\tiny D3};
\draw (2.5,0.9)--(3,0.9) node[black,midway, yshift=0.2cm] {\tiny $2N$};
\draw (4,2)--(4.5,2) node[black,midway, yshift=0.1cm] {\tiny $1$};
\draw (3.5,1.8)--(4,1.8) node[black,midway, yshift=0.1cm] {\tiny $2$};
\draw node at (1,2.8) {\scriptsize NS5};
\draw node at (2,1.5) {$\cdots$ };
\draw node at (3.25,1.5) {\tiny $\mathbf{\cdots}$ }; 
\node at (2.75,1.5) {\scriptsize $\bullet$} node at (2.75,1.7) {\tiny D5};
\draw [decorate, decoration={brace, mirror}](3,-0.1)--(4.5,-0.1) node[black,midway,yshift=-0.3cm] {\scriptsize $2N~\text{NS5s}$};
\draw [decorate, decoration={brace, mirror}](0.5,-0.1)--(2.5,-0.1) node[black,midway,yshift=-0.3cm] {\scriptsize $n-2N~\text{NS5s}$};
\end{tikzpicture}}
\ee
The corresponding quiver theory associated to this system is \cite{Hanany:2001iy, Cremonesi:2014xha}
\be
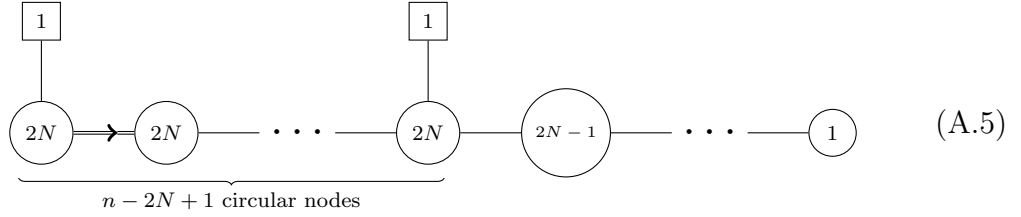
\label{O5braneMirr}
\begin{tikzpicture}[font=\scriptsize,baseline]
\begin{scope}[auto,%
  every node/.style={draw, minimum size=0.5cm}, node distance=0.8cm];
\node[circle] (U2N1) at (0,0) {$2N$};
\node[rectangle, above =of U2N1] (f1)  {$1$};
\node[circle, right =of U2N1] (U2N2) {$2N$};
\node[draw=none, right =of U2N2] (dots) {\Large${\cdots}$};
\node[circle, right =of dots] (U2N3) {$2N$};
\node[circle, right =of U2N3] (UT1) {\tiny $2N-1$};
\node[draw=none, right =of UT1] (dots1) {\Large${\cdots}$};
\node[circle, right =of dots1] (UT) {$1$};
\node[rectangle, above =of U2N3] (f2)  {$1$};
\end{scope}
\draw (f1)--(U2N1);
\draw[->,double] (U2N1)--(1,0); 
\draw[double] (1,0)--(U2N2);
\draw (U2N2)--(dots)--(U2N3)--(UT1)--(dots1)--(UT);
\draw (U2N3)--(f2);
\draw [decorate, decoration={brace, mirror}](-0.3,-0.6)--(5.3,-0.6) node[black,midway,yshift=-0.3cm] {\scriptsize $n-2N+1~\text{circular nodes}$};
\end{tikzpicture}
\ee
The presence of the $\ON^+$ plane gives rise to the double lace at the left end.  This part of the quiver does not have a known Lagrangian description.  However, as explained in \cite{Cremonesi:2014xha}, the part of the quiver that corresponds to a double lace, whose arrow goes from the gauge group $U(N_1)$ to $U(N_2)$, contributes to the dimension of the monopole operator as
\be
	\Delta_{(U(N_1)) \Rightarrow (U(N_2))}(\bm{m}^{(1)},\bm{m}^{(2)})=\frac{1}{2}\sum_{i=1}^{N_1}\sum_{j=1}^{N_2} |2 m^{(1)}_i-m^{(2)}_j|,
\ee
where $\bm{m}^{(1)}$ and $\bm{m}^{(2)}$ are the magnetic fluxes associated with the gauge groups $U(N_1)$ and $U(N_2)$ respectively.
For \eref{O5braneMirr}, the Coulomb branch has dimension
\be
	\text{dim}_\mathbb{H}\, \cC_{(\ref{O5braneMirr})}=2N(n-2N+1)+\sum_{i=1}^{2N-1}i=2Nn-2N^2+N~,
\ee
equal to the Higgs branch dimension of \eref{O5brane}, as expected from mirror symmetry.  Note that we have assumed that the two gauge nodes connected by the double lace contribute as the others.  Since we do not have information about matter associated with the double lace, we cannot compute the Higgs branch dimension of \eref{O5braneMirr} using the quiver description.

Let us consider a specific example by choosing $N=1$ and $n=4$. The unrefined Higgs branch Hilbert series of (\ref{O5brane}) is 
\be
\begin{split}
	H[\cH_{(\ref{O5brane})}] &= \oint_{|z|=1} \frac{d z}{2 \pi i z} (1-z^2) \oint_{|q|=1} \frac{d q}{2 \pi i q}  \PE\Big[ 4 (z+z^{-1})(q+q^{-1}) \\
	& \qquad + (z^2+1+z^{-2})(q^2+q^{-2}) t -(z^2+1+z^{-2}+1)t^2 \Big] \\
	&=\text{PE}\,[16 t^2 + 20 t^3 - 12 t^5 - 32 t^6+\dots]~.
\end{split}
\ee
For the mirror theory (\ref{O5braneMirr}) the unrefined Coulomb branch Hilbert series can be computed in the same way as described in \cite{Cremonesi:2014xha}.  The result is
\be \label{exCoulHS}
\begin{split}
	H[\cC_{(\ref{O5braneMirr})}] &= \sum_{m^{(1)}_1 \geq m^{(1)}_2 > -\infty}\,\, \sum_{m^{(2)}_1 \geq m^{(2)}_2 > -\infty} \,\, \sum_{m^{(3)}_1 \geq m^{(3)}_2 > -\infty} \,\,  \sum_{m \in \BZ} t^{2\Delta({\vec m}^{(1)}, {\vec m}^{(2)}, {\vec m}^{(3)}, m)} \\
	& \qquad \times P_{U(2)}(t, {\vec m}^{(1)}) P_{U(2)}(t, {\vec m}^{(2)}) P_{U(2)}(t, {\vec m}^{(3)}) P_{U(1)} (t, m)\\
	&=\text{PE}\,[16 t^2 + 20 t^3 - 12 t^5 - 32 t^6+\dots],
\end{split}
\ee
where ${\vec m}^{(i)} = (m^{(i)}_1,m^{(i)}_2)$ for $i=1,2,3$ and we define
\be \label{alldefU}
\begin{split}
\Delta({\vec m}^{(1)}, {\vec m}^{(2)}, {\vec m}^{(3)}, m) &= \Delta_{U(2) \Rightarrow U(2)} ({\vec m}^{(1)}, {\vec m}^{(2)}) + \Delta_{U(2)-U(2)}({\vec m}^{(2)}, {\vec m}^{(3)}) \\
& \qquad + \Delta_{U(2)-U(1)}({\vec m}^{(3)}, m) +  \Delta_{U(2)-U(1)}({\vec m}^{(1)}, 0) \\
& \qquad +  \Delta_{U(2)-U(1)}({\vec m}^{(3)}, 0) - \sum_{i=1}^3 \Delta^{\text{vec}}_{U(2)}({\vec m}^{(i)})\\
2\Delta_{U(N_1) \Rightarrow U(N_2)}( \vec m, \vec n) &=\sum_{i=1}^{N_1}\sum_{j=1}^{N_2} |2 m_i-n_j| \\
2\Delta_{U(N_1)- U(N_2)}( \vec m, \vec n) &=\sum_{i=1}^{N_1}\sum_{j=1}^{N_2} |m_i-n_j|  \\
\Delta^{\text{vec}}_{U(2)}(\vec m) &= |m_1-m_2| \\
P_{U(2)} (t; m_1, m_2) &= \begin{cases} 
(1-t^2)^{-2}~, &\quad m_1 \neq m_2 \\
(1-t^2)^{-1}(1-t^4)^{-1}~, &\quad m_1 =m_2
\end{cases} \\
P_{U(1)} (t; m) &=
(1-t^2)^{-1}~.
\end{split}
\ee
The two Hilbert series are equal as expected. 

\subsubsection*{\it The case with an $S$-fold}
One can insist with the insertion of an $S$-fold also for theories involving an $\text{O5}^+$ plane. The brane configuration and the quiver theory are as follows

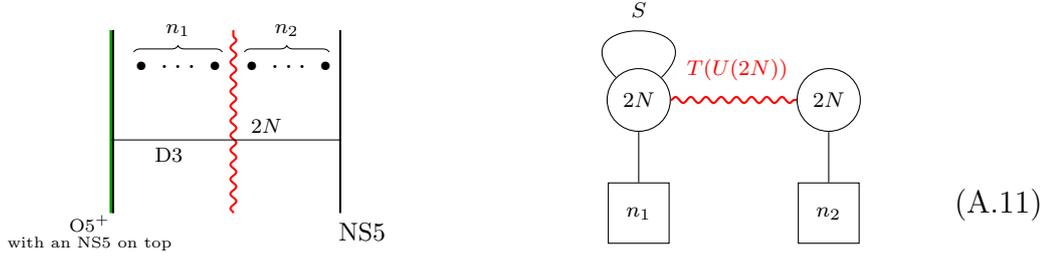
\be\label{SfoldU2Nsym}
\scalebox{0.97}{
\begin{tikzpicture}[baseline, scale=1, transform shape]
\draw [ultra thick,black!40!green] (0,0)--(0,2.5) node[black,midway, xshift =-0.3cm, yshift=-1.5cm] {\footnotesize $\overset{\Of^+}{{\tiny \text{with an NS5 on top}}}$}; 
\draw [thick,black] (0.02,0)--(0.02,2.5);
\node[midway, color=black, xshift=0.4cm, yshift=2cm] {\scriptsize $\bullet$};
\node[midway, color=black, xshift=0.9cm, yshift=2cm] {$~\ldots~$};
\node[midway, color=black, xshift=1.4cm, yshift=2cm] {\scriptsize $\bullet$};
\draw [decorate, decoration={brace}](0.3,2.2)--(1.5,2.2) node[black,midway,yshift=0.3cm] {\scriptsize $n_1$};
\node[midway, color=black, xshift=1.9cm, yshift=2cm] {\scriptsize $\bullet$};
\node[midway, color=black, xshift=2.4cm, yshift=2cm] {$~\ldots~$};
\node[midway, color=black, xshift=2.9cm, yshift=2cm] {\scriptsize $\bullet$};
\draw [decorate, decoration={brace}](1.8,2.2)--(2.95,2.2) node[black,midway,yshift=0.3cm] {\scriptsize $n_2$};
\draw [thick,red,snake it] (1.65,0)--(1.65,2.5);
\draw [thick,black] (3.1,0)--(3.1,2.5) node[black,midway, xshift =0.3cm, yshift=-1.5cm] {\footnotesize NS5};
\draw (0,1)--(3.1,1) node[black,xshift=-1cm, yshift=0.2cm] {\scriptsize $2N$} node[black,near start, yshift=-0.2cm] {\scriptsize D3};
\end{tikzpicture}}
\qquad \qquad \quad
\begin{tikzpicture}[baseline,font=\scriptsize]
\begin{scope}[auto,%
  every node/.style={draw, minimum size=0.8cm}, node distance=0.6cm];
\node[draw, circle, black] (node1) at (0,1.5) {$2N$};
\node[draw, circle, black] (node2) at (2.5,1.5) {$2N$};
\draw[black] (node1) edge [out=45,in=135,loop,looseness=5]  (node1) node[draw=none] at (0,2.7) {$S$} ;
\node[draw, rectangle,black] (sqnode1) at (0,0) {$n_1$};
\node[draw, rectangle,black] (sqnode2) at (2.5,0) {$n_2$};
\end{scope}
\draw (node1)--(sqnode1);
\draw (node2)--(sqnode2);
\draw[draw=red,solid, snake it, thick,-] (node1)--(node2);
\node[draw=none] at (1.3,1.9) {{\red$T(U(2N))$}};
\end{tikzpicture}
\ee
This theory has Coulomb and Higgs branches with the following dimensions
\be \label{HBdimONp}
\begin{split}
	\text{dim}_\mathbb{H}\, \cC_{(\ref{SfoldU2Nsym})} &=0,\\
	\text{dim}_\mathbb{H}\, \cH_{(\ref{SfoldU2Nsym})}&=\text{dim}_\mathbb{H}\, \cH_{(\ref{O5brane})}|_{n=n_1+n_2} + (4N^2-2N)-4N^2 \\
	&= 2N(n_1+n_2) -2N^2 -N~,
\end{split}
\ee 
where the first line follows from the fact that the two circular nodes are connected by the $T$-link and hence do not contribute to the Coulomb branch.
The brane system we get after applying $S$-duality is 
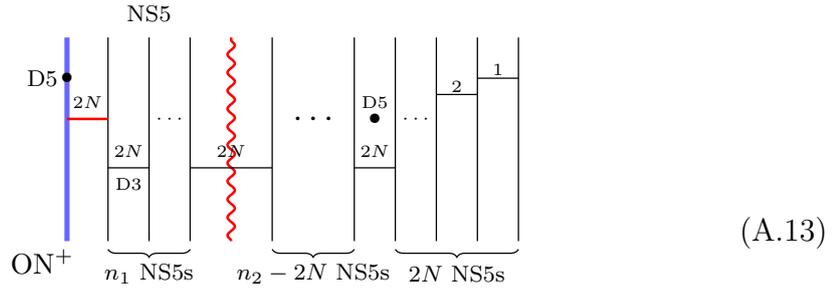
\be 
\scalebox{1.2}{
\begin{tikzpicture} [baseline=0, scale=0.9, transform shape]
\draw [ultra thick, blue!60] (0,0)--(0,2.5) node[black,midway, xshift =-0.3cm, yshift=-1.5cm] {\footnotesize $\ON^+$}; 
\node[midway, color=black, xshift=0cm, yshift=2cm] {\scriptsize $\bullet$} node[midway, color=black, xshift=-0.3cm, yshift=2cm] {\scriptsize D5};
\draw (0.5,0)--(0.5,2.5); \draw (1,0)--(1,2.5); \draw (1.5,0)--(1.5,2.5);
\draw [thick,red,snake it] (2,0)--(2,2.5);

 \draw (3.5,0)--(3.5,2.5); \draw (4,0)--(4,2.5); \draw (4.5,0)--(4.5,2.5); 
\draw (5,0)--(5,2.5); \draw (5.5,0)--(5.5,2.5); \draw (2.5,0)--(2.5,2.5);

\draw[thick,red] (0,1.5)--(0.5,1.5) node[black,midway, yshift=0.2cm] {\tiny $2N$};
\draw (0.5,0.9)--(1,0.9) node[black,midway, yshift=0.2cm] {\tiny $2N$} node[black,midway, yshift=-0.2cm] {\tiny D3}; ;
\draw (1.5,0.9)--(2.5,0.9) node[black,midway, yshift=0.2cm] {\tiny $2N$};
\draw (3.5,0.9)--(4,0.9) node[black,midway, yshift=0.2cm] {\tiny $2N$};
\draw (5,2)--(5.5,2) node[black,midway, yshift=0.1cm] {\tiny $1$};
\draw (4.5,1.8)--(5,1.8) node[black,midway, yshift=0.1cm] {\tiny $2$};
\draw node at (1,2.8) {\scriptsize NS5};
\draw node at (1.25,1.5)  {\tiny $\mathbf{\cdots}$ };
\draw node at (3,1.5) { ${\cdots}$ }; 
\draw node at (4.25,1.5) {\tiny $\mathbf{\cdots}$ }; 
\node at (3.75,1.5) {\scriptsize $\bullet$} node at (3.75,1.7) {\tiny D5};
\draw [decorate, decoration={brace, mirror}](4,-0.1)--(5.5,-0.1) node[black,midway,yshift=-0.3cm] {\scriptsize $2N~\text{NS5s}$};
\draw [decorate, decoration={brace, mirror}](2.5,-0.1)--(3.5,-0.1) node[black,midway,yshift=-0.3cm] {\scriptsize $n_2-2N~\text{NS5s}$};
\draw [decorate, decoration={brace, mirror}](0.5,-0.1)--(1.5,-0.1) node[black,midway,yshift=-0.3cm] {\scriptsize $n_1~\text{NS5s}$};
\end{tikzpicture}}
\ee
whose associated gauge theory reads

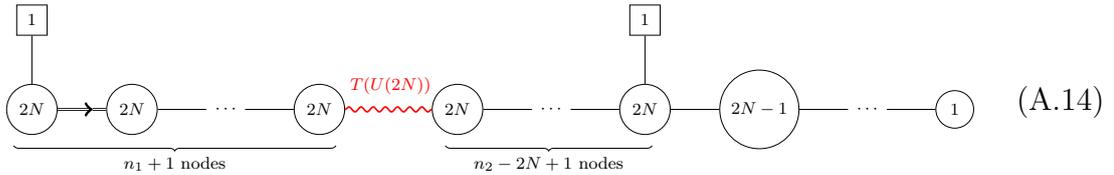
\be\label{O5braneSfoldMirr}
\scalebox{0.8}{
\begin{tikzpicture}[font=\scriptsize,baseline]
\begin{scope}[auto,%
  every node/.style={draw, minimum size=0.5cm}, node distance=0.8cm];
\node[circle] (U2N1) at (0,0) {$2N$};
\node[rectangle, above =of U2N1] (f1)  {$1$};
\node[circle, right =of U2N1] (U2N2) {$2N$};
\node[draw=none, right =of U2N2] (dots) {${\cdots}$};
\node[circle, right =of dots]  (U2NS1) {$2N$};
\node[circle] (U2NS2) at (7,0) {$2N$};
\node[draw=none, right =of U2NS2] (dots2) {${\cdots}$};
\node[circle, right =of dots2] (U2N3) {$2N$};
\node[circle, right =of U2N3] (UT1) {$2N-1$};
\node[draw=none, right =of UT1] (dots1) {${\cdots}$};
\node[circle, right =of dots1] (UT) {$1$};
\node[rectangle, above =of U2N3] (f2)  {$1$};
\draw[draw=red,solid, snake it, thick,-] (U2NS1)--(U2NS2);
\node[draw=none] at (5.9,0.4) {{\red$T(U(2N))$}};
\end{scope}
\draw (f1)--(U2N1);
\draw[->,double] (U2N1)--(1,0); 
\draw[double] (1,0)--(U2N2);
\draw (U2N3)--(UT1)--(dots1)--(UT);
\draw (U2N2)--(dots)--(U2NS1);
\draw (U2NS2)--(dots2)--(U2N3);
\draw (U2N3)--(f2);
\draw [decorate, decoration={brace, mirror}](-0.3,-0.6)--(5,-0.6) node[black,midway,yshift=-0.3cm] {\scriptsize $n_1+1~\text{nodes}$};
\draw [decorate, decoration={brace, mirror}](6.8,-0.6)--(10.2,-0.6) node[black,midway,yshift=-0.3cm] {\scriptsize $n_2-2N+1~\text{nodes}$};
\end{tikzpicture}}
\ee
The Coulomb branch dimension of this theory reads
\be
	\text{dim}_\mathbb{H}\, \cC_{(\ref{O5braneSfoldMirr})}=2N(n_1+1+n_2-2N+1-2)+\sum_{i=1}^{2N-1}i=2N(n_1+n_2)-2N^2-N~,
\ee
which equal to \eref{HBdimONp}.

Let us consider the example of $N=1$, $n_1=2$ and $n_2=2$. The Higgs branch of \eref{SfoldU2Nsym} splits into a product of two hyperK\"ahler spaces as usual.   The right part gives the nilpotent cone of $su(2)$ (which is isomorphic to $\BC^2/\BZ_2$), as pointed out in \eref{orbsuNsunspecial}; the corresponding unrefined Hilbert series is $\PE[3 t^2 -t^4]$.  The left part contributes to the Hilbert series as
\be
\begin{split}
&\oint_{|z|=1} \frac{d z}{2 \pi i z} (1-z^2) \oint_{|q|=1} \frac{d q}{2 \pi i q}  \PE\Big[ 2 (z+z^{-1})(q+q^{-1}) \\
& \qquad + (z^2+1+z^{-2})(q^2+q^{-2}) t + (z^2+1+z^{-2})t^2 - t^4 \\
& \qquad -(z^2+1+z^{-2}+1)t^2 \Big] = \text{PE}\,[4 t^2 + 6 t^3 + 4 t^4+\dots]~.
\end{split}	
\ee
Hence the Higgs branch Hilbert series of (\ref{SfoldU2Nsym}) is
\be
	H[\cH_{(\ref{SfoldU2Nsym})}]=\text{PE}\,[4 t^2 + 6 t^3 + 4 t^4+\dots] \, \text{PE}\,[3 t^2 -t^4] .
\ee

The Coulomb branch Hilbert series of \eref{O5braneSfoldMirr}, with $N=1$, $n_1=2$ and $n_2=2$, can be obtained by taking the circular nodes connected by the $T$-link to be separated flavour nodes.  Hence, the quiver splits into two parts.  The right part contributes as the $U(1)$ gauge theory with $2$ flavours, whose Coulomb branch is $\BC^2/\BZ_2$.  The Coulomb branch Hilbert series of the left part can be computed in a similar way as \eref{exCoulHS}.  The result is therefore
\be
H[\cC_{(\ref{O5braneSfoldMirr})}]= \text{PE}\,[4 t^2 + 6 t^3 + 4 t^4+\dots] \, \text{PE}\,[3 t^2 -t^4] .
\ee
This is equal to the Higgs branch Hilbert series of (\ref{SfoldU2Nsym}) and is, therefore, consistent with mirror symmetry.

\bibliographystyle{ytphys}
\bibliography{ref}

\providecommand{\href}[2]{#2}\begingroup\raggedright\begin{thebibliography}{10}

\bibitem{Intriligator:1996ex}
K.~A. Intriligator and N.~Seiberg, ``{Mirror symmetry in three-dimensional
  gauge theories},'' \href{http://dx.doi.org/10.1016/0370-2693(96)01088-X}{{\em
  Phys. Lett.} {\bfseries B387} (1996) 513--519},
\href{http://arxiv.org/abs/hep-th/9607207}{{\ttfamily arXiv:hep-th/9607207
  [hep-th]}}.

\bibitem{Kapustin:1999ha}
A.~Kapustin and M.~J. Strassler, ``{On mirror symmetry in three-dimensional
  Abelian gauge theories},''
  \href{http://dx.doi.org/10.1088/1126-6708/1999/04/021}{{\em JHEP} {\bfseries
  04} (1999) 021},
\href{http://arxiv.org/abs/hep-th/9902033}{{\ttfamily arXiv:hep-th/9902033
  [hep-th]}}.

\bibitem{Witten:2003ya}
E.~Witten, ``{SL(2,Z) action on three-dimensional conformal field theories with
  Abelian symmetry},''
\href{http://arxiv.org/abs/hep-th/0307041}{{\ttfamily arXiv:hep-th/0307041
  [hep-th]}}.

\bibitem{deBoer:1996mp}
J.~de~Boer, K.~Hori, H.~Ooguri, and Y.~Oz, ``{Mirror symmetry in
  three-dimensional gauge theories, quivers and D-branes},''
  \href{http://dx.doi.org/10.1016/S0550-3213(97)00125-9}{{\em Nucl. Phys.}
  {\bfseries B493} (1997) 101--147},
\href{http://arxiv.org/abs/hep-th/9611063}{{\ttfamily arXiv:hep-th/9611063
  [hep-th]}}.

\bibitem{Porrati:1996xi}
M.~Porrati and A.~Zaffaroni, ``{M theory origin of mirror symmetry in
  three-dimensional gauge theories},''
  \href{http://dx.doi.org/10.1016/S0550-3213(97)00061-8}{{\em Nucl.Phys.}
  {\bfseries B490} (1997) 107--120},
\href{http://arxiv.org/abs/hep-th/9611201}{{\ttfamily arXiv:hep-th/9611201
  [hep-th]}}.

\bibitem{Hanany:1996ie}
A.~Hanany and E.~Witten, ``{Type IIB superstrings, BPS monopoles, and
  three-dimensional gauge dynamics},''
  \href{http://dx.doi.org/10.1016/S0550-3213(97)00157-0}{{\em Nucl.Phys.}
  {\bfseries B492} (1997) 152--190},
\href{http://arxiv.org/abs/hep-th/9611230}{{\ttfamily arXiv:hep-th/9611230
  [hep-th]}}.

\bibitem{Kapustin:1998fa}
A.~Kapustin, ``{D(n) quivers from branes},''
  \href{http://dx.doi.org/10.1088/1126-6708/1998/12/015}{{\em JHEP} {\bfseries
  12} (1998) 015},
\href{http://arxiv.org/abs/hep-th/9806238}{{\ttfamily arXiv:hep-th/9806238
  [hep-th]}}.

\bibitem{Hanany:1999sj}
A.~Hanany and A.~Zaffaroni, ``{Issues on orientifolds: On the brane
  construction of gauge theories with SO(2n) global symmetry},''
  \href{http://dx.doi.org/10.1088/1126-6708/1999/07/009}{{\em JHEP} {\bfseries
  07} (1999) 009},
\href{http://arxiv.org/abs/hep-th/9903242}{{\ttfamily arXiv:hep-th/9903242
  [hep-th]}}.

\bibitem{Feng:2000eq}
B.~Feng and A.~Hanany, ``{Mirror symmetry by O3 planes},''
  \href{http://dx.doi.org/10.1088/1126-6708/2000/11/033}{{\em JHEP} {\bfseries
  11} (2000) 033},
\href{http://arxiv.org/abs/hep-th/0004092}{{\ttfamily arXiv:hep-th/0004092
  [hep-th]}}.

\bibitem{Gaiotto:2008ak}
D.~Gaiotto and E.~Witten, ``{S-Duality of Boundary Conditions In N=4 Super
  Yang-Mills Theory},''
  \href{http://dx.doi.org/10.4310/ATMP.2009.v13.n3.a5}{{\em Adv. Theor. Math.
  Phys.} {\bfseries 13} no.~3, (2009) 721--896},
\href{http://arxiv.org/abs/0807.3720}{{\ttfamily arXiv:0807.3720 [hep-th]}}.

\bibitem{Bergman:1999na}
O.~Bergman, A.~Hanany, A.~Karch, and B.~Kol, ``{Branes and supersymmetry
  breaking in three-dimensional gauge theories},''
  \href{http://dx.doi.org/10.1088/1126-6708/1999/10/036}{{\em JHEP} {\bfseries
  10} (1999) 036},
\href{http://arxiv.org/abs/hep-th/9908075}{{\ttfamily arXiv:hep-th/9908075
  [hep-th]}}.

\bibitem{Aharony:1997bh}
O.~Aharony, A.~Hanany, and B.~Kol, ``{Webs of (p,q) five-branes,
  five-dimensional field theories and grid diagrams},''
  \href{http://dx.doi.org/10.1088/1126-6708/1998/01/002}{{\em JHEP} {\bfseries
  01} (1998) 002},
\href{http://arxiv.org/abs/hep-th/9710116}{{\ttfamily arXiv:hep-th/9710116
  [hep-th]}}.

\bibitem{Gulotta:2011si}
D.~R. Gulotta, C.~P. Herzog, and S.~S. Pufu, ``{From Necklace Quivers to the
  F-theorem, Operator Counting, and T(U(N))},''
  \href{http://dx.doi.org/10.1007/JHEP12(2011)077}{{\em JHEP} {\bfseries 12}
  (2011) 077},
\href{http://arxiv.org/abs/1105.2817}{{\ttfamily arXiv:1105.2817 [hep-th]}}.

\bibitem{Assel:2014awa}
B.~Assel, ``{Hanany-Witten effect and SL(2, $\mathbb{Z}$) dualities in matrix
  models},'' \href{http://dx.doi.org/10.1007/JHEP10(2014)117}{{\em JHEP}
  {\bfseries 10} (2014) 117},
\href{http://arxiv.org/abs/1406.5194}{{\ttfamily arXiv:1406.5194 [hep-th]}}.

\bibitem{Garozzo:2018kra}
I.~Garozzo, G.~Lo~Monaco, and N.~Mekareeya, ``{The moduli spaces of $S$-fold
  CFTs},'' \href{http://dx.doi.org/10.1007/JHEP01(2019)046}{{\em JHEP}
  {\bfseries 01} (2019) 046},
\href{http://arxiv.org/abs/1810.12323}{{\ttfamily arXiv:1810.12323 [hep-th]}}.

\bibitem{Gaiotto:2008sd}
D.~Gaiotto and E.~Witten, ``{Janus Configurations, Chern-Simons Couplings, And
  The theta-Angle in N=4 Super Yang-Mills Theory},''
  \href{http://dx.doi.org/10.1007/JHEP06(2010)097}{{\em JHEP} {\bfseries 06}
  (2010) 097},
\href{http://arxiv.org/abs/0804.2907}{{\ttfamily arXiv:0804.2907 [hep-th]}}.

\bibitem{Assel:2018vtq}
B.~Assel and A.~Tomasiello, ``{Holographic duals of 3d S-fold CFTs},''
  \href{http://dx.doi.org/10.1007/JHEP06(2018)019}{{\em JHEP} {\bfseries 06}
  (2018) 019},
\href{http://arxiv.org/abs/1804.06419}{{\ttfamily arXiv:1804.06419 [hep-th]}}.

\bibitem{Garcia-Etxebarria:2015wns}
I.~García-Etxebarria and D.~Regalado, ``{$ \mathcal{N}=3 $ four dimensional
  field theories},'' \href{http://dx.doi.org/10.1007/JHEP03(2016)083}{{\em
  JHEP} {\bfseries 03} (2016) 083},
\href{http://arxiv.org/abs/1512.06434}{{\ttfamily arXiv:1512.06434 [hep-th]}}.

\bibitem{Aharony:2016kai}
O.~Aharony and Y.~Tachikawa, ``{S-folds and 4d N=3 superconformal field
  theories},'' \href{http://dx.doi.org/10.1007/JHEP06(2016)044}{{\em JHEP}
  {\bfseries 06} (2016) 044},
\href{http://arxiv.org/abs/1602.08638}{{\ttfamily arXiv:1602.08638 [hep-th]}}.

\bibitem{Couzens:2017way}
C.~Couzens, C.~Lawrie, D.~Martelli, S.~Schafer-Nameki, and J.-M. Wong,
  ``{F-theory and AdS$_{3}$/CFT$_{2}$},''
  \href{http://dx.doi.org/10.1007/JHEP08(2017)043}{{\em JHEP} {\bfseries 08}
  (2017) 043},
\href{http://arxiv.org/abs/1705.04679}{{\ttfamily arXiv:1705.04679 [hep-th]}}.

\bibitem{Couzens:2017nnr}
C.~Couzens, D.~Martelli, and S.~Schafer-Nameki, ``{F-theory and
  AdS$_{3}$/CFT$_{2}$ (2, 0)},''
  \href{http://dx.doi.org/10.1007/JHEP06(2018)008}{{\em JHEP} {\bfseries 06}
  (2018) 008},
\href{http://arxiv.org/abs/1712.07631}{{\ttfamily arXiv:1712.07631 [hep-th]}}.

\bibitem{DHoker:2007zhm}
E.~D'Hoker, J.~Estes, and M.~Gutperle, ``{Exact half-BPS Type IIB interface
  solutions. I. Local solution and supersymmetric Janus},''
  \href{http://dx.doi.org/10.1088/1126-6708/2007/06/021}{{\em JHEP} {\bfseries
  06} (2007) 021},
\href{http://arxiv.org/abs/0705.0022}{{\ttfamily arXiv:0705.0022 [hep-th]}}.

\bibitem{DHoker:2007hhe}
E.~D'Hoker, J.~Estes, and M.~Gutperle, ``{Exact half-BPS Type IIB interface
  solutions. II. Flux solutions and multi-Janus},''
  \href{http://dx.doi.org/10.1088/1126-6708/2007/06/022}{{\em JHEP} {\bfseries
  06} (2007) 022},
\href{http://arxiv.org/abs/0705.0024}{{\ttfamily arXiv:0705.0024 [hep-th]}}.

\bibitem{Ganor:2014pha}
O.~J. Ganor, N.~P. Moore, H.-Y. Sun, and N.~R. Torres-Chicon, ``{Janus
  configurations with SL(2,$\mathbb{Z}$)-duality twists, strings on mapping
  tori and a tridiagonal determinant formula},''
  \href{http://dx.doi.org/10.1007/JHEP07(2014)010}{{\em JHEP} {\bfseries 07}
  (2014) 010},
\href{http://arxiv.org/abs/1403.2365}{{\ttfamily arXiv:1403.2365 [hep-th]}}.

\bibitem{Inverso:2016eet}
G.~Inverso, H.~Samtleben, and M.~Trigiante, ``{Type II supergravity origin of
  dyonic gaugings},'' \href{http://dx.doi.org/10.1103/PhysRevD.95.066020}{{\em
  Phys. Rev.} {\bfseries D95} no.~6, (2017) 066020},
\href{http://arxiv.org/abs/1612.05123}{{\ttfamily arXiv:1612.05123 [hep-th]}}.

\bibitem{Martucci:2014ema}
L.~Martucci, ``{Topological duality twist and brane instantons in F-theory},''
  \href{http://dx.doi.org/10.1007/JHEP06(2014)180}{{\em JHEP} {\bfseries 06}
  (2014) 180},
\href{http://arxiv.org/abs/1403.2530}{{\ttfamily arXiv:1403.2530 [hep-th]}}.

\bibitem{Gadde:2014wma}
A.~Gadde, S.~Gukov, and P.~Putrov, ``{Duality Defects},''
\href{http://arxiv.org/abs/1404.2929}{{\ttfamily arXiv:1404.2929 [hep-th]}}.

\bibitem{Assel:2016wcr}
B.~Assel and S.~Schäfer-Nameki, ``{Six-dimensional origin of $\mathcal{N} = 4$
  SYM with duality defects},''
  \href{http://dx.doi.org/10.1007/JHEP12(2016)058}{{\em JHEP} {\bfseries 12}
  (2016) 058},
\href{http://arxiv.org/abs/1610.03663}{{\ttfamily arXiv:1610.03663 [hep-th]}}.

\bibitem{Lawrie:2018jut}
C.~Lawrie, D.~Martelli, and S.~Schäfer-Nameki, ``{Theories of Class F and
  Anomalies},'' \href{http://dx.doi.org/10.1007/JHEP10(2018)090}{{\em JHEP}
  {\bfseries 10} (2018) 090},
\href{http://arxiv.org/abs/1806.06066}{{\ttfamily arXiv:1806.06066 [hep-th]}}.

\bibitem{Terashima:2011qi}
Y.~Terashima and M.~Yamazaki, ``{SL(2,R) Chern-Simons, Liouville, and Gauge
  Theory on Duality Walls},''
  \href{http://dx.doi.org/10.1007/JHEP08(2011)135}{{\em JHEP} {\bfseries 08}
  (2011) 135},
\href{http://arxiv.org/abs/1103.5748}{{\ttfamily arXiv:1103.5748 [hep-th]}}.

\bibitem{Gang:2015wya}
D.~Gang, N.~Kim, M.~Romo, and M.~Yamazaki, ``{Aspects of Defects in 3d-3d
  Correspondence},'' \href{http://dx.doi.org/10.1007/JHEP10(2016)062}{{\em
  JHEP} {\bfseries 10} (2016) 062},
\href{http://arxiv.org/abs/1510.05011}{{\ttfamily arXiv:1510.05011 [hep-th]}}.

\bibitem{Cremonesi:2014xha}
S.~Cremonesi, G.~Ferlito, A.~Hanany, and N.~Mekareeya, ``{Coulomb Branch and
  The Moduli Space of Instantons},''
  \href{http://dx.doi.org/10.1007/JHEP12(2014)103}{{\em JHEP} {\bfseries 12}
  (2014) 103},
\href{http://arxiv.org/abs/1408.6835}{{\ttfamily arXiv:1408.6835 [hep-th]}}.

\bibitem{Witten:1998xy}
E.~Witten, ``{Baryons and branes in anti-de Sitter space},''
  \href{http://dx.doi.org/10.1088/1126-6708/1998/07/006}{{\em JHEP} {\bfseries
  07} (1998) 006},
\href{http://arxiv.org/abs/hep-th/9805112}{{\ttfamily arXiv:hep-th/9805112
  [hep-th]}}.

\bibitem{Cremonesi:2014uva}
S.~Cremonesi, A.~Hanany, N.~Mekareeya, and A.~Zaffaroni, ``{T$_{\rho}^{\sigma}$
  (G) theories and their Hilbert series},''
  \href{http://dx.doi.org/10.1007/JHEP01(2015)150}{{\em JHEP} {\bfseries 01}
  (2015) 150},
\href{http://arxiv.org/abs/1410.1548}{{\ttfamily arXiv:1410.1548 [hep-th]}}.

\bibitem{Hanany:2016gbz}
A.~Hanany and R.~Kalveks, ``{Quiver Theories for Moduli Spaces of Classical
  Group Nilpotent Orbits},''
  \href{http://dx.doi.org/10.1007/JHEP06(2016)130}{{\em JHEP} {\bfseries 06}
  (2016) 130},
\href{http://arxiv.org/abs/1601.04020}{{\ttfamily arXiv:1601.04020 [hep-th]}}.

\bibitem{Cabrera:2018ldc}
S.~Cabrera, A.~Hanany, and R.~Kalveks, ``{Quiver Theories and Formulae for
  Slodowy Slices of Classical Algebras},''
\href{http://arxiv.org/abs/1807.02521}{{\ttfamily arXiv:1807.02521 [hep-th]}}.

\bibitem{Hanany:2008sb}
A.~Hanany, N.~Mekareeya, and G.~Torri, ``{The Hilbert Series of Adjoint
  SQCD},'' \href{http://dx.doi.org/10.1016/j.nuclphysb.2009.09.016}{{\em Nucl.
  Phys.} {\bfseries B825} (2010) 52--97},
\href{http://arxiv.org/abs/0812.2315}{{\ttfamily arXiv:0812.2315 [hep-th]}}.

\bibitem{Hanany:2011db}
A.~Hanany and N.~Mekareeya, ``{Complete Intersection Moduli Spaces in N=4 Gauge
  Theories in Three Dimensions},''
  \href{http://dx.doi.org/10.1007/JHEP01(2012)079}{{\em JHEP} {\bfseries 01}
  (2012) 079},
\href{http://arxiv.org/abs/1110.6203}{{\ttfamily arXiv:1110.6203 [hep-th]}}.

\bibitem{Namikawa2018}
Y.~Namikawa, ``A characterization of nilpotent orbit closures among symplectic
  singularities,'' \href{https://doi.org/10.1007/s00208-017-1572-9}{{\em
  Mathematische Annalen} {\bfseries 370} no.~1, (Feb, 2018) 811--818},
  \href{http://arxiv.org/abs/1603.06105}{{\ttfamily arXiv:1603.06105}}.

\bibitem{Cabrera:2016vvv}
S.~Cabrera and A.~Hanany, ``{Branes and the Kraft-Procesi Transition},''
  \href{http://dx.doi.org/10.1007/JHEP11(2016)175}{{\em JHEP} {\bfseries 11}
  (2016) 175},
\href{http://arxiv.org/abs/1609.07798}{{\ttfamily arXiv:1609.07798 [hep-th]}}.

\bibitem{Hanany:2017ooe}
A.~Hanany and R.~Kalveks, ``{Quiver Theories and Formulae for Nilpotent Orbits
  of Exceptional Algebras},''
  \href{http://dx.doi.org/10.1007/JHEP11(2017)126}{{\em JHEP} {\bfseries 11}
  (2017) 126},
\href{http://arxiv.org/abs/1709.05818}{{\ttfamily arXiv:1709.05818 [hep-th]}}.

\bibitem{Giddings:1995ns}
S.~B. Giddings and J.~M. Pierre, ``{Some exact results in supersymmetric
  theories based on exceptional groups},''
  \href{http://dx.doi.org/10.1103/PhysRevD.52.6065}{{\em Phys. Rev.} {\bfseries
  D52} (1995) 6065--6073},
\href{http://arxiv.org/abs/hep-th/9506196}{{\ttfamily arXiv:hep-th/9506196
  [hep-th]}}.

\bibitem{Sen:1998ii}
A.~Sen, ``{Stable nonBPS bound states of BPS D-branes},''
  \href{http://dx.doi.org/10.1088/1126-6708/1998/08/010}{{\em JHEP} {\bfseries
  08} (1998) 010},
\href{http://arxiv.org/abs/hep-th/9805019}{{\ttfamily arXiv:hep-th/9805019
  [hep-th]}}.

\bibitem{Cabrera:2017njm}
S.~Cabrera and A.~Hanany, ``{Branes and the Kraft-Procesi transition: classical
  case},''
\href{http://arxiv.org/abs/1711.02378}{{\ttfamily arXiv:1711.02378 [hep-th]}}.

\bibitem{Assel:2018exy}
B.~Assel and S.~Cremonesi, ``{The Infrared Fixed Points of 3d $\mathcal{N}=4$
  $USp(2N)$ SQCD Theories},''
  \href{http://dx.doi.org/10.21468/SciPostPhys.5.2.015}{{\em SciPost Phys.}
  {\bfseries 5} no.~2, (2018) 015},
\href{http://arxiv.org/abs/1802.04285}{{\ttfamily arXiv:1802.04285 [hep-th]}}.

\bibitem{Seiberg:1996nz}
N.~Seiberg and E.~Witten, ``{Gauge dynamics and compactification to
  three-dimensions},'' in {\em {The mathematical beauty of physics: A memorial
  volume for Claude Itzykson. Proceedings, Conference, Saclay, France, June
  5-7, 1996}}, pp.~333--366.
\newblock 1996.
\newblock
\href{http://arxiv.org/abs/hep-th/9607163}{{\ttfamily arXiv:hep-th/9607163
  [hep-th]}}.
\newblock

\bibitem{Ferlito:2016grh}
G.~Ferlito and A.~Hanany, ``{A tale of two cones: the Higgs Branch of Sp(n)
  theories with 2n flavours},''
\href{http://arxiv.org/abs/1609.06724}{{\ttfamily arXiv:1609.06724 [hep-th]}}.

\bibitem{Cremonesi:2013lqa}
S.~Cremonesi, A.~Hanany, and A.~Zaffaroni, ``{Monopole operators and Hilbert
  series of Coulomb branches of $3d$ $\mathcal{N} = 4$ gauge theories},''
  \href{http://dx.doi.org/10.1007/JHEP01(2014)005}{{\em JHEP} {\bfseries 01}
  (2014) 005},
\href{http://arxiv.org/abs/1309.2657}{{\ttfamily arXiv:1309.2657 [hep-th]}}.

\bibitem{Hanany:2001iy}
A.~Hanany and J.~Troost, ``{Orientifold planes, affine algebras and magnetic
  monopoles},'' \href{http://dx.doi.org/10.1088/1126-6708/2001/08/021}{{\em
  JHEP} {\bfseries 08} (2001) 021},
\href{http://arxiv.org/abs/hep-th/0107153}{{\ttfamily arXiv:hep-th/0107153
  [hep-th]}}.

\end{thebibliography}\endgroup

\end{document}